\documentclass[12pt]{article}
\usepackage{epsf}
\usepackage{feynmf}
\usepackage{subfigure}
\def\la{\mathrel{\mathpalette\fun <}}

\def\fun#1#2{\lower3.6pt\vbox{\baselineskip0pt\lineskip.9pt
\ialign{$\mathsurround=0pt#1\hfil##\hfil$\crcr#2\crcr\sim\crcr}}}

\title{Theory of $Z$ boson decays}
\author{V.A.Novikov$^*$, L.B.Okun
\thanks{E-mail: novikov@heron.itep.ru; okun@heron.itep.ru;
vysotsky@heron.itep.ru}, \\
ITEP, 117218 Moscow, Russia \\
A.N.Rozanov
\thanks{E-mail: rozanov@cppm.in2p3.fr}, \\
ITEP and CPPM, CNRS/IN2P3 - Univ. Mediterran\'ee, Marseille, France \\
M.I.Vysotsky$^*$,
ITEP and INFN, Sezione di Ferrara, Italy
}
\date{\it Accepted in Reports on Progress in Physics
}
\begin{document}
\maketitle

\begin{abstract}

The precision data on $Z$ boson decays from LEP-I and SLC colliders are
compared with the predictions based on the Minimal Standard Theory. The Born
approximation of the theory is based on three most accurately known
observables: $G_{\mu}$ -- the four fermion coupling constant of muon decay,
$m_Z$ -- the mass of the $Z$ boson, and $\alpha(m_Z)$ -- the value of the
``running fine structure constant" at the scale of $m_Z$. The electroweak
loop corrections are expressed, in addition, in terms of the masses of higgs,
$m_H$, of the top and bottom quarks, $m_t$ and $m_b$, and
of the strong interaction constant $\alpha_s(m_Z)$.
The main emphasis of the review is focused on
the one-electroweak-loop approximation. Two electroweak loops have been
calculated in the literature only partly. Possible manifestations of new
physics are briefly discussed.

\end{abstract}

\vspace{5mm}

\begin{center}
{\bf Contents.}
\end{center}

\vspace{3mm}

\begin{enumerate}
\item {\bf Introduction.}
\item {\bf Basic parameters of the theory.}
\item {\bf Amplitudes, widths, and asymmetries.}
\item {\bf One-loop corrections to hadronless
observables.} \\
\hspace{2cm}  4.1 Four types of Feynman diagrams. \\
\hspace{2cm} 4.2 The asymptotic limit
 at $m_t^2 \gg m_Z^2$. \\
\hspace{2cm} 4.3  The functions $V_m(t,h)$, $V_A(t,h)$ and
 $V_R(t,h)$.\\
\hspace{2cm}  4.4   Corrections $\delta V_i(t)$.\\
\hspace{2cm}  4.5  Accidental (?)
 compensation and the mass of the $t$-quark.\\
\hspace{2cm} 4.6   How to calculate
 $V_i$? `Five steps'.
   \item {\bf One-loop corrections to hadronic decays of the $Z$
boson.}\\
\hspace{2cm} 5.1 The leading quarks and hadrons.\\
\hspace{2cm} 5.2  Decays to pairs of light quarks.\\
\hspace{2cm} 5.3  Decays to $b\bar b$ pair.
  \item {\bf Comparison of one-electroweak-loop results
  and  experimental LEP-I and SLC data.} \\
\hspace{2cm} 6.1   LEPTOP code. \\
\hspace{2cm} 6.2   One-loop general fit.
   \item {\bf Two-loop electroweak corrections and
  theoretical uncertainties.}\\
   \hspace{2cm} 7.1  $\alpha_W^2 t^2$ corrections to
$m_W/m_Z$ , $ g_A$ and
$g_V/g_A$ from reducible diagrams.\\
  \hspace{2cm} 7.2  $\alpha^2_W t^2$ corrections from
  irreducible diagrams.\\
  \hspace{2cm} 7.3 $\alpha^2_Wt$ corrections  and the two-loop fit
  of experimental data.\\
  \item {\bf Extensions of the Standard Model.}\\
  \hspace{2cm}  8.1 Sequential heavy generations in the Standard Model.\\
  \hspace{2cm}  8.2  SUSY extensions of the Standard Model.\\
  \item {\bf  Conclusions.}
\end{enumerate}
\noindent
{ \bf
Appendix A. Regularization of  Feynman integrals.} \\
\\
{\bf  Appendix B. Relation between $\bar{\alpha}$ and
$\alpha(0)$.} \\
\\
{\bf Appendix C.  How $\alpha_W(q^2)$ and
$\alpha_Z(q^2)$ `crawl'.} \\
\\
{\bf Appendix D. General expressions for one-loop corrections to
hadronless observables.} \\
\\
{\bf Appendix E.
 Radiators $R_{Aq}$ and $R_{Vq}$.} \\
\\
{\bf Appendix F.
$\alpha_W^2 t^2$ corrections from reducible diagrams.} \\
\\
{\bf Appendix G.  Oblique corrections from new generations and SUSY.} \\
\\
{\bf Appendix H. Other parametrizations of   radiative
 corrections.} \\
\\
{\bf References.} \\
\\
{ \bf Figure captions.}

\newpage

\section{Introduction}

$Z$ boson, electrically neutral vector boson (its spin equals 1) with mass
$m_Z \simeq 91$ GeV and width $\Gamma_Z\simeq 2.5$ GeV, \footnote{Throughout
the paper we use units in which $\hbar, c =1$.} occupies a unique place in
physics.  This heavy analog of the photon was experimentally discovered in
1983, practically simultaneously with its charged counterparts $W^{\pm}$
bosons with mass $m_W\simeq 80$ GeV and width $\Gamma_W\simeq 2$ GeV \cite{1}.

The discovery was crowned by Nobel Prize to Carlo Rubbia (for the bosons)
and to Simon van der Meer (for the CERN proton-antiproton collider, which was
specially constructed to produce $W$ and $Z$ bosons) \cite{2}.

The extremely short-lived vector bosons ($\tau = 1/\Gamma \simeq 10^{-25}$
sec) were detected by their decays into various leptons and hadrons. The
detectors, in which these decay products were observed, were built
and operated by
collaborations of physicists and engineers the largest in the previous
history of physics.

The discovery of $W$ and $Z$ bosons was a great triumph of experimental
physics, but even more so of theoretical physics. The masses and widths of
the particles, the cross-sections of their production turned out to be in
perfect agreement with the predictions of electroweak theory of Sheldon
Glashow, Abdus Salam and Steven Weinberg \cite{3}. The theory was so
beautiful that its authors received the Nobel Prize already in 1979 \cite{4},
four years before its crucial confirmation.

The electroweak theory unified two types of fundamental interactions:
electromagnetic and weak. The theory of electromagnetic interaction --
quantum electrodynamics, or QED, was cast in its present relativistically
covariant form in the late 1940's and early 1950's and served as a ``role
model" for the relativistic field theories of two other fundamental
interactions: weak and strong.

The main virtue of QED was its renormalizability. Let us explain this
``technical" term by using the example of interaction of photons with
electrons. One can find systematic presentation in modern textbooks
\cite{444}. In the lowest approximation of perturbation theory (the so called
tree approximation in the language of Feynman diagrams) all electromagnetic
phenomena can be described in terms of electric charge and mass of the
electron ($e$, $m$). The small parameter of perturbation theory is the famous
$\alpha = e^2/4\pi \simeq 1/137$.

The problem with any quantum field theory is that in higher orders of
perturbation theory, described by Feynman graphs with loops, the integrals
over momenta of virtual particles have ultraviolet divergences, so that all
physical quantities including electric charge and mass of the electron
themselves become infinitely large. To avoid infinities an ultraviolet cut-off
$\Lambda$ could be introduced. Another, more sophisticated method is to use
dimensional regularization: to calculate the Feynman integrals in momentum
space of $D$ dimensions. These integrals diverge at $D=4$, but are finite,
proportional to $1/\varepsilon$ in vicinity of $D=4$, where by definition
$2\varepsilon = 4-D \to 0$ (see Appendix A).

The theory is called renormalizable, if one can get rid of this cut-off
(or $1/\varepsilon$) by
establishing relations between observables only. In the case of electrons and
photons such basic observables are physical (renormalized) charge and mass of
the electron. This allows to calculate higher order effects in $\alpha$ and
compare the theoretical predictions with the results of high precision
measurements of such observables as e.g. anomalous magnetic moments of
electron or muon.

The renormalizability of electrodynamics is guaranteed by the dimensionless
nature of the coupling constant $\alpha$ and by conservation of electric
current.

After this short description of QED let us turn to the weak interaction.

The first manifestation of the weak interaction was discovered by Henri
Becquerel at the end of the XIX century. Later this type of radioactivity was
called $\beta$-decay.
The first theory of $\beta$-decay was proposed by
Enrico Fermi in 1934 \cite{5}.
The theory was modelled after quantum electrodynamics with two major
differences: first, instead of charge conserving, ``neutral", electrical
current of the type $-\bar{e}\gamma_{\alpha}e + \bar{p}\gamma_{\alpha}p$ there
were introduced two charge changing, ``charged", vector currents: one for
nucleons, transforming neutron into a proton, $\bar{p}\gamma_{\alpha}n$,
another for leptons, transforming neutrino into electron or creating a pair:
electron plus antineutrino, $\bar{e}\gamma_{\alpha}\nu$. (Here $\bar{e}(e)$
denotes operator, which creates (annihilates) electron and annihilates
(creates) positron. The symbols of other particles have similar meaning;
$\gamma_{\alpha}$ are four Dirac matrices, $\alpha = 0,1,2,3$.)

The second difference between the Fermi theory and electrodynamics was that
the charged currents interacted locally via four fermion interaction:
\begin{equation}
G \cdot \bar{p}\gamma_{\alpha} n \cdot \bar{e}\gamma_{\alpha}\nu + \mbox{\rm
h.c.} \;\; ,
\label{1}
\end{equation}
where summation over index $\alpha$ is implied
(in this summation we use Feynman's convention: $+$ for $\alpha =0$ and $-$
for $\alpha = 1,2,3$); h.c. means hermitian
conjugate. The coupling constant $G$ of this interaction was called Fermi
coupling constant.

From simple dimensional considerations it is evident that dimension of $G$ is
(mass)$^{-2}$ and therefore the four-fermion interaction is not
renormalizable, the higher orders being divergent as $G^2 \Lambda^2$,
$G^3 \Lambda^4$ ... . Why
these divergent corrections still allow one to rely on the lowest order
approximation, remained a mystery. But for many years the lowest order
four-fermion interaction served as a successful phenomenological theory of
weak interactions.

It is in the framework of this phenomenological theory that a number of
subsequent experimental discoveries were accommodated. First, it turned out
that $\beta$-decay is one of the large family of weak processes, involving
newly discovered particles, such as pions, muons and muonic neutrinos,
strange particles, etc. Second, it was discovered that all these processes
are caused by selfinteraction of one weak charged current, involving
leptonic and hadronic terms. Later on, when the quark structure of hadrons
was established, the hadronic part of the current was expressed through
corresponding quark current. Third, it was established in 1957 that all weak
interactions violate parity conservation $P$ and charge conjugation
invariance $C$. This violation turned out to have a universal pattern: the
vector form of the current $V$, introduced by Fermi, was substituted
\cite{6} by one half of
the sum of vector and axial vector, $A$, which meant that
$\gamma_{\alpha}$ should be substituted by
$\frac{1}{2}\gamma_{\alpha}(1+\gamma_5)$.

In other words one can say that fermion $\psi$ enters the charged current
only through its left-handed chiral component
\begin{equation}
\psi_L =\frac{1}{2}(1+\gamma_5)\psi
\label{2}
\end{equation}

From such structure of the charged current it follows that the corresponding
antifermions interact only through their right-handed components.

Attempts to construct a renormalizable theory of weak interaction
resulted in a unified theory of electromagnetic and weak interactions -- the
electroweak theory \cite{3,4} with two major predictions. The first
prediction was the existence alongside with the charged weak current also of
a neutral weak current.
The second prediction was the
existence of the vector bosons:  $Z$ coupled to the weak neutral current and
$W^+$ and $W^-$ coupled to the charged current (like
$\frac{1}{2}\bar{e}\gamma_{\alpha}(1+\gamma_5)\nu$) and its hermitian
conjugate current (like $\frac{1}{2}\bar{\nu}\gamma_{\alpha}(1+\gamma_5)e$).

The vector bosons were massive analogues of the photon $\gamma$; their
couplings to the corresponding currents, $f$ and $g$, were the analogues
of the electric charge $e$. Thus $\alpha_Z =f^2/4\pi$ and $\alpha_W
=g^2/4\pi$ were dimensionless like $\alpha =e^2/4\pi$, which was a
necessary (but not sufficient) condition of renormalizability of the weak
interaction.

The first theory, involving charged vector bosons and photon, was proposed by
Oscar Klein just before World War II \cite{7}. Klein based his theory on the
notion of local isotopic symmetry: he considered isotopic doublets $(p,n)$
and $(\nu, e)$, and isotopic triplet $(B^+, A^0, B^-)$. Where $B^{\pm}$
denoted what we call now $W^{\pm}$, while $A^0$ was electromagnetic field. He
also mentioned the possibility to incorporate a neutral massive field $C^0$
(the analogue of our $Z^0$). In fact this was the first attempt to construct
a theory based on a non-abelian gauge symmetry, with vector fields playing the
role of gauge fields. The gauge symmetry was essential for conservation of
the currents. Unfortunately, Klein did not discriminate between weak and
strong interaction and his paper was firmly forgotten.

The non-abelian gauge theory was rediscovered in 1954 by C.N.Yang and R.Mills
\cite{8} and became the basis of the so-called Standard Model with its colour
$SU(3)_c$ group for strong interaction of quarks and gluons and $SU(2)_L
\times U(1)_Y$ group for electroweak interaction (here indices denote: $c$ --
colour, $L$ -- weak isospin of left-handed spinors, and $Y$ -- the weak
hypercharge). The electric charge $Q=T_3 +Y/2$, where $T_3$ is the third
projection of isospin. $Y=1/3$ for a doublet of quarks, $Y=-1$ for a
doublet of leptons. As for the right-handed spinors, they are isosinglets,
and hence
$$
Y(\nu_R) =0 \; , \;\; Y(e_R)=-2 \; , \;\; Y(u_R)=4/3 \; , \;\; Y(d_R)=-2/3
\;\; .
$$
Thus, parity violation and charge conjugation violation were incorporated
into the foundation of electroweak theory.

Out of the four fields (three of $SU(2)$ and one of $U(1)$, usually denoted
by $W^+, W^0, W^-$ and $B^0$, respectively) only two directly correspond to
the observed vector bosons: $W^+$ and $W^-$. The $Z^0$ boson and photon are
represented by two orthogonal superpositions of $W^0$ and $B^0$:
\begin{eqnarray}
Z^0 &=& c W^0 - s B^0 \nonumber \\
&&~~~ \\
A^0 &=& s W^0 + c B^0 \;\; , \nonumber
\label{3}
\end{eqnarray}
where $c=\cos\theta$, $s=\sin\theta$, while
the weak angle $\theta$ is a free parameter of electroweak theory.
The value of $\theta$ is determined from experimental data on $Z$ boson
coupling to neutral current. The ``$Z$-charge", characterizing the coupling
of $Z$ boson to a spinor with definite helicity is given by
\begin{equation}
\bar{f}(T_3 -Q s^2) \;\; ,
\label{4}
\end{equation}
where
\footnote{We denote by $\bar{e}, \bar{f},
\bar{g}$ the values of the corresponding charges at $m_Z$-scale, while
$e$, $f$, $g$
refer to values at vanishing momentum transfer. The same refers to
$\bar{\alpha}$, $\bar{\alpha}_Z$, $\bar{\alpha}_W$ and $\alpha$, $\alpha_Z$,
$\alpha_W$ (see below Eqs. (\ref{13}) - (\ref{18})).}
\begin{equation}
\bar{f} = \bar{g}/c \;\; .
\label{5}
\end{equation}

Note that the ``$Z$-charge" is different for the right- and left-handed
spinors with the same value of $Q$ because they have different values of
$T_3$. The coupling constant of $W$-bosons is also expressed in terms of
$\bar{e}$ and $\theta$:
\begin{equation}
\bar{g} =\bar{e}/s \;\; .
\label{6}
\end{equation}

The theory described above has many nice features, the most important of
which is its renormalizability. But at first sight it looks absolutely
useless: all fermions and bosons in it are massless. This drawback cannot be
fixed by simply adding mass terms to the Lagrangian. The mass terms of
fermions would contain both $\psi_L$ and $\psi_R$ and thus explicitly break
the isotopic invariance and hence renormalizability. The gauge invariance
would also be broken by the mass terms of the vector bosons. All this would
result in divergences of the type $\Lambda^2/m^2$, $\Lambda^4/m^4$, etc.

The way out of this trap is the so called Higgs mechanism \cite{9}. In the
framework of the Minimal Standard Model the problem of mass is solved by
postulating the existence of the doublet $\varphi_H = (\varphi^+, \varphi^0)$
and corresponding antidoublet $(\bar{\varphi}^0, -\varphi^-)$ of spinless
particles. These four bosons differ from all other particles by the form of
their self-interaction, the energy of which is minimal when the neutral field
$\varphi_1 =\frac{1}{\sqrt{2}}(\varphi^0 +\bar{\varphi}^0)$ has a
nonvanishing vacuum expectation value. The isospin of the Higgs doublet is
$1/2$, its hypercharge is $1$. Thus, it interacts with all four gauge
bosons. In particular, it has quartic terms
$\frac{1}{4}\bar{g}^2\bar{W}W\varphi_1^2$, ~~~
$\frac{1}{8}\bar{f}^2\bar{Z}Z\varphi_1^2$, which give masses to the vector
bosons when $\varphi_1$ acquires its vacuum expectation value ({\it vev})
$\eta$:
\begin{equation}
m_W =\bar{g} \eta/2 \; , \;\; m_Z =\bar{f}\eta/2 \;\; .
\label{7}
\end{equation}
The magnitude of $\eta$ can easily be
derived from that of the four-fermion interaction
constant $G_{\mu}$ in muon decay:
\begin{equation}
\frac{G_{\mu}}{\sqrt{2}}\cdot \bar{\nu}_{\mu}\gamma_{\alpha}(1+\gamma_5)\mu
\cdot \bar{e}\gamma_{\alpha}(1+\gamma_5)\nu_e \;\; .
\label{8}
\end{equation}
In the Born approximation of electroweak theory this four-fermion interaction
is caused by an exchange of a virtual $W$-boson (see Figure 1). Hence
\footnote{More on electroweak Born approximation, for which equality
$g=\bar{g}$ holds, see below, Eqs. (\ref{14}) - (\ref{25}).}
\begin{equation}
\frac{G_{\mu}}{\sqrt{2}}=\frac{g^2}{8m_W^2}=\frac{\bar{g}^2}{8m_W^2} \; ,
\;\; \eta =(\sqrt{2}G_{\mu})^{-1/2} = 246 \; \mbox{\rm GeV} \;\; .
\label{9}
\end{equation}

Such mechanism of appearance of masses of $W$ and $Z$ bosons is called
spontaneous symmetry breaking. It preserves renormalizability
\cite{9a}.
(As a hint, one can use the symmetrical form of Lagrangian by not
 specifying the
{\it vev} $\eta$.)

The fermion masses can be introduced also without explicitly breaking the
gauge symmetry. In this case the mass arises from an isotopically invariant
term $f_Y\cdot \varphi_H\bar{\psi}_L\psi_R +{\rm h.c.}$, where $f_Y$ is called
Yukawa coupling. The mass of a fermion $m=f_Y\eta/\sqrt{2}$. There is a
separate Yukawa coupling for each of the known fermions. Their largely varying
values are at present free parameters of the theory and await for the
understanding of this hierarchy.

Let us return for a moment to the vector bosons. A massless vector boson
(e.g. photon) has two spin degrees of freedom -- two helicity states. A
massive vector boson has three spin degrees of freedom corresponding, say,
projections $\pm 1, 0$ on its momentum. Under spontaneous symmetry breaking
three out of four spinless states, $\varphi^{\pm}, \varphi_2^0
=\frac{1}{\sqrt{2}}(\varphi^0 -\bar{\varphi}^0)$ become third components of
the massive vector bosons. Thus, in the Minimal Standard Model there must
exist only one extra particle: a neutral Higgs scalar boson, or simply,
higgs,
\footnote{Throughout this review we consistently use
capital ``H" in such terms as ``Higgs mechanism", ``Higgs boson", ``Higgs
doublet", but low case ``h" of higgs is used as a name of a particle, not a
man.}
representing a quantum of excitation of field $\varphi_1^0$ over its {\it
vev} $\eta$. The discovery of this particle is crucial for testing the
correctness of MSM.

The first successful test of electroweak theory was provided by the discovery
of neutral currents in the interaction of neutrinos with nucleons \cite{10}.
Further study of this deep inelastic scattering (DIS) allowed to extract the
rough value of the $\sin\theta$: $s^2\simeq 0.23$ and thus to
predict the values of $m_W \simeq 80$ GeV and $m_Z \simeq 90$ GeV, which
served as leading lights for the discovery of these particles.

A few other neutral current interactions have been discovered and studied:
neutrino-electron scattering \cite{111}, parity violating electron-nucleon
scattering at high energies \cite{112}, parity violation in atoms \cite{113}.
All of them turned out to be in agreement with electroweak theory. A
predominant part of theoretical work on electroweak corrections prior to the
discovery of the $W$ and $Z$ bosons was devoted to calculating the
neutrino-electron \cite{114} and especially nucleon-electron \cite{115}
interaction cross sections.

After the discovery of $W$ and $Z$ bosons it became evident that the next
level in the study of electroweak physics must consist in precision
measurements of production and decays of $Z$ bosons in order to test the
electroweak radiative correction. For such measurements special
electron-positron colliders SLC (at SLAC) and LEP-I (at CERN) were constructed
and started to operate in the fall of 1989. SLC had one intersection point of
colliding beams and hence one detector (SLD); LEP-I had four intersection
points and four detectors: ALEPH, DELPHI, L3, OPAL.

In connection with the construction of LEP and SLC,
a number of teams of theorists carried out detailed calculations
of the required radiative corrections. These calculations were
discussed and compared at special workshops and meetings. The
result of this work was the publication of two so-called `CERN
yellow reports' \cite{116}, \cite{117}, which, together with
the `yellow report' \cite{118}, became the `must' books for
experimentalists and theoreticians studying the $Z$ boson.
The book \cite{119} (which should be published in 1999) summarizes results of
theoretical studies.

More than 2.000 experimentalists and engineers and hundreds of theorists
participated in this unique collective quest for truth!

The sum of energies of $e^+ +e^-$ was chosen to be equal to the $Z$ boson
mass. LEP-I was terminated in the fall of 1995 in order to give place to
LEP-II,
which will operate in the same tunnel till 2001 with maximal energy $200$
GeV. SLC continued at energy close to $91$ GeV.

The reactions, which has been studied at LEP-I and SLC may be presented in
the form (see Figure 2) :
\begin{equation}
e^+ e^- \to Z \to f\bar{f} \;\; ,
\label{10}
\end{equation}
where
$$
\begin{array}{ll}
f\bar{f} =& \nu\bar{\nu} (\nu_e\bar{\nu}_e, \nu_{\mu}\bar{\nu}_{\mu},
\nu_{\tau}\bar{\nu}_{\tau})  \;\;
\mbox{\rm -- ~ invisible} \;\; , \\
 & l\bar{l}(e\bar{e}, \mu\bar{\mu}, \tau\bar{\tau}) \;\;
\mbox{\rm -- ~ charged~ leptons} \;\; , \\
 & q\bar{q}(u\bar{u}, d\bar{d}, s\bar{s}, c\bar{c}, b\bar{b}) \;\;
\mbox{\rm -- ~~ hadrons} \;\; .
\end{array}
$$

About 20,000,000 $Z$ bosons have been detected at LEP-I and 550
thousands at SLC (but here electrons are polarized, which compensates the
lower number of events).

Experimental data from all five detectors were summarized and analyzed by the
LEP Electroweak Working Group and the SLD Heavy Flavour and Electroweak
Groups which prepared a special report ``A Combination of Preliminary
Electroweak Measurements and Constraints on the Standard Model" \cite{11}.
These data were analyzed in \cite{11} by using
ZFITTER code (see below section 6.1) and independently by J.Erler and
P.Langacker \cite{erler}.

Fantastic precision has been reached in the
measurement of the $Z$ boson mass and width \cite{11}:
\begin{equation}
m_Z = 91,186.7(2.1) \; \mbox {\rm MeV}\; , \;\;
\Gamma_Z = 2,493.9 \pm 2.4 \; \mbox{\rm MeV} \;\; .
\label{11}
\end{equation}
Of special interest is the measurement of the width of invisible decays of
$Z$:
\begin{equation}
\Gamma_{\mbox{\rm invisible}} = 500.1 \pm 1.9 \;\; \mbox{\rm MeV} \;\; .
\label{12}
\end{equation}
By comparing this number with theoretical predictions for neutrino decays it
was established that the number of neutrinos which interact with $Z$ boson is
three ($N_{\nu} = 2.994 \pm 0.011$). This is a result of fundamental
importance. It means that there exist only 3 standard families (or
generations) of leptons and quarks \footnote{Combining eq.(\ref{12}) with
the data on $\nu_{\mu}e^-$ \cite{12a} and $\nu_e e^-$ \cite{12b} scattering
allowed to establish that $\nu_e$, $\nu_{\mu}$ and $\nu_{\tau}$ have equal
values of couplings with $Z$ boson \cite{11c}.}. Extra families (if they
exist) must have either very heavy neutrinos ($m_N > m_Z/2$), or no neutrinos
at all.

This review is devoted to the description of the theory of electroweak
radiative corrections in $Z$ boson decays and to their comparison with
experimental data \cite{11}.  Our approach to the theory of electroweak
corrections differs somewhat from that used in
\cite{116}-\cite{erler}.
We believe that it is simpler and more transparent (see Section 6.1).
In Section 2 we introduce the basic input parameters of
the electroweak Born approximation. In Section 3 we present phenomenological
formulas for amplitudes, decay widths, and asymmetries of numerous decay
channels of $Z$ bosons. The main subject of our review is the calculation of
one-electroweak loop radiative corrections to the Born approximation. In
Section 4 they are calculated to the hadronless decays and the mass of the
$W$ boson, while in Section 5 -- to the hadronic decays. In Section 6 the
results of the one-electroweak loop calculations are compared with the
experimental data. Section 7 gives a sketch of two-electroweak loop
corrections and of their influence on the fit of experimental data. Section 8
discusses possible manifestations of new physics (extra generations of
fermions and supersymmetry). Section 9 contains conclusions.

In order to make easier the reading of the main text, technical details and
derivations are collected in Appendices.

\section{Basic parameters of the theory.}

The first step in the theoretical analysis is to separate genuinely
electroweak effects from purely electromagnetic ones, such as real photons
emitted by initial and final particles in reaction (\ref{10}) and virtual
photons emitted and absorbed by them. The electroweak quantities extracted
in this way are called sometimes \cite{118} pseudoobservables, but for the
sake of brevity we will refer to them as observables.

A key role among purely electromagnetic effects is played by a phenomenon
which is called the running of electromagnetic coupling ``constant"
$\alpha(q^2)$.  The dependence of the electric charge on the square of the
four-momentum transfer $q^2$ is caused by the photon polarization of vacuum,
i.e. by loops of charged leptons and quarks (hadrons)
 (see Figure 3(a)).

As is well known (see e.g. \cite{13})
\begin{equation}
\alpha\equiv \alpha(q^2 =0) =[137.035985(61)]^{-1} \;\; .
\label{13}
\end{equation}
It has a very high accuracy and is very important in the theory of
electromagnetic processes at low energies. As for electroweak processes in
general and $Z$ decays in particular, they are determined by \cite{11}
\begin{equation}
\bar{\alpha}\equiv \alpha(q^2 =m_Z^2) =[128.878(90)]^{-1} \;\; ,
\label{14}
\end{equation}
the accuracy of which is much worse.

It is convenient to denote
\begin{equation}
\bar{\alpha} =\frac{\alpha}{1-\delta\alpha} \;\; ,
\label{15}
\end{equation}
where
\begin{equation}
\delta\alpha =\delta\alpha_l +\delta\alpha_h = 0.031498(0)
+0.05940(66) \;\;
\label{16}
\end{equation}
(for the value of $\delta\alpha_l$ see ref. \cite{17a}, while for the value
of $\delta\alpha_h$ see \cite{17b}).

It is obvious that the uncertainty of $\delta\alpha$ and hence of
$\bar{\alpha}$ stems from that of hadronic contribution $\delta\alpha_h$.

While $\alpha(q^2)$ is running electromagnetically fast, $\alpha_Z(q^2)$ and
$\alpha_W(q^2)$ are ``crawling" electroweakly slow for $q^2 \la m_Z^2$:
\begin{equation}
\alpha_Z \equiv \alpha_Z(0) =1/23.10 \; , \;\;
\bar{\alpha}_Z \equiv \alpha_Z(m_Z^2) = 1/22.91
\label{17}
\end{equation}
\begin{equation}
\alpha_W \equiv \alpha_W(0) =1/29.01 \; , \;\;
\bar{\alpha}_W \equiv \alpha_W(m_Z^2) = 1/28.74 \;\; .
\label{18}
\end{equation}

The small differences $\alpha_Z -\bar{\alpha}_Z$ and $\alpha_W
-\bar{\alpha}_W$ are caused by electroweak radiative corrections. Therefore
one could and should neglect them when defining the electroweak Born
approximation. (We used this recipe ($g=\bar{g}$) when deriving the
relation (\ref{9}) between $G_{\mu}$ and $\eta$.)

The theoretical analysis of electroweak effects in this article is based on
three most accurately known parameters $G_{\mu}$, $\bar{\alpha}$
(Eq.(\ref{14})) and $m_Z$ (Eq.(\ref{11})).
\begin{equation}
G_{\mu} = 1.16639(1) \cdot 10^{-5} \; \mbox{\rm GeV}^{-2} \;\; .
\label{19}
\end{equation}

This value of $G_{\mu}$ \cite{13} is extracted from the muon life-time after
taking into account the purely electromagnetic corrections (including
bremsstrahlung) and kinematical factors \cite{14}:
\begin{equation}
\frac{1}{\tau_{\mu}} \equiv \Gamma_{\mu} =\frac{G_{\mu}^2
m_{\mu}^5}{192\pi^3} f(\frac{m_e^2}{m_{\mu}^2}) [1-\frac{\alpha(m_{\mu})}{2\pi} (\pi^2
-\frac{25}{4})] \;\; ,
\label{20}
\end{equation}
where
$$
f(x) = 1-8x +8x^3 -x^4 -12x^2 \log x \;\; ,
$$
and
$$
\alpha(m_{\mu})^{-1} =\alpha^{-1}
-\frac{2}{3\pi}\log(\frac{m_{\mu}}{m_e})+\frac{1}{6\pi} \approx 136 \;\; .
$$

Now we are ready to express the weak angle $\theta$ in terms of $G_{\mu}$,
$\bar{\alpha}$ and $m_Z$. Starting from Eqs.(\ref{9}), (\ref{5}) and
(\ref{6}), we get in the electroweak Born approximation:
\begin{equation}
G_{\mu} =\frac{\bar{g}^2}{4\sqrt{2}m_W^2} =\frac{\bar{f}^2}{4\sqrt{2}m_Z^2}
=\frac{\pi\bar{\alpha}}{\sqrt{2}m_Z^2 s^2 c^2}
\label{21}
\end{equation}
Wherefrom:
\begin{equation}
\bar{f}^2 =4\sqrt{2} G_{\mu} m_Z^2 = 0.54863(3) \;\; ,
\label{22}
\end{equation}
$$
\bar{f} = 0.74070(2)
$$
\begin{equation}
\sin^2 2\theta = 4\pi\bar{\alpha}/\sqrt{2} G_{\mu} m_Z^2 = 0.71090(50) \;\; ,
\label{23}
\end{equation}
\begin{equation}
s^2 = 0.23116(23) \;\; ,
\label{24}
\end{equation}
\begin{equation}
c = 0.87683(13) \;\; .
\label{25}
\end{equation}

The angle $\theta$ was introduced in mid-1980s \cite{15}. However its
consistent use began only in the 1990s \cite{16}. Using $\theta$
automatically takes into account the running of $\alpha(q^2)$ and makes it
possible to concentrate on genuinely electroweak corrections as will be
demonstrated below.

( In this review we consistently use $m_Z$ as defined by EWWG in accord
 with our eq. (\ref{E.5}). 
Note that a different definition of the $Z$ boson mass $\overline{m_Z}$ is
known in the literature, related to a different parametrization of the shape
of the $Z$ boson peak \cite{232}.
)

The introduction of the Born approximation described above differs from the
traditional approach in which $\bar{\alpha}-\alpha$ is treated as the largest
electroweak correction, masses $m_W$ and $m_Z$ are handled on an equal
footing, and the angle $\theta_W$, defined by
\begin{equation}
c_W \equiv \cos\theta_W = m_W/m_Z \; , \;\;
s_W^2 = 1- c_W^2 \;\; ,
\label{26}
\end{equation}
is considered as one of the basic parameters of the theory. (Note that the
experimental accuracy of $\theta_W$ is much worse than that of $\theta$.)

After discussing our approach and its main parameters we are prepared to
consider various decays of $Z$ bosons.

\section{Amplitudes, widths, and asymmetries.}

Phenomenologically the amplitude of the $Z$ boson decay into a
fermion-antifermion pair $f\bar{f}$ can be presented in the form:
\begin{equation}
M(Z\to f\bar{f})
=\frac{1}{2}\bar{f}\bar{\psi}_f(g_{Vf}\gamma_{\alpha}+
g_{Af}\gamma_{\alpha}\gamma_5)\psi_f Z_{\alpha} \;\; ,
\label{27}
\end{equation}
where coefficient $\bar{f}$ is given by Eq.(\ref{22}).\footnote{$Z$ boson
couplings are diagonal in flavor unlike that of $W$ boson, where
Cabibbo-Kobayashi-Maskawa  mixing matrix \cite{233} should be accounted for
in the case of couplings with quarks.}
In the case of neutrino
decay channel there is no final state interaction or bremsstrahlung.
Therefore the width into any pair of neutrinos is given by
\begin{equation}
\Gamma_{\nu} \equiv \Gamma(Z\to \nu\bar{\nu}) =4\Gamma_0(g^2_{A\nu}
+g^2_{V\nu}) = 8\Gamma_0 g^2_{\nu} \;\; ,
\label{28}
\end{equation}
where neutrino masses are assumed to be negligible, and $\Gamma_0$ is the
so-called standard width:
\begin{equation}
\Gamma_0 = \frac{\bar{f}^2 m_Z}{192\pi} = \frac{G_{\mu}m_Z^3}{24\sqrt{2}\pi} =
82.940(6) \; \mbox{\rm MeV} \;\; .
\label{29}
\end{equation}

For decays to any of the pairs of charged leptons $l\bar{l}$ we have:
\begin{equation}
\Gamma_l \equiv \Gamma(Z\to l\bar{l}) =4\Gamma_0\left[g_{Vl}^2
(1+\frac{3\bar{\alpha}}{4\pi})+g_{Al}^2(1+\frac{3\bar{\alpha}}{4\pi}
-6\frac{m_l^2}{m_Z^2})\right] \;\; .
\label{30}
\end{equation}

The QED ``radiator" ($1+3\bar{\alpha}/4\pi$) is due to bremsstrahlung of real
photons and emission and absorption of virtual photons by $l$ and $\bar{l}$.
Note that it is expressed not through $\alpha$, but through $\bar{\alpha}$.

For the decays to any of the five pairs of quarks $q\bar{q}$ we have
\begin{equation}
\Gamma_q \equiv\Gamma(Z\to q\bar{q}) = 12\Gamma_0[g_{Aq}^2 R_{Aq} +g_{Vq}^2
R_{Vq}] \;\; .
\label{31}
\end{equation}
Here an extra factor of 3 in comparison with leptons takes into account the
three colours of each quark. The radiators $R_{Aq}$ and $R_{Vq}$ contain
contributions from the final state gluons and photons. In the crudest
approximation
\begin{equation}
R_{Vq} =R_{Aq} = 1+\frac{\hat{\alpha}_s}{\pi} \;\; ,
\label{32}
\end{equation}
where $\alpha_s(q^2)$ is the QCD running coupling constant:
\begin{equation}
\hat{\alpha}_s \equiv \alpha_s(q^2 =m_Z^2) \simeq 0.12 \;\; .
\label{33}
\end{equation}
(For additional details on $\hat{\alpha}_s$ and radiators see Appendix E.)

The full hadron width (to the accuracy of very small corrections,
see Sections 5 and 7) is the sum
of widths of five quark channels:
\begin{equation}
\Gamma_h =\Gamma_u +\Gamma_d +\Gamma_s +\Gamma_c +\Gamma_b \;\; .
\label{34}
\end{equation}

The total width of the $Z$ boson:
\begin{equation}
\Gamma_Z = \Gamma_h +\Gamma_e +\Gamma_{\mu} +\Gamma_{\tau} +3\Gamma_{\nu}
\;\; .
\label{35}
\end{equation}

The cross section of annihilation of $e^+ e^-$ into hadrons at the $Z$ peak
is given by the Breit-Wigner formula
\begin{equation}
\sigma_h =\frac{12{\pi}}{m_Z^2} \frac{\Gamma_e \Gamma_h}{\Gamma_Z^2} \;\; .
\label{36}
\end{equation}

Finally the following notations for the ratio of partial widths are widely
used:
\begin{equation}
R_b =\frac{\Gamma_b}{\Gamma_h} \; , \;\; R_c =\frac{\Gamma_c}{\Gamma_h} \; ,
\;\; R_l = \frac{\Gamma_h}{\Gamma_l} \;\; .
\label{37}
\end{equation}
(Note that $\Gamma_l$ in the numerator of $R_l$ refers to a single charged
lepton channel, whose lepton mass is neglected.)

Parity violating interference of $g_{Af}$ and $g_{Vf}$ leads to a number of
effects: forward-backward asymmetries $A_{FB}$, longitudinal polarization
of $\tau$-lepton $P_{\tau}$, dependence of the total cross-section
at $Z$-peak on the longitudinal
polarization of the initial electron beam $A_{LR}$, etc. Let us define for
the channels of charged lepton and light quark ($u, d, s, c$) whose mass may
be neglected the quantity
\begin{equation}
A_f = \frac{2g_{Af}g_{Vf}}{g_{Af}^2 + g_{Vf}^2} \;\; .
\label{38}
\end{equation}

For $f=b$:
\begin{equation}
A_b = \frac{2g_{Ab}g_{Vb}}{v_b^2 g_{Ab}^2 + (3-v_b^2)g_{Vb}^2/2} \;\; ,
\label{39}
\end{equation}
where $v_b$ is the velocity of the $b$-quark:
\begin{equation}
v_b =\sqrt{1-\frac{4\hat{m}_b^2}{m_Z^2}} \;\; .
\label{40}
\end{equation}

Here $\hat{m}_b$ is the value of the running mass of the $b$-quark at scale
$m_Z$ calculated in $\overline{MS}$ scheme \cite{18}.

The forward-backward charge asymmetry in the decay to $f\bar{f}$ equals:
\begin{equation}
A_{FB}^f =\frac{N_F -N_B}{N_F +N_B} =\frac{3}{4}A_e A_f \;\; ,
\label{41}
\end{equation}
where $N_F(N_B)$ -- the number of events with $f$ going into forward
(backward) hemisphere; $A_e$ refers to the creation of $Z$ boson in $e^+
e^-$-annihilation, while $A_f$ refers to its decay in $f\bar{f}$.

The longitudinal polarization of the $\tau$-lepton in the decay $Z\to
\tau\bar{\tau}$ is $P_{\tau} =-A_{\tau}$. If, however, the polarization is
measured as a function of the angle $\theta$ between the momentum of a
$\tau^-$ and the direction of the electron beam, this allows the
determination of not only $A_{\tau}$, but $A_e$ as well:
\begin{equation}
P_{\tau}(\cos\theta) =-\frac{A_{\tau}(1+\cos^2\theta)+2A_e\cos\theta}
{1+\cos^2\theta +2A_{\tau}A_e} \;\; .
\label{42}
\end{equation}

The polarization $P_{\tau}$ is found from $P_{\tau}(\cos\theta)$ by
separately integrating the numerator and the denominator in Eq.(\ref{42})
over the total solid angle.

The relative difference between total cross-section at the $Z$-peak for the
left- and right-polarized electrons that collide with non-polarized
positrons (measured at the SLC collider) is
\begin{equation}
A_{LR} \equiv \frac{\sigma_L -\sigma_R}{\sigma_L +\sigma_R} =A_e \;\; .
\label{43}
\end{equation}
The measurement of parity violating effects allows one to determine
experimentally the ratios $g_{Vf}/g_{Af}$, while the measurements of leptonic
and hadronic widths allow to find $g_{Af}$ and $\hat{\alpha}_s$.

\vspace{5mm}
\begin{center}

{\bf Table 1}

\vspace{3mm}

\begin{tabular}{|l|r|r|r|}
\hline
Observable & Experiment & Born & Pull \\
\hline
$m_W$ [GeV] & 80.390(64) & 79.956(12) & 6.8 \\
$m_W/m_Z$ & 0.8816(7) & 0.8768(1) & 6.8 \\
$s_W^2$ & 0.2228(12) & 0.2312(2) & -6.8 \\
$\Gamma_l$ [MeV] & 83.90(10) & 83.57(1) & 3.3 \\
$g_{Al}$ & -0.5010(3) & -0.5000(0) & -3.3 \\
$g_{Vl}/g_{Al}$ & 0.0749(9) & 0.0754(9) & -0.5 \\
$s_l^2$ & 0.2313(2) & 0.2312(2) & 0.5 \\
\hline
\end{tabular}

\end{center}

\vspace{5mm}

Table 1 compares the experimental and the Born values of the so-called
``hadronless" observables $m_W$, $g_{Al}$ and $g_{Vl}$. For the reader's
convenience the table lists different representations of the same observable
known in the literature
\begin{equation}
s_W^2 = 1-\frac{m_W^2}{m_Z^2} \;\; ,
\label{44}
\end{equation}
\begin{equation}
s_l^2 \equiv s_{eff}^2 \equiv \sin^2
\theta_{eff}^{lept} \equiv
\frac{1}{4}(1-\frac{g_{Vl}}{g_{Al}})
\label{45}
\end{equation}

The experimental values in the table are taken from ref. \cite{11}, assuming
that lepton universality holds. The pull shown in the last column is obtained
by dividing the difference ``Exp - Born" by experimental uncertainty (shown
in brackets). One can see that the discrepancy between experimental data and
Born values are very large for $m_W$ and substantial for $g_{Al}$.
That means that electroweak radiative corrections are essential. As for
$g_{Vl}/g_{Al}$, its experimental and Born values coincide. Moreover the
theoretical uncertainty is the same as the experimental one; thus the
pull is vanishing practically. Such high experimental accuracy for
$g_{Vl}/g_{Al}$ has been achieved only recently. As for $m_W$ and
$\Gamma_l$, their experimental uncertainties are much larger than the
theoretical ones.

We would like to mention that in 1991, when we published our first paper on
 electroweak corrections to $Z$-decays, the LEP experimental data were in perfect agreement with Born predictions of Table1. This demonstrates the remarkable progress in experimental accuracy.

\section{One-loop corrections to hadronless observables.}

\subsection{Four types of Feynman diagrams.}

Four types of Feynman diagrams contribute to electroweak corrections
for the observables of interest to us here, $m_W/m_Z,
g_{Al}, g_{Vl}/g_{Al}$:
\begin{enumerate}
\item{Self-energy loops for $W$ and $Z$ bosons with
virtual $\nu, l, q, H, W$ and $Z$ in loops  (see Figures 3(b)-3(n)).
}
\item{Loops of charged particles that result in transition of $Z$ boson into
a virtual photon  (see Figures 3(o)-3(r)).}
\item{Vertex triangles with virtual leptons and a virtual $W$ or
$Z$ boson  (see Figures 4(a)-4(c)).}
\item{Electroweak corrections to lepton and Z boson
wavefunctions  (see Figures 4(d)-4(g)).}
\end{enumerate}
It must be emphasized that $Z$ boson self-energy
loops contribute not only to the mass $m_Z$ and,
consequently, to the $m_W/m_Z$ ratio but also to the $Z$ boson decay to
$l\bar{l}$, to which $Z \leftrightarrow \gamma$ transitions also
contribute because these diagrams give corrections to the $Z$-boson
wavefunction. Moreover, there is no simple one-to-one correspondence between
Feynman diagrams and amplitudes. This is caused by the choice of $G_{\mu}$ as
an input observable which enters the expression for $s$ and $c$. As a result
e.g.  there is a contribution to $m_W/m_Z$ coming from the box and vertex
diagrams in the one-loop amplitude of the muon decay. In a similar way the
self-energy of the $W$ boson enters the amplitudes for decay $Z\to l\bar{l}$.
More on that see in Appendix D.

Obviously, electroweak corrections to $m_W/m_Z, g_{Al}$ and
$g_{Vl}/g_{Al}$ are dimensionless and thus can be expressed in
terms of $\bar{\alpha}, c, s$ and the dimensionless
parameters
\begin{equation}
t =(\frac{m_t}{m_Z})^2\;\; , \;\; h = (\frac{m_H}{m_Z})^2\;\; ,
\label{46}
\end{equation}
where $m_t$ is the mass of the $t$-quark and $m_H$ is the higgs
mass. (Masses of leptons and all quarks except $t$ give only very small
corrections.)

\subsection{The asymptotic limit at $\mbox{\boldmath$m^2_t \gg m^2_Z$}$.}

Starting from papers by Veltman \cite{19a} it became clear that
in the limit $t\gg 1$ electroweak radiative corrections are dominated by
terms proportional to $t$. These terms stem from the violation of weak
isotopic invariance by large difference of $m_t$ and $m_b$
 (see Figures 3(c), 3(i), and 3(j)).

After the discovery of the top quark it turned out that experimentally
$t\simeq 3.7$. As we will demonstrate in this review, for such value of $t$
the contributions of the terms, which are not enhanced by the factor $t$ are
comparable to the enhanced ones. Still it is convenient to split the
calculation of corrections into a number of stages and begin with
calculating the asymptotic limit for $t \gg 1$.

The main contribution comes from diagrams that contain $t$- and
 $b$-quarks because large difference of $m_t$ and $m_b$ strongly breaks
 isotopic invariance.  A simple calculation (see Appendix D) gives the
 following result for the sum of the Born and one-loop terms:
 \begin{equation}
 m_W/m_Z = c + \frac{3c}{32\pi s^2(c^2 - s^2)}
 \bar{\alpha}t\;\; .
 \label{47}
 \end{equation}
 \begin{equation} g_{Al} =
 -\frac{1}{2} - \frac{3}{64\pi s^2c^2} \bar{\alpha} t\;\; ,
 \label{48}
\end{equation}
 \begin{equation}
 R \equiv g_{Vl}/g_{Al} = 1 - 4s^2 + \frac{3}{4\pi(c^2 - s^2)}
 \bar{\alpha} t\;\; ,
 \label{49}
 \end{equation}
 \begin{equation}
 g_{\nu} = \frac{1}{2} + \frac{3}{64\pi s^2 c^2}  \bar{\alpha} t\;\; .
 \label{50}
 \end{equation}

The presence of $t$-enhanced terms in radiative corrections to $Z$ boson
decays allowed to predict the mass of the top quark before its actual
discovery \cite{19b, 19c}.

\subsection{The functions $\mbox{\boldmath$V_m(t,h)$}$,
$\mbox{\boldmath$V_A(t,h)$}$ and
$\mbox{\boldmath$V_R(t,h)$}$.}

If we now switch from the asymptotic case of $t \gg 1$ to the
realistic value of $t$, then one should make the substitution
in the eqs.(\ref{47}) - (\ref{50}):
\begin{equation}
t \to t + T_i(t)\;\; ,
\label{51}
\end{equation}
in which the index $i = m, A, R $, $\nu$ denotes $m_W/m_Z,\;\;
g_{Al},\;\; R \equiv g_{Vl}/g_{Al}$ and $g_{\nu}$, respectively. 

The functions $T_i$ are relatively simple combinations of
algebraic and logarithmic functions. Their numerical values for a range of
values of $m_t$ are given in Table 2. The functions $T_i(t)$
thus describe the contribution of the quark doublet $t,b$
to $m_W/m_Z$, $g_A$, $R = g_{Vl}/g_{Al}$ and
$g_{\nu}$. If, however, we now take into account
the contributions of the remaining virtual particles, then the
result can be given in the form
\begin{equation}
t \to V_i(t,h) = t + T_i(t) + H_i(h) + C_i + \delta V_i(t)\;\; .
\label{52}
\end{equation}

Here $H_i(h)$ contain the contribution of the virtual vector and higgs
bosons $W, Z$ and $H$ and are  functions of the higgs mass $m_H$.
(The mass of the $W$-boson  enters  $H_i(h)$ via
the parameter $c$, defined by equation (\ref{25})).
The explicit form of the functions $H_i$ is given in refs. \cite{20,21}
and their numerical values for various values of $m_H$
are given in Table 3.

\newpage
\begin{center}
{\bf Table 2}
\vspace{2mm}

\begin{tabular}{|r|r|r|r|r|}
\hline
$m_t$ & $t$ & $T_m$ & $T_A$ & $T_R$ \\
(GeV) & & & & \\ \hline
120 & 1.732 & 0.323 & 0.465 & 0.111 \\
130 & 2.032 & 0.418 & 0.470 & 0.154 \\
140 & 2.357 & 0.503 & 0.473 & 0.193 \\
150 & 2.706 & 0.579 & 0.476 & 0.228 \\
160 & 3.079 & 0.649 & 0.478 & 0.261 \\
170 & 3.476 & 0.713 & 0.480 & 0.291 \\
180 & 3.896 & 0.772 & 0.481 & 0.319 \\
190 & 4.341 & 0.828 & 0.483 & 0.345 \\
200 & 4.810 & 0.880 & 0.484 & 0.370 \\
210 & 5.303 & 0.929 & 0.485 & 0.393 \\
220 & 5.821 & 0.975 & 0.485 & 0.415 \\
230 & 6.362 & 1.019 & 0.486 & 0.436 \\
240 & 6.927 & 1.061 & 0.487 & 0.456 \\
\hline
\end{tabular}
\end{center}
\vspace{3mm}

\vspace{3mm}

\begin{center}

{\bf Table 3}
\vspace{2mm}

\begin{tabular}{|r|r|r|r|r|}
\hline
$m_H$ & $h$ & $H_m$ & $H_A$ & $H_R$ \\
(GeV) & & & & \\ \hline
0.01 & 0.000 & 1.120 & -8.716 & 1.359 \\
0.10 & 0.000 & 1.119 & -5.654 & 1.354 \\
1.00 & 0.000 & 1.103 & -2.652 & 1.315 \\
10.00 & 0.012 & 0.980 & -0.133 & 1.016 \\
50.00 & 0.301 & 0.661 & 0.645 & 0.360 \\
100.00 & 1.203 & 0.433 & 0.653 & -0.022 \\
150.00 & 2.706 & 0.275 & 0.588 & -0.258 \\
200.00 & 4.810 & 0.151 & 0.518 & -0.430 \\
250.00 & 7.516 & 0.050 & 0.452 & -0.566 \\
300.00 & 10.823 & -0.037 & 0.392 & -0.679 \\
350.00 & 14.732 & -0.112 & 0.338 & -0.776 \\
400.00 & 19.241 & -0.178 & 0.289 & -0.860 \\
450.00 & 24.352 & -0.238 & 0.244 & -0.936 \\
500.00 & 30.065 & -0.292 & 0.202 & -1.004 \\
550.00 & 36.378 & -0.341 & 0.164 & -1.065 \\
600.00 & 43.293 & -0.387 & 0.128 & -1.122 \\
650.00 & 50.809 & -0.429 & 0.095 & -1.175 \\
700.00 & 58.927 & -0.469 & 0.064 & -1.223 \\
750.00 & 67.646 & -0.506 & 0.035 & -1.269 \\
800.00 & 76.966 & -0.540 & 0.007 & -1.311 \\
850.00 & 86.887 & -0.573 & -0.019 & -1.352 \\
900.00 & 97.410 & -0.604 & -0.044 & -1.390 \\
950.00 & 108.534 & -0.633 & -0.067 & -1.426 \\
1000.00 & 120.259 & -0.661 & -0.090 & -1.460 \\
\hline
\end{tabular}
\end{center}
\vspace{3mm}
The constants $C_i$ in eq.(\ref{52}) include the contributions
of light fermions to the self-energy of the $W$ and $Z$ bosons,
and also to the Feynman diagrams, describing
the electroweak corrections to the muon decay,
as well as triangle diagrams, describing the $Z$ boson decay. The
constants $C_i$ are relatively complicated functions of $s^2$
(see \cite{20, 21}).
We list
here their numerical values for $s^2 = 0.23110 - \delta s^2$:
\begin{equation}
C_m = -1.3497 + 4.13 \delta s^2\;\;,
\label{53}
\end{equation}
\begin{equation}
C_A = -2.2621 - 2.63 \delta s^2\;\;,
\label{54}
\end{equation}
\begin{equation}
C_R = -3.5045 - 5.72 \delta s^2\;\;,
\label{55}
\end{equation}
\begin{equation}
C_{\nu} = -1.1641 - 4.88 \delta s^2\;\;.
\label{56}
\end{equation}

\subsection{Corrections $\mbox{\boldmath$\delta V_i(t)$}$.}

Finally, the last term in equation (\ref{52}) includes the sum
of corrections of three different types.
Their common feature is
that they do not contain more than one electroweak loop.
\begin{equation}
\delta V_i =\delta_1 V_i +\delta_2 V_i +\delta_3 V_i
\label{57}
\end{equation}

\begin{enumerate}
\item{The corrections $\delta_1 V_i$ are extremely small. They contain
contributions of the $W$-boson and the $t$-quark to the polarization of the
electromagnetic vacuum $\delta_W \alpha$ and $\delta_t\alpha$,
which traditionally are not included into the running of $\alpha(q^2)$, i.e.
into $\bar{\alpha}$ (see Figure 5). It is reasonable to treat them as electroweak
corrections.  This is especially true for the $W$-loop that depends on the
gauge chosen for the description of the $W$ and $Z$ bosons. Only after this
loop is taken into account, the resultant electroweak corrections become
gauge-invariant, as it should indeed be for physical observables. Here and
hereafter in the calculations the 't Hooft--Feynman gauge is used.
\begin{equation}
\delta_1 V_m(t,h) = -\frac{16}{3} \pi s^4
\frac{1}{\alpha}(\delta_W\alpha + \delta_t \alpha) = -0.055 \;\; ,
\label{58}
\end{equation}
\begin{equation}
\delta_1 V_R(t,h) = -\frac{16}{3}
\pi s^2 c^2 \frac{1}{\alpha}(\delta_W\alpha + \delta_t\alpha) = -0.181 \;\; ,
\label{59}
\end{equation}
\begin{equation}
\delta_1 V_A(t,h) = \delta_1
V_{\nu}(t,h) = 0\;,
\label{60}
\end{equation}
where
\begin{equation}
\delta_W\alpha = 0.00050 \; ,
\label{61}
\end{equation}
\begin{equation}
\delta_t\alpha  \simeq  - 0.00005(1) \;\; .
\label{62}
\end{equation}
(See eqs.(B.18) and (B.17) from Appendix B. Unless specified otherwise,
we  use $m_t = 175$ GeV in numerical evaluations.)}

\item{The corrections $\delta_2V_i$ are the largest ones. They are
caused in the order $\bar{\alpha}\hat{\alpha}_s$ by virtual
gluons in electroweak loops
of light quarks
$q = u, d, s, c, b$ and heavy quark $t$ (see Figure 6):
\begin{equation}
\delta_2 V_i(t) = \delta^q_2 V_i + \delta^t_2 V_i(t)\;\;.
\label{63}
\end{equation}
Due to asymptotic freedom of QCD \cite{299} these corrections were
calculated in perturbation theory.
The analytical expressions for corrections $\delta^q_2 V_i$ and
$\delta^t_2 V_i(t)$ are given in \cite{20, 21}. Here we only give
numerical estimates for them,
\begin{equation}
\delta^q_2 V_m =
-0.377\frac{\hat{\alpha}_s}{\pi}\;\;,
\label{64}
\end{equation}
\begin{equation}
\delta^q_2 V_A = 1.750 \frac{\hat{\alpha}_s}{\pi}\;\;,
\label{65}
\end{equation}
\begin{equation}
\delta^q_2 V_R = 0\;\;,
\label{66}
\end{equation}
\begin{equation}
\delta^t_2 V_m(t) = -11.67 \frac{\hat{\alpha}_s(m_t)}{\pi}
= -10.61 \frac{\hat{\alpha}_s}{\pi} \;\; ,
\label{67}
\end{equation}
\begin{equation}
\delta^t_2 V_A(t) = -10.10 \frac{\hat{\alpha}_s(m_t)}{\pi} = -9.18
\frac{\hat{\alpha}_s}{\pi} \;\; ,
\label{68}
\end{equation}
\begin{equation}
\delta^t_2 V_R(t) = -11.88 \frac{\hat{\alpha}_s(m_t)}{\pi} = -10.80
\frac{\hat{\alpha}_s}{\pi} \;\; ,
\label{69}
\end{equation}
where
\cite{299}
\begin{equation}
\hat{\alpha}_s(m_t)
=\frac{\hat{\alpha}_s}{1+\frac{23}{12\pi}\hat{\alpha}_s\log t}\;\;.
\label{70}
\end{equation}
(For numerical evaluation, we use $\hat{\alpha}_s\equiv
\hat{\alpha}_s(m_Z) = 0.120$.)}
We have mentioned already that the corrections $\delta^t_2 V_i(t)$,
whose numerical values were given in (\ref{67})--(\ref{69}),
are much larger than all other terms included in $\delta V_i$.
We emphasize that the term in $\delta^t_2 V_i$ that is leading
for high $t$ is universal: it is independent of $i$. As shown in
\cite{22}, this leading term is obtained by multiplying the
Veltman asymptotics $t$ by a factor
\begin{equation}
1 - \frac{2\pi^2 + 6}{9} \frac{\hat{\alpha}_s(m_t)}{\pi}\;\;,
\label{71}
\end{equation}
or, numerically,
\begin{equation}
t \to t(1 - 2.86 \frac{\hat{\alpha}_s(m_t)}{\pi}) \;\; .
\label{72}
\end{equation}
Qualitatively the factor (\ref{71}) corresponds to the fact
that the running mass of the $t$-quark at momenta $p^2 \sim m^2_t$
that circulate in the $t$-quark loop is lower than the
"on the mass-shell"
mass  of the $t$-quark. It is interesting to compare the
correction (\ref{72}) with the quantity
\begin{equation}
\tilde{m}^2_t \equiv m^2_t(p^2_t =
- m^2_t) = m^2_t(1 - 2.78 \frac{\hat{\alpha}_s(m_t)}{\pi}) \;\; ,
\label{73}
\end{equation}
calculated in the Landau gauge in \cite{23}, p.~102. The
agreement is overwhelming. There is, therefore, a simple
mnemonic rule for evaluating the main gluon corrections for the
$t$-loop.
\item{Corrections $\delta_3 V_i$ of the order of
$\bar{\alpha}\hat{\alpha}^2_s$ are extremely small. They
were calculated in the literature
\cite{24} for the term leading in $t$ (i.e.
$\bar{\alpha}\hat{\alpha}^2_s t$).
They are independent of $i$ (in numerical estimates we use for
the number of light quark flavors $N_f = 5$):
\begin{equation}
\delta_3 V_i(t)\simeq -(2.38 - 0.18 N_f) \hat{\alpha}_s^2(m_t)t
\simeq - 1.48 \hat{\alpha}_s^2(m_t)t = -0.07 \;\; .
\label{74}
\end{equation}
}
\end{enumerate}

\subsection{Accidental (?) compensation and the mass of the
$\mbox{\boldmath$t$}$-quark.}

Now that we have expressions for all terms in eq.(\ref{52}),
it will be convenient to analyze their roles and the general
behaviour of the functions $V_i(t,h)$.
As functions of $m_t$ at three fixed values of $m_H$, they
are shown in Figures 7, 8, 9. On all these figures, we see
a cusp at $m_t = m_Z/2$. This is a typical
threshold singularity that arises when the channel $Z \to t\bar{t}$
is opened. It is of no practical significance since
experiments give $m_t \simeq 175$ GeV. What really impresses is that
the function $V_R$ vanishes at this value of $m_t$.
This happens because of the compensation of the leading
term $t$ and the rest of the terms which produce a negative
aggregate contribution, the main negative
contribution coming from the light fermions (see eq.(\ref{55}) for the
constant $C_R$).

In the one-electroweak loop approximation  each
function $V_i(t,h)$ is a sum of two functions one of which is
$t$-dependent but independent of $h$, while the other is
$h$-dependent but independent of $t$ (plus, of course, a
constant which is independent of both $t$ and $h$). Therefore
the curves for $m_H =100$ and $800$ GeV in Figures 7, 8, 9
are produced by the parallel transfer of the curve
for $m_H= 200$ GeV.

We see in Figure 9 that if the $t$-quark were light,
radiative corrections would be large and
negative, and if it were very
heavy, they would be large and positive. This looks like a
conspiracy of the
observable mass of the
$t$-quark and all other parameters of the electroweak theory,
as a result of which the electroweak correction $V_R$ becomes
anomalously small.

One should specially note the dashed parabola in Figures 7, 8
and 9 corresponding to the Veltman term $t$. We see that in
the interval $0 < m_t < 250$ GeV it lies much higher than
$V_A$ and $V_R$ and approaches $V_m$ only in the right-hand side
of Figure 7. Therefore, the so-called non-leading `small' corrections
that were typically replaced with ellipses in standard texts,
are found to be comparable with the leading term $t$.

A glance at Figure 9 readily explains how the experimental
analysis of electroweak corrections allowed, despite their
smallness, a prediction, within the framework of the minimal
standard model, of the $t$-quark mass. Even when the
experimental accuracy of LEP-I and SLC experiments was not
sufficient for detecting electroweak corrections, it was
sufficient for establishing the $t$-quark mass using the points
at which the curves $V_R(m_t)$ intersect the horizontal line
corresponding to the experimental value of $V_R$ and the
parallel to it thin lines that show the band of one
standard deviation. The accuracy in determining $m_t$ is imposed
by the band width and the slope of $V_R(m_t)$.

The dependence $V_i(m_H)$ for three fixed values of
$m_t = 150$, 175 and 200 GeV can be
presented in a similar manner. As follows from the explicit
form of the terms $H_i(m_H)$, the dependence $V_i(m_H)$ is
considerably less steep (it is logarithmic). This is the reason why
the prediction of the higgs mass extracted from electroweak corrections
has such a high uncertainty.
The accuracy of prediction of
$m_H$ greatly depends on the value of the $t$-quark mass.
If $m_t = 150 \pm 5$ GeV, then $m_H < 80$ GeV
at the $3 \sigma$ level. If $m_t = 200 \pm 5$ GeV, then
$m_H > 150$ GeV at the $3 \sigma$ level. If, however,
$m_t = 175 \pm 5$ GeV, as
given by FNAL experiments \cite{13},
we are hugely unlucky:
the constraint on $m_H$ is rather mild (see Figure 10).

Before starting a discussion of hadronic decays of the $Z$ boson,
let us `go back to the roots' and recall how the equations for
$V_i(m_t, m_H)$ were derived.

\subsection{How to calculate $\mbox{\boldmath$V_i$}$? `Five steps'.}

An attentive reader should have already come up with the
question: what makes the amplitudes of the lepton decays of the $Z$ boson
in the one-loop approximation depend on the self-energy of the $W$
boson?  Indeed, the loops describing the self-energy of the $W$
boson appear in the decay diagrams of the $Z$ boson
only beginning with the two-loop approximation. The
answer to this question was already given at the beginning of Section 4. We
have already emphasized that we find expressions for radiative corrections to
$Z$-boson decays in terms of $\bar{\alpha}$, $m_Z$ and $G_{\mu}$.  However,
the expression for $G_{\mu}$ includes the self-energy of the $W$ boson even
in the one-loop approximation. The point is that we express some
observables (in this particular case, $m_W/m_Z$, $g_{Al}$, $g_{Vl}/g_{Al}$)
in terms of other, more accurately measured observables ($\bar{\alpha}$,
$m_Z$, $G_{\mu}$).

Let us trace how this is achieved, step by step.
There are altogether `five steps to happiness',
based on the one-loop approximation. All necessary formulas can be found in
Appendix D.

{\bf Step I.} We begin with the electroweak Lagrangian after it
had undergone the spontaneous violation of the
$SU(2) \times U(1)$-symmetry by the higgs vacuum condensate
(vacuum expectation value -- {\it vev})
$\eta$ and the $W$ and $Z$ bosons became massive. Let us
consider the bare coupling constants (the bare charges $e_0$ of the photon,
$g_0$ of the $W$-boson and $f_0$ of the $Z$-boson) and the bare
masses of the vector bosons:
\begin{equation}
m_{Z0} = \frac{1}{2} f_0 \eta \;\; ,
\label{75}
\end{equation}
\begin{equation}
m_{W0} = \frac{1}{2} g_0 \eta \;\; ,
\label{76}
\end{equation}
and also bare masses: $m_{t0}$ of the $t$-quark and $m_{H0}$ of
the higgs.

{\bf Step II.} We express $\bar{\alpha}$, $G_{\mu}$, $m_Z$ in
terms of $f_0$, $g_0$, $e_0$, $\eta$, $m_{t0}$, $m_{H0}$ and
$1/\varepsilon$ (see Appendix D). Here $1/\varepsilon$ appears
because we use the dimensional regularization, calculating
Feynman integrals in the space of $D$ dimensions (see Appendix A).
These integrals diverge at $D=4$ and are finite
in the vicinity of  $D = 4$.
By definition,
\begin{equation}
2\varepsilon = 4 - D \to 0 \;\;.
\label{77}
\end{equation}
Note that in the one-loop approximation $m_{t0} = m_t$, $m_{H0} = m_H$,
since we neglect the electroweak corrections to the masses of the
$t$-quark and the higgs.

Step II is almost physics: we calculate Feynman diagrams
(we say `almost' to emphasize that observables are
expressed in terms of nonobservable, `bare', and generally
infinite quantities).

{\bf Step III.} Let us invert the expressions derived at step II
and write $f_0$, $g_0$, $\eta$ in terms of $\bar{\alpha}$,
$G_{\mu}$, $m_Z$, $m_t$, $m_H$ and $1/\varepsilon$. This step is a
pure algebra.

{\bf Step IV.} Let us express $V_m$, $V_A$, $V_R$ (or the electroweak
one-loop correction to any other electroweak observable, all of
them being treated on an equal basis) in terms of $f_0$, $g_0$,
$\eta$, $m_t$, $m_H$ and $1/\varepsilon$. (Like step II, this
step is again almost physics.)

{\bf Step V.} Let us express $V_m$, $V_A$, $V_R$ (or any other
electroweak correction) in terms of $\bar{\alpha}$, $G_{\mu}$, $m_Z$,
$m_t$, $m_H$ using the results of steps III and IV. Formally
this is pure algebra, but in fact pure physics, since now we
have expressed certain physical observables in terms of other
observables. If no errors were made on the way, the terms
$1/\varepsilon$ cancel out. As a result, we arrive at formula
(\ref{52}) which gives $V_i$ as elementary functions of
$t$, $h$ and $s$.

The five steps outlined above are very simple and visually clear.
We obtain the main relations without using the `heavy artillery'
of quantum field theory with its counterterms in the Lagrangian
and the renormalization procedure. This simplicity and visual
clarity became possible owing to the one-loop electroweak
approximation (even though this approach to renormalization is
possible in multiloop calculations, it becomes more cumbersome than
the standard procedures). As for the QCD-corrections to quark electroweak
loops hidden in the terms
$\delta V_i$ in equation (\ref{52}), we take the relevant formulas
from the calculations of other authors.

\section{One-loop corrections to hadronic decays of the ${\bf{Z}}$ boson.}

\subsection{The leading quarks and hadrons.}

As discussed above (see formulas (\ref{31})--(\ref{37})), an analysis
of hardronic
decays reduces to the calculation of decays to pairs of quarks:
$Z \to q\bar{q}$. The key role is played by the concept of leading
hadrons that carry away the predominant part of the energy.
For example, the $Z \to c\bar{c}$ decay
mostly produces two hadron jets flying in opposite directions,
in one of which the leading hadron is the one containing the
$\bar{c}$-quark, for example, $D^- = \bar{c}d$, and in the other
the hadron with the c-quark, for example, $D^0 = c\bar{u}$
or $\Lambda^+_c = udc$. Likewise, $Z \to b\bar{b}$ decays are
identified by the presence of high-energy $B$ or $\bar{B}$
mesons. If one selects only particles with energy close to
$m_Z/2$, the identification of the initial quark channels is
unambiguous. The total number of such cases will, however, be
small. If one takes into account as a signal less energetic
$B$-mesons, one faces the problem of their origin. Indeed, a pair
$b\bar{b}$ can be created not only directly by a $Z$ boson but also
by a virtual gluon in, say, a $Z \to c\bar{c}$ decay
or $Z \to u\bar{u}$, or $s\bar{s}$. This example shows the
sort of difficulty encountered by experimentalists trying to
identify a specific quark--antiquark channel. Furthermore, owing
to such secondary pairs, the total hadron width is not strictly equal to
the sum of partial quark widths.

We remind the reader that for the partial width $\Gamma_q$
of the $Z \to q\bar{q}$
decay we had eq.(\ref{31}),
where the standard width $\Gamma_0$ was given by eq.(\ref{29})
and the radiators $R_{Aq}$ and $R_{Vq}$ are given in Appendix E.
As for the electroweak corrections, they are included in the
coefficients $g_{Aq}$ and $g_{Vq}$. The sum of the Born and one-loop
terms has the form
\begin{equation}
g_{Aq} = T_{3q}[1 + \frac{3\bar{\alpha}}{32 \pi s^2c^2}
V_{Aq}(t,h)] \;\;,
\label{78}
\end{equation}
\begin{equation}
R_q \equiv g_{Vq}/g_{Aq} = 1 - 4|Q_q| s^2 + \frac{3|Q_q|}{4\pi(c^2
- s^2)} \bar{\alpha} V_{Rq}(t,h) \;\;.
\label{79}
\end{equation}

\subsection{Decays to pairs of light quarks.}

Here, as in the case of hadronless observables, the quantities $V$
that characterize corrections are normalized in the standard
way: $V \to t$ as $t \gg 1$. Naturally, those terms in $V$ that
are due to the self-energies of vector bosons are identical for both
leptons and quarks. The deviation of the differences $V_{Aq}- V_{Al}$
and $V_{Rq} - V_{Rl}$ from zero are caused by the differences in
radiative corrections to vertices $Z \to q\bar{q}$ and $Z \to
l\bar{l}$.  For four light quarks we have
\begin{equation}
V_{Au}(t,h) = V_{Ac}(t,h) = V_{Al}(t,h)
 + [\frac{128\pi s^3 c^3}{3 \bar{\alpha}} (F_{Al}+F_{Au}) =
0.2634] \;\; ,
\label{80}
\end{equation}
\begin{equation}
V_{Ad}(t,h) = V_{As}(t,h) = V_{Al}(t,h) + [\frac{128\pi s^3 c^3}{3
\bar{\alpha}} (F_{Al}-F_{Ad}) = 0.6295] \;\;,
\label{81}
\end{equation}
\begin{eqnarray}
V_{Ru}(t,h) = V_{Rc}(t,h) = V_{Rl}(t,h)+ [\frac{16\pi sc(c^2 - s^2)}{3
\bar{\alpha}} \times  \nonumber \\
\times [F_{Vl}-(1-4s^2)F_{Al} + \frac{3}{2}(-(1-\frac{8}{3}s^2)
F_{Au}+F_{Vu})] = 0.1220] \;\; ,
\label{82}
\end{eqnarray}
\begin{eqnarray}
V_{Rd}(t,h) = V_{Rs}(t,h) = V_{Rl}(t,h)+ [\frac{16\pi sc(c^2 - s^2)}{3
\bar{\alpha}} \times  \nonumber \\
\times [F_{Vl}-(1-4s^2)F_{Al} + 3((1-\frac{4}{3}s^2)F_{Ad}-F_{Vd})] =
0.2679] \;\; ,
\label{83}
\end{eqnarray}
where (see \cite{20, 21}):
\begin{equation} F_{Al} =
\frac{\bar{\alpha}}{4\pi}(3.0099 + 16.4 \delta s^2)\;,
\label{84}
\end{equation}
\begin{equation}
F_{Vl} = \frac{\bar{\alpha}}{4\pi}(3.1878 + 14.9 \delta s^2)\;,
\label{85}
\end{equation}
\begin{equation}
F_{Au} = - \frac{\bar{\alpha}}{4\pi}(2.6802 + 14.7 \delta s^2)\;,
\label{86}
\end{equation}
\begin{equation}
F_{Vu} = - \frac{\bar{\alpha}}{4\pi}(2.7329 + 14.2 \delta s^2)\;,
\label{87}
\end{equation}
\begin{equation}
F_{Ad} = \frac{\bar{\alpha}}{4\pi}(2.2221 + 13.5 \delta s^2)\;,
\label{88}
\end{equation}
\begin{equation}
F_{Vd} = \frac{\bar{\alpha}}{4\pi}(2.2287 + 13.5 \delta s^2)\;.
\label{89}
\end{equation}
The values of $F$ are given here for $s^2 = 0.23110 - \delta s^2$.
The accuracy to five decimal places is purely arithmetic. The
physical uncertainties introduced by neglecting higher-order
loops manifest themselves already in the third decimal place.

In addition to the changes given by eqs.(\ref{80}) - (\ref{83}), one has to
take into account also emission of virtual or ``free"
gluon from a vertex quark triangle.

The corresponding effect cannot be parametrized in terms $V_{Aq}$ and
$V_{Rq}$, because it contributes also to the radiators $R_{Aq}$ and $R_{Vq}$.
The change of $\Gamma_h$ caused by it has been calculated only recently
\cite{25} and turned out to be rather small:
\begin{equation}
\delta\Gamma_h(Z\to u, d, s, c) =-0.59(3) {\rm MeV}
\label{900}
\end{equation}

\subsection{Decays to $\mbox{\boldmath$b\bar{b}$}$ pair.}

In the $Z \to b\bar{b}$ decay it is necessary to take into account
additional $t$-dependent vertex corrections:
\begin{equation}
V_{Ab}(t,h) =
V_{Ad}(t,h)-\left[\frac{8s^2 c^2}{3(3-2s^2)}(\phi (t) +
\delta_{\alpha_s} \phi (t))=5.03\right] \;\; ,
\label{90}
\end{equation}
\begin{equation}
V_{Rb}(t,h) = V_{Rd}(t,h)-
\left[\frac{4s^2 (c^2 -s^2)}{3(3-2s^2)}(\phi (t) + \delta_{\alpha_s}
\phi (t))=1.76\right] \;\;  .
\label{91}
\end{equation}
Here the term $\phi(t)$ calculated in \cite{26} corresponds to
a $t\bar{t}W$ vertex triangle (see Figure 11 (a)), while
 the term $\delta_{\alpha_s} \phi(t)$ calculated in \cite{27},
corresponds to the leading gluon corrections
to the term $\phi(t)$ (see Figure 11 (b)):
 $\delta_{\alpha_s}\phi(t) \sim \alpha_s t$.
Expressions for $\phi(t)$ and
$\delta_{\alpha_s}\phi(t)$ are given in refs. \cite{20, 21}.
For $m_t = 175$ GeV, $\hat{\alpha}_s(m_Z) = 0.120$
\begin{equation}
\phi(t) = 29.96 \;\; ,
\label{92}
\end{equation}
\begin{equation}
\delta_{\alpha_s} \phi(t) = -3.02 \;\; ,
\label{93}
\end{equation}
and correction terms in equations (\ref{90}) and (\ref{91}) are
very large. The subleading gluon corrections to $\phi(t)$ calculated recently
\cite{311} are very small: $\delta\Gamma_h(Z\to b) = -0.04$ MeV.

\section{Comparison of one-electroweak-loop results and experimental
LEP-I and SLC data.}

\subsection{LEPTOP code.}

A number of computer programs (codes) were written for
comparing high-precision data of LEP-I and SLC. The best known of
these programs in Europe is ZFITTER \cite{28}, which takes into
account not only electroweak radiative corrections but also all
purely electromagnetic ones, including, among others, the
emission of photons by colliding electrons and positrons. Some of
the first publications in which the $t$ quark mass was
predicted on the basis of precision measurements \cite{29},
were based on the code ZFITTER. Other European codes, BHM, WOH \cite{30},
TOPAZO \cite{31}, somewhat differ from ZFITTER. The best
known in the USA are the results generated by the code used by
Erler and Langacker \cite{32}, \cite{erler}.

The original idea of the authors of this review in 1991--1993
was to derive simple analytical formulas for electroweak
radiative corrections, which would make it possible to predict
the $t$-quark mass using no computer codes, just by analyzing
experimental data on a sheet of paper. Alas, the diversity of
hadron decays of $Z$ bosons, depending on the constants of strong
gluon interaction $\hat{\alpha}_s$, was such that it was
necessary to convert analytical formulas into a computer program
which we jokingly dubbed LEPTOP \cite{33}. The LEPTOP
calculates the electroweak observables in the framework of the
Minimal Standard Model and fits experimental data so as to
determine the quantities $m_t$, $m_H$ and $\hat{\alpha}_s(m_Z)$.
The logical structure of LEPTOP is clear from the preceding
sections of this review and is shown in the Flowchart. The code of
LEPTOP can be downloaded from the Internet home page:
http://cppm.in2p3.fr./leptop/intro$\_$leptop.html

A comparison of the codes ZFITTER, BHM, WOH, TOPAZO and LEPTOP
carried out in 1994--95 \cite{118} has demonstrated that their
predictions for all electroweak observables coincide
with accuracy that is much better than the accuracy of
the experiment.
The Flowcharts of LEPTOP and ZFITTER are compared
on pages 25 and 27 of \cite{118}; numerical comparison of five codes
(their 1995 versions) 
for twelve observables is presented in figures 11-23 of the same 
reference.
 The results of processing the experimental data
using LEPTOP are shown below.

\subsection{One-loop general fit.}

Second column of Table 4 shows experimental values of the electroweak
observables, obtained by averaging the results of four LEP
detectors (part a), and also SLC data (part b) and the data on
$W$ boson mass (part c). (The data on the W boson mass
from  the $p\bar{p}$-colliders and LEP-II
are also shown, for the reader's convenience, in the form
of $s^2_W$, while the data on $s_W^2$ from $\nu N$-experiments
are also shown in the form of $m_W$. These two numbers are
given in italics, emphasizing that they are not independent
experimental data. The same refers to $s^2_l$ ($A_{LR}$).)
We take experimental data from the paper
\cite{11}.
The experimental data of Table 4 are used for determining
(fitting) the parameters of the standard model
in one-electroweak-loop approximation: $m_t$, $m_H$,
$\hat{\alpha}_s(m_Z)$ and  $\bar{\alpha}$.
(In fitting $m_t$ the direct
measurements of $m_t$ by CDF and D0 [collaborations] \cite{13} are also
used. In fitting $\bar{\alpha}$ its value from eq.(\ref{14}) was used.)
The third column shows the results of the fit of electroweak observables
with one loop electroweak formulas.
The last column shows the value of the `pull'. By
definition, the pull is the difference between the experimental and the
theoretical values divided by experimental uncertainty.  The pull values show
that for most observables the discrepancy is less than $1\sigma$. The number
of degrees of freedom is $18-4=14$.

\newpage
\begin{center}
{\bf Table 4. Fit of the experimental data \cite{11} with one-electroweak-loop
 formulas.}

\vspace{3mm}

$m_Z =91.1867(21)$ GeV is used as an input.

Output of the fit: $m_H =139.1^{+134.2}_{-76.5}$ GeV, $\hat{\alpha}_s =
0.1195 \pm 0.0030$, $\chi^2/n_{d.o.f.} =15.1/14$

\vspace{2mm}

\begin{tabular}{|l|c|c|c|}
\hline
& Experimental & Fit & \\
Observable & data & standard & Pull \\
& & model & \\ \hline
a) \underline{LEP} & & & \\
& & & \\
shape of $Z$-peak & & & \\
and lepton asymmetries: & & & \\
$\Gamma_Z$ [GeV] & 2.4939(24) & 2.4959(18) & -0.8 \\
$\sigma_h [nb]$ & 41.491(58) & 41.472(16) & 0.3 \\
$R_l$ & 20.765(26) & 20.747(20) & 0.7 \\
$A^l_{FB}$ & 0.0168(10) & 0.0161(3) & 0.8 \\
$\tau$-polarization: & & & \\
$A_{\tau}$ & 0.1431(45) & 0.1465(14) & -0.8 \\
$A_e$ & 0.1479(51) & 0.1465(14) & 0.3 \\
Results for $b$ and $c$ & & & \\
quarks: & & & \\
$R_b$$^*$
 & 0.2166(7) & 0.2158(2) & 1 \\
$R_c$$^*$ & 0.1735(44) & 0.1723(1) & 0.3 \\
$A^b_{FB}$ & 0.0990(21) & 0.1027(10) & -1.8 \\
$A^c_{FB}$ & 0.0709(44) & 0.0734(8) & -0.6 \\
Charge asymmetry for pairs & & & \\
of light quarks $q\bar{q}$: & & & \\
$s^2_l(Q_{FB})$ & 0.2321(10) & 0.2316(2) & 0.5 \\
\hline
b) \underline{SLC} & & & \\
$A_{LR}$ & 0.1504(23) & 0.1465(14) & 1.7 \\
$s^2_l(A_{LR})$ & {\it 0.2311(3)} & {\it 0.2316(2)} & {\it -1.7} \\
$R_b$$^*$ & 0.2166(7) & 0.2158(2) & 0.9 \\
$R_c$$^*$ & 0.1735(44) & 0.1723(1) & 0.3 \\
$A_b$ & 0.8670(350) & 0.9348(1) & -1.9 \\
$A_c$ & 0.6470(400) & 0.6676(6) & -0.5 \\
\hline
c) \underline{$p\bar{p} +$ LEP-II $+ \nu N$} & & & \\
& & & \\
$m_W$ [GeV] $(p\bar{p})$  + LEP-II & 80.39(6) &80.36(3) & 0.5 \\
&{\it 0.2228(13)} & & \\
$s_W^2$ $(\nu N)$ & 0.2254(21) & 0.2234(6) & 0.9 \\
&{\it 80.2547(1089)} & & \\
$m_t$ [GeV]    & 173.8(5.0)  & 171.6(4.9) & 0.4\\
\hline
\end{tabular}
\end{center}
\noindent
$^*$ {\footnotesize Experimental values of $R_b$ and $R_c$ correspond to the
average of LEP-I and SLC results.}

\newpage

\begin{center}
{\bf Table 5}
\vspace{3mm}

\begin{tabular}{|l||c|c|cc|}
\hline
Observable & $s^2_l$ & Average over & Cumulative & $\chi^2/n_{d.o.f.}$\\
& & groups of observations & average &
\\ \hline \hline
$A^l_{FB}$ & 0.23117(55) & & &\\
$A_{\tau}$ & 0.23202(57) & & &\\
$A_e$ & 0.23141(65) & 0.23153(34) & 0.23153(34)
&1.2/2 \\ \hline
$A^b_{FB}$ & 0.23226(38) & & & \\
$A^c_{FB}$ & 0.23223(112) & 0.23226(36) & 0.23187(25)
& 3.4/4 \\ \hline
$<Q_{FB}>$ & 0.23210(100) & 0.23210(100) & 0.23189(24) & 3.4/5 \\
\hline
$A_{LR}$ (SLD) & 0.23109(30) & 0.23109(30) & 0.23157(19)
& 7.8/6 \\ \hline
\end{tabular}
\end{center}

Table 5
\footnote{Table 5 is our recalculation with LEPTOP program of the
Table 30 of EWWG report \cite{11}.
The numbers for $A_e$ and $A_{\tau}$ in tables 4 and 5 agree with
each other, while they disagree in EWWG report in tables 30 and 31. In
order to restore the agreement one has to interchange $A_e$ and $A_{\tau}$
in table 30 in EWWG report.}
 gives experimental values of $s^2_l$. The third column was
obtained by averaging of the second column, and the fourth by
cumulative averaging of the third; it also lists the values of
$\chi^2$ over the number of degrees of freedom.
\vspace{3mm}

\section{Two-loop electroweak corrections and theoretical uncertainties.}

In this Section we will discuss
heavy top corrections of the  second order in $\alpha_W$ to
$m_W$ and to coupling constants of $Z$-boson with fermions. Full calculation
of $\alpha_W^2$ corrections is still absent. What
have been
calculated are corrections of the order $\alpha_W^2$ $t^2 =\alpha_W^2
(m_t/m_Z)^4$ \cite{260, 270} and corrections
$\sim\alpha_W^2 t$ \cite{280} - \cite{300}.

There are two sources of $\alpha_W^2 t^2$ corrections in our approach.
The first source are reducible diagrams with top quark in each loop.
The second
source are irreducible
two-loop Feynman diagrams which contain top quark
\cite{260, 270}. We start our consideration with the first source the
contribution of which is proportional to $(\Pi_Z(0)-\Pi_W(0))^2$. Detailed
 calculations are presented in Appendix F.

\subsection{$\mbox{\boldmath$\alpha_W^2 t^2$}$ corrections to
$\mbox{\boldmath$m_W/m_Z\; , \;\; g_A$}$ and
$\mbox{\boldmath$g_V/g_A$}$ from reducible diagrams.}

We start our consideration from the ratio of vector boson masses. From
eq.(\ref{4.13}) and (\ref{4.14}) we obtain:
\begin{equation}
\frac{m_W}{m_Z} =c [1+\frac{c^2}{2(c^2 -s^2)}\delta +\frac{3c^4 -10c^4
s^2}{8(c^2 -s^2)^3}\delta^2] \;\; .
\label{4.15}
\end{equation}

Substituting the expression for $\delta$ from (\ref{4.11})
and using definition of $V_m$ eq.(\ref{47}), (\ref{52}) we obtain the
following correction to the function $V_m$:
\begin{equation}
\delta'_4 V_m =
\frac{4\pi s^2 c^4(3-10s^2)}{3\bar{\alpha}(c^2 -s^2)^2} \delta^2 =
\frac{3(3-10s^2)\bar{\alpha}t^2}{64\pi s^2(c^2 -s^2)^2} \;\; ,
\label{4.17}
\end{equation}

The correction to axial coupling constant $g_{Al}$ is easily derived from
equations (\ref{4.18}) (since $g_{Al} \sim f_0$), (\ref{4.11}) and definition of $V_{Al}$, equations
(\ref{48}), (\ref{52}):
\begin{equation}
g_{Al} = -\frac{1}{2} -\frac{1}{4}\delta -\frac{3}{16}\delta^2 \;\; ,
\label{4.19}
\end{equation}
\begin{equation}
\delta'_4 V_A = \frac{9\bar{\alpha}t^2}{64\pi s^2 c^2} \;\; .
\label{4.21}
\end{equation}

Finally, taking into account definition of $V_R$, equations (\ref{49}),
(\ref{52}), and equations (\ref{4.22}), (\ref{4.7})  we get:
\begin{eqnarray}
g_{Vl}/g_{Al} &=&  1-4[1-c^2
-\frac{c^2 s^2}{c^2 -s^2}\delta +\frac{c^4 s^4}{(c^2 -s^2)^3}\delta^2] =
\nonumber \\
&=& 1-4s^2 +\frac{4c^2 s^2}{c^2 -s^2}\delta -\frac{4c^4 s^4}{(c^2
-s^2)^3}\delta^2 \;\; ,
\label{1000}
\end{eqnarray}
\begin{equation}
\delta'_4 V_R = -\frac{3\bar{\alpha}t^2}{16\pi(c^2 -s^2)^2}
\label{4.24}
\end{equation}

Formulas (\ref{4.17}), (\ref{4.21}) and (\ref{4.24}) contain corrections to
the functions $V_i$ which come from the squares of polarization operators
and are proportional to $\bar{\alpha}t^2$ -- so, it is leading ($\sim t^2$)
parts of $(\Pi_Z -\Pi_W)^2$ corrections. Numerically they are several times
smaller than $\bar{\alpha}t^2$ corrections which originate from irreducible
diagrams.

\subsection{$\mbox{\boldmath$\alpha_W^2 t^2$}$ corrections from irreducible
diagrams.}

The major part of $\alpha_W^2 t^2$ corrections comes from the
irreducible
two-loop Feynman diagrams \cite{260, 270}.
The key observation in performing their
calculation is that these corrections are of the order of
$[\alpha_W(\frac{m_t}{m_Z})^2]^2 \sim \lambda_t^4$, where $\lambda_t$ is the
coupling constant of higgs doublet with the top quark. That is why they can
be calculated in a theory without vector bosons, taking into account only
top-higgs interactions \cite{260}. Corresponding pieces of vector boson
self-energies can be extracted from the self-energies of would-be-goldstone
bosons which enter Higgs doublet (those components which after mixing with
massless vector bosons form massive $W$- and $Z$-bosons). Correction of the
order of $\alpha_W^2 t^2$ is contained in the difference $\Pi_Z(0)
-\Pi_W(0)$ (see Figure 12), so it is universal, i.e. one and the same for $V_m$, $V_A$ and
$V_R$. In \cite{21} we call these corrections $\delta_4 V_i$:
\begin{equation}
\delta_4 V_i(t,h) =-\frac{\bar{\alpha}}{16\pi s^2 c^2}A(h/t)\cdot t^2
\;\; ,
\label{4.25}
\end{equation}
where function $A(h/t)$ is given in the Table 6.
To obtain this Table for $m_H/m_t < 4$ we use a Table from the paper
\cite{270}, and for $m_H/m_t > 4$  we use expansion over $m_t/m_H$ from the
paper \cite{260}.
For $m_t =175$ GeV and $m_H =150$ GeV we get $A=6.4$ and
$\delta_4 V_i(t,h) =-0.08$. This corresponds to the shifts: $-12$ MeV for
$m_W$, $7\cdot 10^{-5}$ for $s_l^2$ and $5\cdot 10^{-5}$ for $g_{Al}$. One
should compare these shifts with one-loop results: $\delta_{1 {\rm loop}}
m_W = 400$ MeV, $\delta_{1 {\rm loop}} s_l^2 = 50 \cdot 10^{-5}$ and
$\delta_{1 {\rm loop}} g_A = 100 \cdot 10^{-5}$. Let us remind that present
experimental accuracy in $m_W$ is $64$ MeV, in $s_l^2$ is $20\cdot
10^{-5}$ and in $g_{Al}$ is $30 \cdot 10^{-5}$.

There is one more place from which corrections $\sim\alpha_W^2 t^2$ appear:
this is the $Z\to b\bar{b}$ decay. At one electroweak loop $t$-quark can
propagate in the vertex triangle ($t\bar{t}W$)
(see Section 5). That is why at two loops correction
of the order $\alpha_W^2 t^2$ emerges. Due to this correction functions
$V_{Ab}(t,h)$ and $V_{Rb}(t,h)$ differ from the corresponding functions
describing $Z\to d\bar{d}$ decay:
\begin{equation}
V_{Ab}(t,h) =
 V_{Ad}(t,h)-\frac{8s^2 c^2}{3(3-2s^2)}(\phi (t) + \delta \phi (t,h))
\;\; ,
\label{4.27}
\end{equation}
\begin{equation}
V_{Rb}(t,h) = V_{Rd}(t,h)-\frac{4s^2 (c^2 -s^2 )}{3(3-2s^2)}(\phi (t) + \delta
 \phi (t,h)) \;\; ,
\label{4.28}
\end{equation}
where function $\phi(t)$ was discussed in Section 5 and
\begin{equation}
\delta\phi(t,h) = \delta_{\alpha_s}(t)\phi +\delta_H\phi(t,h) =
\frac{3-2s^2}{2s^2c^2}\left
\{-\frac{\pi^2}{3}(\frac{\hat{\alpha}_s(m_t)}{\pi})
t +\frac{1}{16s^2c^2}(\frac{\bar{\alpha}}{\pi})t^2
\tau_b^{(2)}(\frac{h}{t})\right\} \;\; ,
\label{4.29}
\end{equation}

First term
in curly braces,
$\delta_{\alpha_s}\phi$, was taken into account earlier, see Section 5, and
the new correction $\delta_H\phi(t,h)$
is proportional to function $\tau_b^{(2)}(h/t)$.
Function $\tau_b^{(2)}$ is given in Table 6.
To obtain this Table for $m_H/m_t < 4$ we use a Table from the paper
\cite{270}, and for $m_H/m_t > 4$  we use expansion over $m_t/m_H$ from the
paper \cite{260}
in full analogy with function $A(h/t)$.

For $m_t =175$ GeV, $m_H =150$ GeV we have $\tau_b^{(2)}=1.6$.

The change of $\Gamma_b$ due to $\tau_b^{(2)} = 1.6$ equals $0.03$ MeV, which
corresponds to $2 \cdot 10^{-5}$ shift in $R_b$, while experimental accuracy
in $R_b$ is $7\cdot 10^{-4}$ (the one e-w loop correction in $R_b$ is $-3.9
\cdot 10^{-3}$). The influence of $\tau_b^{(2)}$ on $A_{FB}^b$ and $A_b$ is
even smaller (by a few orders of magnitude).

\begin{center}

{\bf Table 6: Functions $\mbox{\boldmath$A(m_H/m_t)$}$ and
$\mbox{\boldmath$\tau_b^{(2)}(m_H/m_t)$}$.}

\vspace{5mm}

\begin{tabular}{|rrr|rrr|}
\hline
 $m_H/m_t$ & $A(m_H/m_t)$ & $\tau_b^{(2)}(m_H/m_t)$ &
 $m_H/m_t$ & $A(m_H/m_t)$ & $\tau_b^{(2)}(m_H/m_t)$
 \\ \hline
  .00  &    .739  & 5.710 &
 2.60  &  10.358  & 1.661 \\
  .10  &   1.821  & 4.671 &
 2.70  &  10.473  & 1.730 \\
  .20  &   2.704  & 3.901 &
 2.80  &  10.581  & 1.801 \\
  .30  &   3.462  & 3.304 &
 2.90  &  10.683  & 1.875 \\
  .40  &   4.127  & 2.834 &
 3.00  &  10.777  & 1.951 \\
  .50  &   4.720  & 2.461 &
 3.10  &  10.866  & 2.029 \\
  .60  &   5.254  & 2.163 &
 3.20  &  10.949  & 2.109 \\
  .70  &   5.737  & 1.924 &
 3.30  &  11.026  & 2.190 \\
  .80  &   6.179  & 1.735 &
 3.40  &  11.098  & 2.272 \\
  .90  &   6.583  & 1.586 &
 3.50  &  11.165  & 2.356 \\
 1.00  &   6.956  & 1.470 &
 3.60  &  11.228  & 2.441 \\
 1.10  &   7.299  & 1.382 &
 3.70  &  11.286  & 2.526 \\
 1.20  &   7.617  & 1.317 &
 3.80  &  11.340  & 2.613 \\
 1.30  &   7.912  & 1.272 &
 3.90  &  11.390  & 2.700 \\
 1.40  &   8.186  & 1.245 &
 4.00  &  11.396  & 2.788 \\
 1.50  &   8.441  & 1.232 &
 4.10  &  11.442  & 2.921 \\
 1.60  &   8.679  & 1.232 &
 4.20  &  11.484  & 3.007 \\
 1.70  &   8.902  & 1.243 &
 4.30  &  11.523  & 3.094 \\
 1.80  &   9.109  & 1.264 &
 4.40  &  11.558  & 3.181 \\
 1.90  &   9.303  & 1.293 &
 4.50  &  11.590  & 3.268 \\
 2.00  &   9.485  & 1.330  &
 4.60  &  11.618  & 3.356 \\
 2.10  &   9.655  & 1.373 &
 4.70  &  11.644  & 3.445 \\
 2.20  &   9.815  & 1.421 &
 4.80  &  11.667  & 3.533 \\
 2.30  &   9.964  & 1.475 &
 4.90  &  11.687  & 3.622 \\
 2.40  &  10.104  & 1.533 &
 5.00  &  11.704  & 3.710 \\
 2.50  &  10.235  & 1.595 & &  & \\
\hline
\end{tabular}
\end{center}

\subsection{$\mbox{\boldmath$\alpha_W^2 t$}$ corrections and the
two-loop fit of experimental data.}

Corrections of the order $\alpha_W^2 t$ originate from the top loop
contribution to $W$- and $Z$-boson self-energies with higgs or vector boson
propagating inside the loop and are of the order of $g^2\lambda_t^2$. We take
into account these corrections in our code LEPTOP using results of the papers
\cite{280} - \cite{300}.

Before we will present results of electroweak precision data fit which take
into account $\alpha_W^2$ corrections, described in this Section, we must
discuss how good the approximation which takes into account $\alpha_W^2 t^2$
and $\alpha_W^2 t$ terms but neglects (still not calculated) $\alpha_W^2$
terms should be. For $m_t =175$ GeV we obtain
$t\simeq 3.7$, thus at first glance we have
good expansion parameter so that $\alpha_W^2$ terms could be safely
neglected.  To check this let us consider first the one electroweak loop,
where the enhanced $\alpha_W t$ terms can be compared with non-enhanced
$\alpha_W$ terms.

By using eqs.(\ref{47}) and (\ref{49}) and by comparing them with
experimental data one sees that for $m_W/m_Z$ the $\alpha_W t$ term is equal
to $0.0057$, while the $\alpha_W$ term is $-0.0014$. As for $g_{Vl}/g_{Al}$,
the two terms are $0.0122$ and $-0.0142$. Thus for $m_W/m_Z$ the $\alpha_W t$
term dominates, while for $g_{Vl}/g_{Al}$ it is practically cancelled by the
$\alpha_W$ term.

Coming back to two-loop corrections we observe, that $\alpha_W^2 t^2$
correction to $m_W$ is not  larger than $\alpha_W^2 t$ correction;
for $m_t =175$ GeV and $m_H =150$ GeV it diminishes $m_W$ by $23$ MeV
(compare with Section 7.2).

In the Table 7 we present results of the fit of the data where we use
theoretical formulas which include  two-loop
electroweak corrections described in this Section.
Comparing Table 7 with Table 4 where the fit of the one-loop electroweak
corrected formulas was presented, we see that the fitted values of all
physical observables are practically the same with one (very important)
exception: the central value of the higgs mass becomes
$\sim 70$ GeV lower. In view
of the previous discussion it seems reasonable to consider this shift as
a cautious estimate of the theoretical uncertainty in $m_H$.

We have a simple qualitative explanation why $\alpha_W^2 t$ corrections
reduce the higgs mass by $\sim 70$ GeV. The point is that these
corrections shift the theoretical value of $s_l^2$ by $+0.0002$, which
is close to experimental error in $s_l^2$. In order to compensate
the shift the fitted mass of the higgs changes. This change can be
easily derived. Indeed, from eqs.(\ref{49}), (\ref{52}), (\ref{45})
we get:
\begin{equation}
\delta s^2 =-\frac{3}{16\pi(c^2 -s^2)}\bar{\alpha}\delta H_R = -0.00086
\delta H_R \;\; ,
\label{1014}
\end{equation}
while from Table 3 we see that changing  $m_H$ from $150$ GeV to $100$
GeV gives $\delta H_R = +0.236$ and $\delta s^2 = -0.0002$.

In Figure 13 the dependence of $\chi^2$ on the value of higgs mass is
shown separately with and without inclusion of SLD data ($Z$-decays into
heavy quark pairs are taken into account on both plots). When all existing
data are taken into account we get central value of higgs mass $m_H =71$ GeV
which is twenty GeV below direct bound \cite{11} from LEP-II search: $m_H >
95$ GeV. However, uncertainty in the value of $m_H$ extracted
from radiative corrections is
quite large, thus there is no contradiction between these two numbers.

At the end of this Section we would like to make two remarks demonstrating
that one should not take too seriously the central values of $m_H$ extracted
from the global fits.

First, if one disregards the FNAL measurements of
$m_t$, then one obtains from the fit:
$$
m_t = 160.7^{+7.7}_{-6.8} \; \mbox{\rm GeV} \;\; ,
$$
$$
m_H = 30.3^{+38.8}_{-14.4} \; \mbox{\rm GeV} \;\; .
$$

Such value of $m_H$ is by 1.5 standard deviations below the lower bound from
direct searches of LEP-II.( Note also that the fitted value of the top
mass $m_t$ is substantially lower than measured at FNAL).

Second, as it was stressed in ref. \cite{Chan}, the values of $s_l^2$
extracted from different observables lead to very different central values of
$m_H$. For example, from SLAC data on $A_{LR}$ it follows that $m_H = 25$
GeV with 90\% confidence interval from $6$ GeV to $100$ GeV. Even smaller values
of $m_H$ follow from LEP measurement of $A_{FB}^{\tau}$: $m_H =4$ GeV ($0.2$
GeV $< m_H < 95$ GeV at 90\% C.L.). As for other asymmetries measured at
LEP, they lead to much heavier higgs: from $A_{FB}^b$, for example, $m_H =
370$ GeV ($100$ GeV $< m_H < 1400$ GeV at 90\% C.L.). That is why the average
of all these values of $m_H$ seems to be not very reliable.

\newpage

\begin{center}

{\bf Table 7. Fit of experimental data \cite{11} with two-electroweak-loop
 formulas.}

\vspace{2mm}

$m_Z =91.1867(21)$ GeV is used as an input.

Output of the fit: $m_H =70.8^{+82}_{-43}$ GeV$^*$,
$\hat{\alpha}_s = 0.1194 \pm
0.0029$, $\chi^2/n_{d.o.f.} =15.0/14$

\vspace{2mm}

\begin{tabular}{|l|r|r|r|}
\hline
Observable & Experimental & Fit Standard & Pull \\
 & data & Model & \\
\hline
a) LEP-I & & & \\
shape of $Z$-peak and & & & \\
lepton asymmetries: & & & \\
$\Gamma_Z$ [GeV] &    2.4939(24)  &  2.4960(18)  & -0.9   \\
$\sigma_h$ [nb] &   41.491(58)  & 41.472(16)  & 0.3    \\
$R_l$ &   20.765(26)  & 20.746(20)  & 0.7   \\
$A_{FB}^l$ & 0.0168(10)  &  0.0161(4)  & 0.7   \\
$\tau$-polarization: & & & \\
$A_{\tau}$ & 0.1431(45) &   0.1467(16)  & -0.8   \\
$A_e$ &  0.1479(51) &    0.1467(16) & 0.2  \\
results for heavy quarks: & & & \\
$R_b$$^{**}$ &    0.2166(7)  &   0.2158(2)  & 1.0   \\
$R_c$$^{**}$    &    0.1735(44)  &   0.1723(1)  & 0.3   \\
$A_{FB}^b$  &    0.0990(21)  &   0.1028(12)  & -1.8   \\
$A_{FB}^c$  &    0.0709(44)  &   0.0734(9)  & -0.6   \\
charge asymmetry for pairs of  & & & \\
light quarks $q\bar{q}$: & & & \\
$s_l^2(Q_{FB})$   &    0.2321(10)  &   0.2316(2)  & 0.5   \\
\hline
b) SLC & & & \\
$s_l^2$ ($A_{LR}$) & {\it 0.2311(3)}  &  {\it  0.2316(2)}  & -1.6   \\
$A_{LR}$   & 0.1504(23)  &   0.1467(16)  & 1.6   \\
$R_b$$^{**}$ &    0.2166(7)  &   0.2158(2)  & 0.9   \\
$R_c$$^{**}$ &  0.1735(44) & 0.1723(1)  & 0.3 \\
$A_b$ &  0.8670(350)  &  0.9348(2)  & -1.9   \\
$A_c$ &    0.6470(400)  &   0.6677(7)  & -0.5   \\
\hline
c) $p\bar{p} +$ LEP-II $+ \nu N$ & & & \\
$m_W$ [GeV] ($p\bar{p} +$ LEP-II) &
80.3902(64)  & 80.3659(34)  & 0.4   \\
 & {\it 0.2228(13)}  &    &    \\
$s_W^2$ ($\nu N$) & 0.2254(21) & 0.2233(7) & 1.0 \\
  & {\it 80.255(109)} &   &  \\
$m_t$ [GeV]    & 173.8(5.0)  & 170.8(4.9) & 0.6\\
\hline
\end{tabular}
\end{center}
\noindent
$^*$ {\footnotesize The most optimistic errors on $M_H$ are obtained in the
fit including $\bar{\alpha}(DH)^{-1} = 128.923(36)$ \cite{DH} and
$\alpha_s(PDG) = 0.1178(23)$ from low energy data \cite{13}. Such a fit gives
$m_H = 93^{+63}_{-41}$ GeV, $m_t = 171.3 \pm 4.8$ GeV, $\alpha_s = 0.1184 \pm
0.0018$, $\chi^2/n_{d.o.f.} = 15.2/14$. However the systematic errors
due to the model assumptions used in the calculations of $\alpha_s(PDG)$ and
$\bar{\alpha}(DH)$ are not easy to estimate. That is why we prefer to use the
result with less optimistic assumptions leading to bigger error in $m_H$.}
\\
$^{**}$ {\footnotesize Experimental values of $R_b$ and $R_c$ correspond to
the average of LEP-I and SLC results.}

\newpage

As can be seen from the Table 8, the LEPTOP fit is very close to the
ZFITTER fit \cite{11} and to the fit by Erler and Langacker \cite{erler}.
This indicates that theoretical uncertainties are very small, 
except for the non-calculated part of the corrections, that is 
common to all three programs.

\begin{center}

{\bf Table 8.  Comparison of the LEPTOP fit with the ZFITTER fit
\cite{11} and with the fit by Erler and Langacker \cite{erler}.}

\vspace{2mm}

\begin{tabular}{|l|r|r|r|r|r|}
\hline
Observable & Experimental         & Fit LEPTOP   & Fit EWWG & Fit Erler- \\
 & data &  & ZFITTER & Langacker$^{*}$  \\
\hline
a) LEP-I & & & & \\
$M_Z$ [GeV] &    91.1867(21)     & 91.1867 fix. & 91.1865 & 91.1865(21)  \\
$\Gamma_Z$ [GeV] &    2.4939(24)  &  2.4960(18)  &  2.4958 & 2.4957(17)  \\
$\sigma_h$ [nb] &   41.491(58)    & 41.472(16)   & 41.473  & 41.473(15)  \\
$R_l$ &   20.765(26)              & 20.746(20)   & 20.748  & 20.748(19)  \\
$A_{FB}^l$ & 0.0168(10)           &  0.0161(4)   & 0.01613 & 0.0161(3)  \\
$A_{\tau}$ & 0.1431(45)           &  0.1467(16)  & 0.1467  & 0.1466(15)    \\
$A_e$ &  0.1479(51)               &  0.1467(16)  & 0.1467  & 0.1466(13)  \\
$R_b$ &    0.2166(7)       &  0.2158(2)   & 0.2159  & 0.2158(2)  \\
$R_c$    &    0.1735(44)   &  0.1723(1)   & 0.1722  & 0.1723(1)  \\
$A_{FB}^b$  &    0.0990(21)       &  0.1028(12)  & 0.1028  & 0.1028(10)  \\
$A_{FB}^c$  &    0.0709(44)       &  0.0734(9)   & 0.0734  & 0.0734(8)  \\
$s_l^2(Q_{FB})$   &    0.2321(10) &  0.2316(2)   & 0.23157 & 0.2316(2)  \\
\hline
b) SLC & & & & \\
$s_l^2$($A_{LR}$)& 0.2311(3)      &  0.2316(2)   & 0.23157 & --  \\
$A_{LR}$   & 0.1504(23)           &  0.1467(16)  & ---     & 0.1466(15)  \\
$A_b$ &  0.8670(350)              &  0.9348(2)   & 0.935   & 0.9347(1)  \\
$A_c$ &    0.6470(400)            &  0.6677(7)   & 0.668   & 0.6676(6)  \\
\hline
c) $p\bar{p} +$ LEP-II $+ \nu N$ & & & & \\
$m_W$ [GeV] ($p\bar{p} +$ LEP-II) &
80.3902(64)                       & 80.3659(34)  & 80.37   & 80.362(23)  \\
 & {\it 0.2228(13)}  &    &    & \\
$s_W^2$ ($\nu N$) & 0.2254(21)    & 0.2233(7)    & 0.2232  &    \\
  & {\it 80.255(109)} &   & & \\
\hline
$m_t$ [GeV]    & 173.8(5.0)       & 170.8(4.9)   & 171.1(4.9)  & 171.4(4.8) \\
 & & & & \\
$m_H$ [GeV] & & $ 71.^{+82}_{-43}$ & $76.^{+85}_{-47}$ & $107.^{+67}_{-45}$ \\
 & & & & \\
$\alpha_s$     &                  &   0.1194(29) & 0.119(3)    & 0.1206(30) \\
$\bar\alpha^{-1}$   & 128.878(90) &  128.875     & 128.878     & \\
 & & & & \\
\hline
\end{tabular}
\end{center}

$^*$ {\footnotesize
Erler-Langacker use slightly different experimental dataset for their fit.
This may cause some of the discrepancies with LEPTOP and ZFITTER.
}
\\

\newpage

\section{Extensions of the Standard Model.}

The Standard Model works well at the energy scale of the order
of the vector bosons masses. We see that the SM description of the electroweak
observables in this energy region is in perfect agreement with the precision
measurements.

 However there are many natural physical questions that have no satisfactory
answers
 within the framework of the SM. So it is hard to believe that the Standard
Model is the Final Theory. The common expectation is that there should be
New Physics beyond the Standard Model.

  Direct accelerator searches did not find yet any trace of New Physics.
Their negative  results gave lower bounds on the masses and upper
bounds  on the production cross sections for the new particles.
In this section we are going to study the indirect bounds on
New Physics that can be theoretically derived from the precision
measurements at low energy of the order of $Z$ and $W$ boson masses. Loops
with hypothetical new particles change the predictions of the SM for
electroweak observables.  Since the SM gives a very good description of the
data there is little room for such new contributions.  In this way one can
get some kind of constraints on  new theory.

  Any possible generalizations of the SM are naturally divided into two
classes: theories with and without decoupling.
In the first class the contribution of new particles into $W$ and
$Z$ boson parameters  are suppressed as  positive powers of $ (m_Z^2/m^2)^n$
when the masses  of new particles $m$ become larger than electroweak scale.
One cannot exclude such  theory by studying loop corrections to low-energy
observables. In this way one may have hopes to bound the masses of new
particles from below. The most famous example of such theory are
supersymmetric extensions of the SM.

  In the second class of theories the contribution of new particles into
low-energy observables
does not decouple even when their masses become very large. Such
 SM generalizations can be excluded if the additional
nondecoupled contributions exceed the discrepancy between SM fit and
experimental data. The example of such generalization is the SM
with additional sequential generations of quarks and leptons.

\subsection{Sequential heavy generations in the Standard Model.}

 We start the discussion of New Physics with the simplest extension
of the SM, namely with
the SM with additional sequential generations of leptons and quarks
(\cite{aaa}- \cite{cc}).
Nobody knows any deep reason for the number of generations to be equal
to three.
So it is interesting to study whether it is allowed to have four and more
generations. Certainly these new generations should be
heavy enough  not to be produced in the $Z$ decays and at LEP-II.

  We consider the case of no mixing between the known generations and
  the new ones.
In this case the new fermion generations affect the ratio $m_W/m_Z$ and the
widths and the
decay asymmetries of the $Z$ boson only through the vector bosons
self-energies.
Such kind of corrections have been dubbed \cite{64}
"oblique corrections".  We start
their study  with the case of $SU(2)$ degenerate
fourth generation:
\begin{equation}
m_U = m_D =  m_Q \;\; , \quad\quad m_N = m_E = m_L
\label{1.1}
\end{equation}

New terms in the self-energies modify the  functions $V_m, V_A, V_R$ , i.e. the
radiative corrections to $m_W/m_Z$, $g_{Al}$ and $g_{Vl}/g_{Al}$.
The contribution to $V_i$ from the fourth generation can be written in the form:
\begin{equation}
V_m \to V_m +\delta^4 V_m \; , \;\; V_A\to V_A +\delta^4 V_A\; , \;\;
V_R\to V_R
+\delta^4 V_R \;\; .
\label{1.2}
\end{equation}
The analytical expressions for
$\delta^4 V_i$ for quark or lepton doublets  (neglecting
gluonic corrections) can be found in Appendix G, eqs (G.1)-(G.3).

 In the limit of a very heavy fourth generation of leptons and quarks
one has :
\begin{equation}
\Sigma\delta^4 V_m \to -\frac{16}{9}s^2 \; , \;\;\Sigma \delta^4 V_R \to
-\frac{8}{9} \; , \;\; \Sigma\delta^4 V_A \to 0 \;\; ,
\label{1.7}
\end{equation}
where $\Sigma$ denotes sum over leptons and quarks with $m_Q = m_L = m_4$,
$s^2 \simeq 0.23$.
Equations (\ref{1.7}) reflect the non-decoupling of
the heavy
degrees of freedom in electroweak theory, caused by the axial current. It is
interesting that the contribution of degenerate generation to $V_m, V_R$ has
negative sign.

   The fourth generation with strong violation of $SU(2)$ symmetry (i.e. with
very large mass difference in the doublet) gives universal contribution to
functions   $\delta^4 V_i$  (similar to the universal contribution of
$t$- and $b$-quarks from the third generation to $V_i$)

\begin{equation}
\delta^4 V_i = 4|m^2_T-m^2_B|/3m^2_Z .
\label{1.8}
\end{equation}

In the case of large mass splitting  $\delta^4 V_i$ are positive.
From eqs.(\ref{1.7}) and (\ref{1.8}) it is clear that somewhere in the
intermediate region
 of mass splitting the functions $\delta^4 V_m$ and $\delta^4 V_R$
intersect zero.
In the vicinities of these zeroes the contribution of new generation
to these specific observables is negligible and one can not exclude these
regions of masses studying only
one of the observables. Fortunately for different
observables  these zeroes are located in the different
places and the general fit overcomes such conspiracy of new physics.

 For different up- and down- quark (and lepton)
masses analytical expressions for
$\delta^4 V_i$ are given in Appendix G, eqs.(\ref{1.9}) - (\ref{1.11}).

 Figure 14 demonstrates 2-dimensional exclusion plot for the case of $n$
 extra generations, where $n$ is formally considered as a continuous
parameter.  We see from this plot that at 90\% c.l. we have less than one
 extra generation and at 99\% c.l.-- less than two extra generations for any
 differences of up- and down- quarks masses.

\subsection{SUSY extensions of the Standard Model.}

  In this section we consider another example of new physics: supersymmetric
extensions of the SM. There are certain aesthetic and conceptual merits
of such SUSY generalization of the SM. Here are some of them:

1) Supersymmetry gives a solution for the problem of fine tuning, i.e. it
prevents the electroweak scale of the SM from mixing with the Planck scale.

2) The problem of unification of electroweak and strong
coupling constants seems to have solution in the
framework of SUSY extensions.

3) Finally, any ambitious "Theory of Everything" inevitably includes SUSY
as the basic element of the construction.

 To make systematic introduction into SUSY extensions of the SM one needs a
separate review paper ( for the review papers see e.g. \cite{81} ). Here we
are going to make a short sketch of this well developed branch of physics in
applications to the theory of $Z$ boson. To construct SUSY extensions one has
to introduce a lot of new particles. For example minimal $N=1$ supersymmetry
automatically doubles the number of degrees of freedom of the SM: any
fermionic degree of freedom has to be coupled with bosonic degree of freedom
and vice versa. Thus the left ( right ) leptons have to be accompanied by
scalars: ``left" (``right" ) sleptons, quarks by squarks, gauge bosons by
spinor particles -- gauginos, etc. The
Higgs mechanism of mass generation for up and down quarks
requests for two Higgs boson doublets ( and two
higgsino doublets respectively).

 Not a one  of these numerous new particles has been observed
yet. If they do exist they are too heavy  to be produced at the working
accelerators. On the other hand, these heavy supersymmetric
particles (again if they do exist) are produced in the virtual states, i.e. in
the loops. Loops with new particles change the predictions of the SM for
the low-energy
observables. ( Under ``low-energy" we mean here $E\la m_Z $ ). In this
 indirect way one can get some information about existence or nonexistence of
SUSY.

 The SUSY extensions of the SM belong to the class of new physics
that decouples from the low-energy observables when the mass scale of this new
physics becomes very large. It means that the additional contribution into
electroweak observables due to the supersymmetric particles are of the order of
$\alpha_W(m_W/m_{SUSY})^2$ or $\alpha_W(m_t/m_{SUSY})^2$ , where $m_{SUSY}$
characterizes the mass scale of
superpartners. Since the fit of the precision data in the framework of the SM
statistically is very good these new additional contributions have to be small.
So in this way one expects to get
strong restrictions on the value of $m_{SUSY}$.

 Supersymmetric contributions into low-energy observables were studied in
papers \cite{82} -
\cite{85}. The results depend on the model and on the pattern of SUSY
violation. Within a given model the results for low-energy observables are
formulated in terms of the functions that depend on the fundamental
parameters of the SUSY Lagrangian that are fixed at the high energy scale of
SUSY violation. The fit of experimental data in the framework of a given
SUSY model imposes certain restrictions on the allowed region of these
high-energy scale parameters of the model. As for the masses of sparticles
their values are calculated by numerical solution of the
renormalization group equations. They also depend on the fundamental SUSY
parameters at high energy scale. In this rather indirect way one gets
restrictions on the physical masses of sparticles in general case.

To give the reader the taste of exploration of the new supersymmetric
physics we consider
in this section only that part of the multi-dimensional space of SUSY
parameters for which all sparticles have more or less
the same masses, i.e. when we have no light sparticles.
(It seems reasonable to start the study of the unknown field
 with such kind of the simplest assumptions). In this case one can find the
class of enhanced oblique corrections that are universal, i.e. that are
the same
for any model. Another merit of these corrections is that they directly depend
on the masses of sparticles.

As will be shown,
the enhanced electroweak radiative SUSY corrections are
induced by
 the large violation of
$SU(2)_L$ symmetry in the third generation of squarks. Therefore
 we start the discussion of the SUSY corrections to the functions $V_i$
with the brief description of the stop $(\tilde{t}_L ,\tilde{t}_R)$  and
sbottom
$(\tilde{b}_L ,\tilde{b}_R)$ sector of the theory. The following relation
between masses of quarks $q$ and diagonal masses of left squarks $\tilde{q}_L$
takes place in a wide class of SUSY models:
\begin{equation}
m^2_{\tilde{q}_L} =m_q^2 + m_{SUSY}^2 + m_Z^2 \cos(2\beta)(T_3 - s^2 Q_q)
\;\; ,
\label{8.0}
\end{equation}
where $s^2 \simeq 0.23$,
 $Q_q$ is the charge and $T_3$ is the third projection of weak isospin
of quark
and ${\rm tg}
 \beta$ is equal to
the ratio of the {\it vev}s of two Higgs fields, introduced in SUSY
models. The second term in the r.h.s of the eq.(\ref{8.0}) violates
supersymmetry.  It is some universal SU(2)-blind SUSY violating soft mass
term. The third term in r.h.s of the eq.(\ref{8.0}) also violates SUSY. It
originates from quartic $D$ term in the effective potential and is different
for $up$ and $down$ components of the doublets. The only hypothesis that is
behind this relation is that the origin of the large breaking of this
$SU(2)_L$ is in the quark-higgs interaction.

 Therefore from eq.(\ref{8.0}) we get the following
relation between masses of stop $m^2_{\tilde{t}_L}$, of sbottom
$m^2_{\tilde{b}_L}$ and of top $m^2_{t}$ ( we neglect $m_b$ ):
\begin{equation}
m^2_{\tilde{t}_L} - m^2_{\tilde{b}_L} =m_t^2 +m_Z^2 \cos(2\beta) c^2 \;\; ,
\label{8.1}
\end{equation}
 Relation (\ref{8.1}) is central for this approach. It demonstrates the
large violation of $SU(2)_L$ symmetry in the third generation of squarks.
On the other hand it demonstrates that in the
limit of very large mass  the left stop and left sbottom  become  degenerate
and the parameter $(m^2_{\tilde{t}_L} - m^2_{\tilde{b}_L})/ m^2_{\tilde{b}_L}$
goes to zero when $m_{SUSY}$ goes to infinity. That
is why the physical observables can depend on this decoupling parameter.

 As for the right sparticles from the third generation, they are
$SU(2)_L$ singlets. But they can mix with the
left sparticles and in this way they contribute into enhanced corrections.
The mixing between $\tilde{b}_L ,\tilde{b}_R$  has to be
 proportional to $m_b$ and can be neglected. The
  $\tilde{t}_L\tilde{t}_R$ mass matrix in general has the following form:
\begin{equation}
\left(
\begin{array}{cc}
m^2_{\tilde{t}_L} & m_t A'_t \\
m_t A'_t & m^2_{\tilde{t}_R}
\end{array}
\right) \;\; ,
\label{8.2}
\end{equation}
where $\tilde{t}_L \tilde{t}_R$ mixing is proportional to $m_t$ and
therefore is not small. Coefficient $A'_t$ depends on the model.
Diagonalizing matrix (\ref{8.2}) we get the
following eigenstates:
\begin{equation}
\left\{
\begin{array}{l}
\tilde{t}_1 = c_u\tilde{t}_L +s_u\tilde{t}_R \\
\tilde{t}_2 = -s_u\tilde{t}_L +c_u\tilde{t}_R \;\; ,
\end{array}
\right.
\label{8.3}
\end{equation}
where $c_u\equiv \cos\theta_{LR}$, $s_u\equiv \sin\theta_{LR}$,
$\theta_{LR}$ is
the $\tilde{t}_L
\tilde{t}_R$ mixing angle, and
\begin{equation}
{\rm tg}^2\theta_{LR} =
\frac{m_1^2 -m_{\tilde{t}_L}^2}{m_{\tilde{t}_L}^2 -m_2^2} \;\; ,
\;\; m_1^2 \geq m_{\tilde{t}_L}^2 \geq m_2^2 \;\; .
\label{8.4}
\end{equation}
Parameters $m_1$ and $m_2$ are the mass eigenvalues:
\begin{equation}
m_{1,2}^2 =
\frac{m^2_{\tilde{t}_L} +m^2_{\tilde{t}_R}}{2} \pm
\frac{|m^2_{\tilde{t}_L} -m^2_{\tilde{t}_R}|}{2}
\sqrt{1+\frac{4m_t^2 A'^{2}_t}{(m^2_{\tilde{t}_L} -m^2_{\tilde{t}_R})^2}}
\;\; .
\label{8.5}
\end{equation}

The enhanced electroweak radiative corrections are induced by the contribution
of the third generation of squarks into self-energy operators of vector bosons.
Nondiagonal
vector currents of squarks are not conserved only because of violation
of SU(2) by mass terms.
Thus one should expect that the self-energy operators are
proportional to the divergency of the currents. To calculate these enhanced
terms it is sufficient to expand the
operators of the vector bosons $\Sigma_V(k^2)$ at $k^2 =0$. The terms enhanced
as $m_t^4/ M_{SUSY}^2$ come from
$\Sigma_W(0)$, while those enhanced as $m_t^2M_W^2/M_{SUSY}^2$ come from
$\Sigma'_{W,Z}(0)$ (see Figure 15).
 These simple self-energy corrections are obviously
universal since stop and sbottom should exist in any SUSY model and the
coupling constants are universal since they are fixed by gauge invariance
only. The higher-order derivatives of self energies are suppressed
as $(m_{W,Z}/m_{SUSY})^2$. They are of the same order of magnitude as the
numerous model-dependent terms coming from vertex and box diagrams. If there
are no very light sparticles the first two universal terms have rather large
enhancement factor of the order of  $t^2\simeq 14$  and $t\simeq
3.7$ respectively.(The presence of terms $\sim m_t^4$ in SUSY models was
recognized long ago \cite{86} ). We neglect the  non-enhanced terms.  The
accuracy of such approximation may be of the order of ten percent
if we are lucky, but it may be as well of the order of unity (see
discussion of $V_R$ in Sections 5 and of the two-loop corrections in Section
7).  As for the stop contributions to the vertex corrections there is only
one relevant case - the amplitude of $Z\to b\bar{b}$ decay.  For vertex with
stop exchange there are no terms enhanced as $(m_t/m_W)^4$ \cite{87}.
Thus we will neglect corresponding corrections as well.

The calculation of the enhanced two terms is a rather trivial exercise. The only subtle point is the diagonalization of the stop propagators. The result of calculations depends on 3 parameters: $m_1$, $m_2$ and $m_{\tilde{b}_L}$.
The dependence on angle $\beta$ is very moderate and in numerical fits we will
use rather popular value ${\rm tg} \beta =2$. In what follows instead of
$m_{\tilde{b}_L}$ we will write $m_{\tilde{b}}$, bearing in mind that
$\tilde{b}_L \tilde{b}_R$ mixing is proportional to $m_b$ and can
be neglected. The formulas that describe
the enhanced SUSY
corrections to the functions $V_i$ can be found in the Appendix G,
eqs.(\ref{8.12}) - (\ref{8.16}).

There is also another source of the potentially large SUSY corrections:
vertices with gluino exchange  of
the order $\hat \alpha_s(m_Z/m_{SUSY})^2$.

These corrections  shift the radiators $R_{V_q}$
and $R_{A_q}$ in eq.(\ref{31}) \cite{89}:
\begin{equation}
\delta R_{V_q} = \delta R_{A_q} =
1+ \frac{\hat{\alpha}_s (m_Z)}{\pi} \Delta_1(x,y) \;\; ,
\label{8.55}
\end{equation}
\begin{equation}
\Delta_1(x,y) =
-\frac{4}{3}\int\limits_0^1 dz_1 \int\limits_0^{1-z_1} dz_2 \log
[1- \frac{xyz_1 z_2}{x+(z_1 +z_2)(y-x)}] \;\; ,
\label{8.66}
\end{equation}
where $x =(m_Z/m_{\tilde{q}})^2$, $y=(m_Z/m_{\tilde{g}})^2$, and
$\Delta_1(x,x)\simeq \frac{1}{18}x + ...$.
We take these gluino corrections into account in our analysis.
The electroweak SUSY corrections to $g_{A_q}$ and $g_{V_q}$ are generated
by the corrections to the function $V_A$
eq.(\ref{8.12}) and $V_R$  eq.(\ref{8.13}).

 Having all the necessary formulas in hands we start the new fit of
the data with the simplest case of the absence of $\tilde{t}_L \tilde{t}_R$
mixing, $\sin\theta_{LR} =0$. In this case we have only one additional mass
parameter. Thus we
expect that this mass should be heavy enough not to destroy the perfect SM fit
of the experimental data. First let us take the value of the lightest
neutral Higgs boson mass as a free parameter and take the masses of the other
three Higgs bosons to
be very heavy. The results of the fit are shown in Table 9. We see that
 to fit the data with light sbottom one has  to take the mass of the Higgs
boson much larger
than its Standard Model fit value. Even in this case the quality of
the fit is worse than the SM one. For very heavy sbottom one reproduces
the SM fit. (To reduce the number of parameters we take $m_{\tilde{g}}
= m_{\tilde{b}}$ in this fit. Let us stress that light squarks
with masses of the order of $100-200$ GeV are usually allowed only if gluinos
are heavy, $m_{\tilde{g}} \ge 500$ GeV \cite{810}. In the case of heavy gluino
the correction $\Delta_1$ (eq. (\ref{8.66})) becomes power suppressed and we
return to the Standard Model fit value of $\hat \alpha_s=0.119(3)$).

\newpage
\begin{center}

{\bf Table 9}

\vspace{3mm}

\begin{tabular}{|c|c|c|c|}
\hline
$m_{\tilde{b}}$ (GeV) & $m_h$ (GeV) & $\hat \alpha_s$ & $\chi^2/n_{d.o.f.}$ \\
\hline
100 & $850^{+286}_{-320}$ & $0.113 \pm 0.003$ & 20.3/14 \\
150 & $484^{+364}_{-235}$ & $0.116 \pm 0.003$ & 18.1/14  \\
200 & $280^{+240}_{-144}$ & $0.117 \pm 0.003$ & 17.3/14 \\
300 & $152^{+145}_{-87}$ & $0.118 \pm 0.003$ & 16.3/14 \\
400 & $113^{+115}_{-68}$ & $0.119 \pm 0.003$ & 15.8/14 \\
1000 & $77^{+87}_{-47}$ & $0.119 \pm 0.003$ & 15.2/14 \\
\hline
\end{tabular}
\end{center}


{\it Fit of the precision data with SUSY corrections taken into account in the
case of the absence of $\tilde{t}_L \tilde{t}_R$ mixing, $\sin\theta_{LR} =0$
and
$m_h$ taken as a free parameter. For $m_{\tilde{b}} > 300$ GeV SUSY
corrections become negligible and SM fit of the data is reproduced.}

\vspace{3mm}

We see that to get reasonably good fit of the data in the framework of the
 SUSY extensions with light squarks one has to put the lightest Higgs mass in
 the TeV region. It is time to remind that
 in SUSY models the mass of Higgs boson is not an absolutely free parameter.
In MSSM (Minimal Supersymmetric Standard Model)
among three neutral Higgs bosons the lightest one should have mass less than
approximately $120 - 135$ GeV \cite{631}.
If other higgses are considerably heavier the
lightest scalar
 boson has the same couplings with gauge bosons as in the Standard
Model. As a result the same SM formulas for radiative corrections can be used
in the SUSY extensions of the SM.
(Deviations from the SM formulas are suppressed as $(m_h/m_A)^2$,
where $m_A$ -- the mass of the heavier higgs. We will assume
in our analysis that $m_A$ is large). For maximal allowed value
$m_h \simeq 120$ GeV the results of the fit are
shown in Table 10. (In what follows we will always take $m_h\simeq 120$ GeV
since for $90$ GeV $< m_h < 135 $ GeV results of the fit are practically
the same.) This table demonstrates that superpartners should be heavy
if we want to have good quality fit of the data.

\begin{center}

{\bf Table 10}

\vspace{3mm}

\begin{tabular}{|c|c|c|}
\hline
$m_{\tilde{b}}$ (GeV) & $\hat \alpha_s$ & $\chi^2/n_{d.o.f.}$ \\
\hline
100 & $0.110 \pm 0.003$ & 30.2/15\\
150 & $0.115 \pm 0.003$ & 21.9/15 \\
200 & $0.116 \pm 0.003$ & 18.6/15  \\
300 & $0.118 \pm 0.003$ & 16.4/15  \\
400 & $0.119 \pm 0.003$ & 15.8/15 \\
1000 & $0.119 \pm 0.003$ & 15.5/15 \\
\hline
\end{tabular}

\end{center}

{\it The same as Table 9 but with the value of the lightest Higgs boson mass
$m_h = 120$ GeV which is about the maximum allowed value in the simplest SUSY
models.}

\vspace{3mm}

The next step is to take into account $\tilde{t}_L \tilde{t}_R$ mixing. In
the Figure 16 we show the dependence of SUSY corrections $\delta_{SUSY} V_i$ on
$m_1$ and $m_2$ for $m_{\tilde{b}} = 200$ GeV.
One clearly sees from this Figure that even for
this small value of $m_{\tilde{b}}$ there exist the domain of low $m_2$
values where the enhanced radiative corrections are suppressed.
In Figure 16 one sees the
valley where $\delta_{SUSY} V_i$ reaches its minimum values which are
considerably smaller than 1. The valley starts at $m_2 \simeq
m_{\tilde{b}}$, $m_1\simeq 1000$ GeV and goes to $m_2\simeq 100$ GeV, $m_1
\simeq 400$ GeV. The smallness of the radiative corrections at the point $m_2
\simeq m_{\tilde{b}}$, $m_1\simeq 1000$ GeV can be easily understood: here
$\theta_{LR} \simeq \pi/2$, $\tilde{t}_2 \simeq \tilde{t}_L$,
$\tilde{t}_1 \simeq \tilde{t}_R$. Thus nondiagonal charged left current of squarks is
conserved and the main enhanced term vanishes. Indeed in
$\delta_{SUSY} V_A$ only the term proportional to $g(m_2, m_{\tilde{b}})$
remains in eq.(\ref{8.12}),
but for $m_2 = m_{\tilde{b}}$ it is equal to zero. At this end point of
the valley $\tilde{t}_2 \simeq \tilde{t}_L$, $\tilde{t}_1 \simeq \tilde{t}_R$, so $m^2_{\tilde{t}_R} \gg
m^2_{\tilde{t}_L}$, which is opposite to the relation between
$m_{\tilde{t}_R}$ and $m_{\tilde{t}_L}$ occurring in a wide class of models.
In these models (e.g. in the MSSM) the left and the
right squark masses are equal at high energy scale. When renormalizing
them to low energies one gets $m^2_{\tilde{t}_L} >
m^2_{\tilde{t}_R}$. Almost along the whole valley we have
${\rm tg}^2\theta_{LR} >1$,
which means that $m^2_{\tilde{t}_R} > m^2_{\tilde{t}_L}$.
This possibility to suppress radiative
corrections was discussed in \cite{770} .
However, in the vicinity of the end point $m_1\simeq 300$ GeV, $m_2\simeq 70$
GeV the value of ${\rm tg}^2\theta_{LR}$ becomes smaller than 1 and
$m^2_{\tilde{t}_R} < m^2_{\tilde{t}_L}$.

In Table 11 we show values of $\chi^2$ along the valley of its minimum, which
is formed for $m_{\tilde{b}} = 200$ GeV. We observe that good
quality of fit is possible for light superpartners if $\tilde{t}_L
\tilde{t}_R$ mixing is taken into account.

\vspace{5mm}
\begin{center}

{\bf Table 11}

\vspace{3mm}

\begin{tabular}{|c|c|c|c|}
\hline
$m_1$ (GeV) & $m_2$ (GeV) & $\hat \alpha_s$ & $\chi^2/n_{d.o.f.}$ \\
\hline
1296 & 193 & $0.118 \pm 0.003$   & 15.6/15 \\
888 & 167 & $0.118 \pm 0.003$   & 15.8/15 \\
387 & 131 & $0.118 \pm 0.003$   & 16.1/15 \\
296 & 72 & $0.117 \pm 0.003$   & 16.7/15 \\
\hline
\end{tabular}
\end{center}

{\it Results of fit along
the valley of minimum of $\chi^2$ for fixed value $m_{\tilde{b}}\simeq 200$
GeV and $m_h\simeq 120$ GeV .}

\vspace{3mm}

The main lesson of this subsection is the following.
The fit of the precision data on electroweak observables (i.e. of $Z$-boson
decay parameters from LEP and SLC and
the values of $m_W$ and the $m_t$ from Tevatron) in the framework of
 SUSY extension of the SM assuming small value of $m_{\tilde{b}}$, the
absence of $\tilde{t}_L \tilde{t}_R$
mixing and $m_h = 120$ GeV leads to the growth of $\chi^2$ value. For heavy
squarks the SUSY sector of the theory decouples from low-energy observables
and the results of Standard Model fit are reproduced. On the other hand  even
for light sbottom and for small mass of
one of two stops, one can find the values of $\tilde{t}_L
\tilde{t}_R$ mixing where supersymmetric corrections appear to be small and
not excluded by experimental data. In this case the quality of the fit
(i.e. the value of $\chi^2$) is almost the same as in the Standard Model.

\section{Conclusions.}

The comparison of LEP-I and SLC precision data on  $Z$ boson decays
with calculations based on the Minimal Standard Model has confirmed
the predictive power of the latter:
\begin{enumerate}
\item
It was proved that there exist only three generations of quarks
and leptons with light neutrinos.
\item
The $Z$ boson couplings of quarks, charged leptons and
neutrinos are in accord with the theory.
\item
From the analysis of the radiative corrections the mass of the top
quark had been correctly predicted before this particle was
discovered at Tevatron.
\item
All electroweak observables (except for the mass of the higgs)
are perfectly fitted by one loop electroweak corrections (with
virtual and "free" gluons being taken into account).
\item
The dependence of the radiative corrections on the mass of the
higgs is feeble when higgs is heavy. Therefore the value of the higgs
mass extracted from LEP-I and SLC data has  rather large  error bars.
Within one
standard  deviation the  central  fitted value of the  higgs mass
becomes smaller when the leading two-electroweak-loop corrections
are taken into account. In this case it is close to 90 GeV -- its
direct lower limit from the LEP-II  search. However the non-leading two
loop corrections may change this result. Calculation of these corrections
is a challenge to theorists. Better understanding of systematic
discrepancies between various asymmetries in $Z$-decays is a challenge
to experimentalists.
\item
One of the main conclusions of the one-electroweak-loop case is
that in this case the value of the leading
and non-leading corrections are comparable and even may cancel each
other (in the case of leptonic parity violating parameter
 $g_{Vl}/g_{Al}$).
\item
The remarkable  agreement between the Minimal Standard Model and
experimental data on $Z$-decays puts strong limits on the
hypothetical "new physics", such as extra generations  of heavy
quarks and leptons and/or properties of supersymmetric particles.
\end{enumerate}

The discovery
of the higgs, more precise measurements of the mass of the $W$ boson at LEP-II
and Tevatron, and more accurate prediction of the value of electric charge at
the scale of $m_Z$, may substantially improve the sensitivity of the
$Z$-decay data to the possible manifestations of the new physics.

We conclude this review with a flowchart summarizing the theoretical approach
used by us.

\vspace{7mm}

{\bf Acknowledgements}

\vspace{5mm}

L.B.O. is grateful to Alexander von Humboldt Stiftung for an award, A.N.R. is
grateful to \\
CPPM/IN2P3/CNRS for supporting this work.
The grants of RFBR No. 96-15-96578, \\
98-07-90076, 98-02-17372 and 98-0217453
and of INTAS-RFBR 95-05678 are acknowledged.
The authors would like to thank D.Bardin, A.Olshevski and P.Renton
for correspondence.
We are grateful to our collaborators on various topics on $Z$-boson decays.

\newpage
\begin{center}
{\bf Flowchart.}
\end{center}
\vspace{2mm}

\setlength{\unitlength}{1mm}
\begin{picture}(160,210)
\put(0,195){\framebox(160,15)}
\put(80,205){\makebox(0,0){\large
Choose three observables measured with the highest accuracy:}}
\put(80,199){\makebox(0,0){\large $G_{\mu}\;,
m_Z\;, \alpha(m_Z) \equiv \bar{\alpha}$ }}
\put(80,190){\line(0,1){5}}

\put(0,175){\framebox(160,15)}
\put(80,185){\makebox(0,0){\large
Determine angle $\theta$ ( $s \equiv \sin \theta, c \equiv \cos
\theta$)}}
\put(80,180){\makebox(0,0){\large in terms of $G_{\mu}\;, m_Z\;,
\bar{\alpha}$:
$G_{\mu} = (\pi / \sqrt{2}) \bar{\alpha} / s^2 c^2 m^2_Z$}}
\put(80,170){\line(0,1){5}}

\put(0,150){\framebox(160,20)}
\put(80,165){\makebox(0,0){\large
Introduce bare coupling constants in the framework of MSM}}
\put(80,160){\makebox(0,0){\large ($\alpha_0\;, \alpha_{Z0}\;,
\alpha_{W0}$), bare masses ($m_{Z0}, m_{W0}, m_{H0}, m_{t0},
m_{q0}$) }}
\put(80,155){\makebox(0,0){\large
and the vacuum expectation value ({\it vev}) of the higgs field, $\eta$.}}
\put(80,145){\line(0,1){5}}

\put(0,130){\framebox(160,15)}
\put(80,140){\makebox(0,0){\large
Express $\alpha_0\;, \alpha_{Z0}\;, m_{Z0}$ in terms of
$G_{\mu}\;, m_Z\;, \bar{\alpha}$ in one-loop}}
\put(80,135){\makebox(0,0){\large
approximation, using dimensional
regularization $(1/\varepsilon\;, \mu$).}}
\put(80,125){\line(0,1){5}}

\put(0,105){\framebox(160,20)}
\put(80,120){\makebox(0,0){\large
Express one-electroweak-loop corrections to all electroweak}}
\put(80,115){\makebox(0,0){\large
observables in terms of
$\alpha_0\;, \alpha_{Z0}\;, m_{Z0}\;, m_t\;, m_H$, and hence, in terms of}}
\put(80,110){\makebox(0,0){\large
of $G_{\mu}\;, m_Z\;, \bar{\alpha}\;,
m_t\;, m_H$. Check cancellation of the terms $(1/\varepsilon\;, \mu)$.}}
\put(80,100){\line(0,1){5}}

\put(0,75){\framebox(160,25)}
\put(80,95){\makebox(0,0){\large
Introduce gluon corrections to quark loops and QED (and QCD)}}
\put(80,90){\makebox(0,0){\large
final state interactions, as well as the two-electroweak-loop }}
\put(80,85){\makebox(0,0){\large
corrections, calculated by other authors, in terms of}}
\put(80,80){\makebox(0,0){\large
$\bar{\alpha}\;,
\alpha_s(m_Z)\;, m_H\;, m_t\;, m_b$.}}
\put(80,70){\line(0,1){5}}

\put(0,35){\framebox(160,35)}
\put(80,65){\makebox(0,0){\large
Compare the predictions of the successive approximations }}
\put(80,60){\makebox(0,0){\large
(Born, one loop, two loops) with the experimental data }}
\put(80,55){\makebox(0,0){\large
on $Z$-decays. Perform the global fits for
$m_H\;, m_t\;, \alpha_s(m_Z)\;, \bar{\alpha}$ }}
\put(80,50){\makebox(0,0){\large
for one and two loops. Derive theoretical predictions }}
\put(80,45){\makebox(0,0){\large
of the central values for all electroweak observables}}
\put(80,40){\makebox(0,0){\large
and of the corresponding uncertainties (`errors').}}
\put(80,30){\line(0,1){5}}

\put(0,10){\framebox(160,20)}
\put(80,25){\makebox(0,0){\large
Check the sensitivity of the fits to the possible}}
\put(80,20){\makebox(0,0){\large
existence of new heavy particles: extra generations, }}
\put(80,15){\makebox(0,0){\large
SUSY, technicolor, $Z'$ etc.}}
\end{picture}

\newpage

\setcounter{equation}{0}
\renewcommand{\theequation}{A.\arabic{equation}}
\begin{center}
{\large \bf Appendix A.}\\

\vspace{3mm}

{\Large\bf Regularization of Feynman integrals.}

\end{center}

\vspace{3mm}

Integrals corresponding to diagrams with loops formally diverge and
thus need regularization. Note that there does not exist yet a consistent
regularization of electroweak theory in all loops. A dimensional
regularization can be used in the first several loops; this corresponds
to a transition to a $D$-dimensional spacetime in which the
following finite expression is assigned to the diverging
integrals:
\begin{eqnarray}
\int\frac{d^D p}{\mu^{D-4}} \frac{(p^2)^s}{(p^2 +m^2)^{\alpha}}
& = &
\frac{\pi^{\frac{D}{2}}}{\Gamma(\frac{D}{2})}
\frac{\Gamma(\frac{D}{2} +s)\Gamma(\alpha - \frac{D}{2}
-s)}{\Gamma(\alpha)} \times \nonumber \\
& \times & \frac{(m^2)^{\frac{D}{2} -\alpha +s}}{\mu^{D-4}} \;\;,
\label{A.1}
\end{eqnarray}
where $\mu$ is a parameter with mass dimension, introduced to
conserve the dimension of the original integral.

This formula holds in the range of convergence of the integral.
In the range of divergence, a formal expression (\ref{A.1})
is interpreted as the analytical continuation. Obviously, the
integral allows a shift in integration variable in the
convergence range as well. Therefore, a shift $p\to p+q$ for
arbitrary $D$ can also be done in (\ref{A.1}). This factor is
decisive in proving the gauge invariance of dimensional
regularization.

At $D=4$ the integrals in (\ref{A.1}) contain a pole term
\begin{equation}
\Delta =\frac{2}{4-D} +\log 4\pi -\gamma -\log\frac{m^2}{\mu^2} \;\;,
\label{A.2}
\end{equation}
where $\gamma =0.577...$ is the Euler constant.
Choice of  constant
terms in (\ref{A.2}) is a matter of convention.

The algebra of $\gamma$-matrices in the $D$-dimensional space is
defined by the relations
\begin{equation}
\gamma_{\mu}\gamma_{\nu} +\gamma_{\nu}\gamma_{\mu} =2g_{\mu\nu}
\times I \;\;,
\label{A.3}
\end{equation}
\begin{equation}
g_{\mu\mu} =D \;\;,
\label{A.4}
\end{equation}
\begin{equation}
\gamma_{\mu}\gamma_{\nu}\gamma_{\mu} =(2-D)\gamma_{\nu} \;\;,
\label{A.5}
\end{equation}
where $I$ is the identity matrix.

As for the dimensionality of spinors, different approaches can
be chosen in the continuation to the $D$-dimensional space.
One possibility is to assume that the $\gamma$ matrices are $4\times 4$
matrices, so that
\begin{equation}
Sp I =4 \;\;.
\label{A.6}
\end{equation}

The $D$-dimensional regularization creates difficulties when one
has to define the absolutely antisymmetric tensor and (or)
$\gamma_5$ matrix. For calculations in several first loops,
a formal definition of $\gamma_5$,
\begin{equation}
\gamma_5 \gamma_{\mu} +\gamma_{\mu}\gamma_5 = 0 \;\;,
\label{A.7}
\end{equation}
\begin{equation}
\gamma_5^2 =I
\label{A.8}
\end{equation}
does not lead to contradictions.

Thus, the amplitudes of physical processes, once they are
expressed in terms of bare charges and bare masses, contain
pole terms $\sim 1/(D-4)$.

If we eliminate bare quantities and express some physical
observables in terms of other physical observables, then all
pole terms cancel out. The general property of renormalizability
guarantees this cancellation. (We have verified this cancellation
directly in \cite{16}.) The ``Five steps" described in Section 4.6 are
 based on this renormalization procedure.

In order to avoid divergences in intermediate expressions, one
can agree to subtract from each Feynman integral the pole
terms $\sim 1/(4-D)$, since they will cancel out anyway in the
final expressions. Depending on which constant terms (in
addition to pole terms) are subtracted from the diagrams,
different subtraction schemes arise: the $\overline{MS}$ scheme
corresponds to subtracting the universal combination
$$
\frac{2}{4-D} -\gamma + \log 4\pi \;\;.
$$

\newpage

\setcounter{equation}{0}
\renewcommand{\theequation}{B.\arabic{equation}}
\begin{center}
{\large \bf Appendix B.}\\

\vspace{3mm}

{\Large\bf Relation between {\boldmath $\bar{\alpha}$} and
{\boldmath $\alpha(0)$}.}
\end{center}

\vspace{2mm}

We begin with the following famous relation of quantum
electrodynamics \cite{58}:
\begin{equation}
\alpha(q^2) = \frac{\alpha(0)}{1 + \Sigma_{\gamma}(q^2)/q^2 -
\Sigma'_{\gamma}(0)}\;\;.
\label{B.1}
\end{equation}

Here the fine structure constant $\alpha \equiv \alpha(0)$ is a
physical quantity. It can be measured as a residue of the
Coulomb pole $1/q^2$ in the scattering amplitude of charged
particles. As for the running coupling constant $\alpha(q^2)$,
it can be measured from the scattering of particles
with large masses $m$
at low momentum transfer: $m \gg \sqrt{|q^2|}$. In the standard model
we have the $Z$-boson, and the contribution of the photon cannot be
identified unambiguously if $q^2 \neq 0$. Therefore, the
definition of the running constant $\alpha(q^2)$ becomes
dependent on convention and on details of calculations.

At $q^2 = m^2_Z$, the contribution of $W$-bosons to $\bar{\alpha}
\equiv \alpha(m^2_Z)$ is not large, so it is convenient to make use
of the definition accepted in QED:  \begin{equation} \bar{\alpha} =
\frac{\alpha}{1 - \delta\alpha}\;\;, \label{B.2} \end{equation} where
\begin{eqnarray}
\nonumber
\delta\alpha = -\Pi_{\gamma}(m^2_Z) + \Sigma'_{\gamma}(0)\;\;, \\
\Pi_{\gamma}(m^2_Z) = \frac{1}{m^2_Z} \Sigma_{\gamma}(m^2_Z)\;\;.
\label{B.3}
\end{eqnarray}

One-loop expression for the self-energy of the photon can be
rewritten as \cite{59}:
\begin{eqnarray}
\nonumber
\Sigma_{\gamma}(s) &=& (\alpha/3\pi) \sum_f N^f_c Q^2_f [s\Delta_f +
(s + 2m^2_f) F (s, m_f, m_f) - s/3] - \\
& - & (\alpha/4\pi) [3s\Delta_W + (3s + 4m^2_W) F (s, m_W, m_W)]\;\;,
\label{B.4}
\end{eqnarray}
where $s \equiv q^2$, the subscript $f$ denotes fermions, the sum
$\Sigma_f$ runs through lepton and quark flavors, and $N^f_c$ is the
number of colors. The contribution of fermions to $\Sigma_{\gamma}(q^2)$
is independent of gauge. The last term in (\ref{B.4})
refers to the gauge-dependent contribution of $W$-bosons; the
 't~Hooft--Feynman gauge was used in equation
(\ref{B.4}).

The singular term $\Delta_i$ is:
\begin{equation}
\Delta_i = \frac{1}{\epsilon} - \gamma + \log 4\pi - \log
\frac{m^2_i}{\mu^2}\;\;,
\label{B.5}
\end{equation}
where $2\epsilon = 4 - D$ ($D$ is the variable dimension of
spacetime, $\epsilon \to 0$), $\gamma = -\Gamma'(1) =
0.577...$ is the Euler constant and $\mu$ is an arbitrary parameter.
Both $1/\epsilon$ and $\mu$ vanish in relations between
observables.

The function $F(s, m_1, m_2)$ is defined by the contribution
to self-energy of a scalar particle at $q^2 = s$, owing to a loop with
two scalar particles (with masses $m_1$ and $m_2$) and with the
coupling constant equal to unity:
\begin{eqnarray}
\nonumber
F(s,m_1, m_2) &=& -1 + \frac{m^2_1 + m^2_2}{m^2_1 - m^2_2} \log
\frac{m_1}{m_2} - \\
&-& \int\limits^1_0 dx \log \frac{x^2s - x(s + m^2_1 - m^2_2) + m^2_1
-i\epsilon}{m_1 m_2}\;\;.
\label{B.6}
\end{eqnarray}
The function $F$ is normalized in such a way that it vanishes at $q^2 = 0$,
which corresponds to subtracting the self-energy at $q^2 = 0$:
\begin{equation}
F(0, m_1, m_2) = 0\;\;,
\label{B.7}
\end{equation}

The following formula holds for $m_1 = m_2 = m$:
\begin{eqnarray}
F(s,m,m)\equiv F(\tau) = \left\{ \begin{array}{ll}
2\left[1-\sqrt{4\tau -1}\arcsin\frac{1}{\sqrt{4\tau}}\right] \;\;,
& 4\tau > 1 \;\;,  \\
2\left[ 1-\sqrt{1-4\tau}\log
\frac{1+\sqrt{1-4\tau}}{\sqrt{4\tau}}\right] \;\;,
& 4\tau < 1 \;\;,
\end{array}
\right.
\label{B.77}
\end{eqnarray}
where $\tau = m^2/s$.

  Let us present the following useful equality which holds for $F(\tau)$
derivative:
\begin{equation}
F'(\tau) \equiv -\tau \frac{d}{d\tau} F(\tau)=\frac{1-2\tau F(\tau)}{4\tau -1}
\label{B.777}
\end{equation}

To calculate the contributions of light fermions, the $t$-quark and
the $W$-boson to $\delta\alpha$, we need the asymptotics $F(\tau)$
for small and large $\tau$:
\begin{equation}
F(\tau) \simeq \log \tau + 2 +...,\;\; |\tau | \ll 1\;\;,
\label{B.8}
\end{equation}
\begin{equation}
F(\tau) \simeq \frac{1}{6\tau} + \frac{1}{60 \tau^2} +..., \;\;
|\tau | \gg 1\;\;,
\label{B.9}
\end{equation}

As a result we obtain
\begin{eqnarray}
\Pi_{\gamma}(m^2_Z) &\equiv & \frac{\Sigma_{\gamma}(m^2_Z)}{m^2_Z} =
\frac{\alpha}{3\pi} \sum_8 N^f_c Q^2_f (\Delta_Z + \frac{5}{3}) +
\nonumber \\
&+& \frac{\alpha}{\pi} Q^2_f \left[ \Delta_t + (1 + 2t) F(t) -
\frac{1}{3} \right] - \nonumber \\
&-& \frac{\alpha}{4\pi} [3\Delta_W + (3 + 4c^2) F(c^2)]\;\;,
\label{B.11}
\end{eqnarray}
where $t = m^2_t/m^2_Z$, and
\begin{equation}
\Sigma'_{\gamma}(0) = \frac{\alpha}{3\pi} \sum_9 N^f_c Q^2_f \Delta_f
- \frac{\alpha}{4\pi} (3\Delta_W + \frac{2}{3})\;\;,
\label{B.12}
\end{equation}
\begin{eqnarray}
\nonumber
\delta\alpha &=& \frac{\alpha}{\pi} \left\{
\sum_8 \frac{N^f_c Q^2_f}{3} \left( \log \frac{m^2_Z}{m^2_f} -
\frac{5}{3} \right) - \right. \\
&-& \left. Q^2_t \left[ (1 + 2t) F(t) - \frac{1}{3} \right] +
\left[ \left( \frac{3}{4} + c^2 \right) F(c^2) - \frac{1}{6} \right]
\right\}\;\;.
\label{B.13}
\end{eqnarray}
Therefore, $\delta \alpha$ is found as a sum of four terms,
\begin{equation}
\delta\alpha = \delta\alpha_l + \delta\alpha_h + \delta\alpha_t +
\delta\alpha_W \;\;,
\label{B.14}
\end{equation}

In one-loop approximation:

\begin{equation}
\delta\alpha_l = \frac{\alpha}{3\pi} \sum_3 \left[ \log
\frac{m^2_Z}{m^2_l} - \frac{5}{3} \right] = 0.03141\;\;.
\label{B.15}
\end{equation}

Higher loops \cite{17a} give:
\begin{equation}
\delta\alpha_l = 0.031498 \;\; .
\label{B.15'}
\end{equation}

Loops with top quarks give:
\begin{equation}
\delta\alpha_t \simeq -\frac{\alpha}{\pi} \frac{4}{45} \left(
\frac{m_Z}{M_t} \right)^2 = -0.00005(1)\;\;,
\label{B.16}
\end{equation}
where we have used that $m_t = 175 \pm 10$ GeV. Note that
$\delta\alpha_t$ is negligible  and has the antiscreening sign
(the screening of the $t$-quark loops in QED begins at $q^2 \gg m^2_t$,
while in our case $q^2 = m^2_Z < m^2_t)$.

Finally, the $W$ loop gives
\begin{equation}
\delta\alpha_W = \frac{\alpha}{2\pi} \left[ (3 + 4c^2) \left( 1 -
\sqrt{4c^2 - 1} \arcsin \frac{1}{2c} \right) - \frac{1}{3} \right] =
0.00050\;\;.
\label{B.17}
\end{equation}
The value of $\delta\alpha_W$ depends on gauge \cite{60};
here we give the result of calculations in the 't~Hooft--Feynman
gauge. Traditionally, the definition of $\bar{\alpha}$ takes
into account the contributions
of leptons and
five light quarks only. The terms $\delta\alpha_t$
and $\delta\alpha_W$ are taken into account in the electroweak
radiative corrections. In our approach, these terms give
the corrections $\delta_1 V_i$. In the same way the loops of not yet
discovered heavy new particles (``New Physics") should be accounted for.

\newpage

\setcounter{equation}{0}
\renewcommand{\theequation}{C.\arabic{equation}}
\begin{center}
{\large\bf Appendix C.}

\vspace{3mm}

{\Large\bf How {\boldmath$\alpha_W(q^2)$} and
{\boldmath$\alpha_Z(q^2)$} `crawl'.}
\end{center}
\vspace{2mm}

The effect of `running' of electromagnetic coupling constants $\alpha(q^2)$
(logarithmic dependence of the effective charge on
momentum transfer $q^2$) is known for more than four decades) \cite{58}.
In contrast to $\alpha(q^2)$, the effective constants of $W$ and
$Z$ bosons $\alpha_W(q^2)$ and $\alpha_Z(q^2)$ in the region $0<q^2 \la
m_Z^2$ `crawl' rather than run \cite{61}.

If we define the effective gauge coupling constant
$g^2(q^2)$ in terms of the bare charge $g_0^2$ and the bare mass $m_0$,
and sum up the geometric series with the self-energy $\Sigma(q^2)$
inserted in the gauge boson propagator, this gives the expression
\begin{equation}
g^2(q^2) = \frac{g_0^2}{1+g_0^2 \frac{\Sigma(q^2)-\Sigma(m^2)}
{q^2 -m^2}} \;\;.
\label{C.1}
\end{equation}
Here $m$ is the physical mass, and $\Sigma(q^2)$ contains the
contribution of fermions only, since loops with $W$, $Z$ and $H$ bosons
do not contain large logarithms in the region $|q^2|\leq m_Z^2$.

The bare coupling constant in the difference $1/g^2(q^2) - 1/g^2(0)$
is eliminated, which gives a finite expression.
The result is
\begin{equation}
1/\alpha_Z(q^2) - 1/\alpha_Z(0) =b_Z F(x) \;\;, \;\;
{\rm where} \;\; x=q^2/m_Z^2 \;\;,
\label{C.2}
\end{equation}
\begin{equation}
1/\alpha_W(q^2) - 1/\alpha_W(0) =b_W F(y) \;\;, \;\;
{\rm where} \;\; y=q^2/m_Z^2 \;\;,
\label{C.3}
\end{equation}
\begin{equation}
F(x) =\frac{x}{1-x} \log |x|
\label{C.4}
\end{equation}

If $x\gg 1$, equations (\ref{C.2}) and (\ref{C.3}) define the
logarithmic running of charges owing to leptons and quarks, and
$b_Z$ and $b_W$ represent the contribution of fermions to the
first coefficient of the Gell-Mann--Low function:
\begin{eqnarray}
b_Z & = & \frac{1}{48\pi}\{N_u 3[1+(1-\frac{8}{3}s^2)^2] +N_d 3[1+(-1
+ \frac{4}{3}s^2)^2] + N_l[2+(1+(1-4s^2)^2)]\} \;\;, \nonumber \\
b_W & = & \frac{1}{16\pi} \ [6N_q +2N_l] \;\;, 
\label{C.5}
\end{eqnarray}
where $N_{u,d,q,l}$ are the numbers of quarks and leptons with
masses that are considerably lower than $\sqrt{q^2}$.

For $q^2 \la m_Z^2$, the numerical values of the coefficients $b_{Z,W}$ are
\cite{61}:
$$
b_Z \simeq 0.195
$$
$$
b_W \simeq 0.239
$$
The massive propagator $\frac{1}{q^2 -m^2}$ in (\ref{C.1})
greatly suppresses the running of $\alpha_W(q^2)$ and $\alpha_Z(q^2)$.
Thus, according to (\ref{C.2}) and (\ref{C.3}), the constant
$\alpha_Z(q^2)$ grows by 0.85\% from $q^2 =0$ to $q^2 = m_Z^2$,
$$
1/\alpha_Z(m_Z^2) = 22.905
$$
\begin{equation}
1/\alpha_Z(m_Z^2) -1/\alpha_Z(0) = -0.195 \;\;,
\label{C.6}
\end{equation}
and the constant $\alpha_W(q^2)$ grows by 0.95\%,
$$
1/\alpha_W(m_Z^2) = 28.74
$$
\begin{equation}
1/\alpha_W(m_Z^2) -1/\alpha_W(0) = -0.272 \;\;,
\label{C.7}
\end{equation}
while the electromagnetic constant $\alpha(q^2)$ increases by
6.34\%:
\begin{equation}
1/\alpha(m_Z^2) -1/\alpha_W(0) = 128.90 - 137.04 = -8.14
\label{C.8}
\end{equation}

With the accuracy indicated above, we can thus assume
\begin{eqnarray}
\alpha_Z(m_Z^2) \simeq \alpha_Z(0) \nonumber \\
\alpha_W(m_Z^2) \simeq\alpha_W(0) .
\label{C.9}
\end{eqnarray}

At the same time, $\alpha(m_Z^2)$ differs greatly from $\alpha(0)$;
therefore the latter has no connection to the electroweak physics
but only to the purely electromagnetic physics.

\newpage

\setcounter{equation}{0}
\renewcommand{\theequation}{D.\arabic{equation}}
\begin{center}
{\large\bf Appendix D.}

\vspace{3mm}
{\Large\bf General  expressions for one-loop corrections
to hadronless observables.}
\end{center}

\vspace{3mm}

The bare quantities are marked by the subscript `0'. In the electroweak
theory, three bare charges $e_0$, $f_0$ and $g_0$ that describe
the interactions of $\gamma$, $Z$ and $W$ are related by a single
constraint:
\begin{equation}
(e_0/g_0)^2 +(g_0/f_0)^2 =1 \;\;.
\label{E.1}
\end{equation}

The bare masses of the vector bosons are defined by the bare
vacuum expectation value of the higgs field $\eta$:
\begin{equation}
m_{Z0}=\frac{1}{2}f_0 \eta \;\;, \;\;
m_{W0} =\frac{1}{2}g_0\eta \;\;.
\label{E.2}
\end{equation}

The fine structure constant $\alpha = e^2/4\pi$ is related to
the bare charge $e_0$ by the formula
\begin{equation}
\alpha\equiv\alpha(q^2 =0) =\frac{e_0^2}{4\pi}
\left(1-\Sigma'_{\gamma}(0)-2\frac{s}{c} \frac{\Sigma_{\gamma Z}(0)}
{m_Z^2} \right) \;\;,
\label{E.3}
\end{equation}
where $\Sigma'(0) =\lim_{q^2\to 0} \Sigma(q^2)/q^2$. In the Feynman
gauge $\Sigma_{\gamma Z}(0)\approx -(\alpha/2\pi)
(m^2_W/cs)(1/\epsilon)$, where the dimension of spacetime
is $D=4-2\varepsilon$. In the unitary gauge $\Sigma_{\gamma Z}(0)=0$.

The simplest way  to verify the presence of the term
$2(s/c)\Sigma_{\gamma Z}(0)/m_Z^2$ is by considering the
interaction of a photon with the right-handed electron $e_R$.
Note that in this case
there are no weak vertex corrections due to the $W$-boson
exchange. (Note also that the left-handed neutrino remains neutral
even when loop corrections are taken into account, since the diagram
with the $\gamma -Z - \nu_L \bar{\nu}_L$ interaction is compensated
for by the vertex diagram with the $W$ exchange).

The relation between $\bar{\alpha}=\alpha(q^2 =m_Z^2)$ and
$\alpha_0$ has the following form:
\begin{equation}
\bar{\alpha}=\alpha_0 \left[1-\tilde{\Pi}_{\gamma}(m_Z^2)
-\Sigma'_{\gamma}(0) +
\tilde{\Sigma}'_{\gamma}(0) -2\frac{s}{c}
\Pi_{\gamma Z}(0) \right] \;\;,
\label{E.4}
\end{equation}
where $\tilde{\Pi}_{\gamma}(q^2)=\tilde{\Sigma}_{\gamma}(q^2)/m_Z^2$,
$\Pi_{\gamma Z}(q^2)=\Sigma_{\gamma Z}(q^2)/m_Z^2$, while
$\tilde{\Sigma}_{\gamma}$ mean that contributions of $W$-boson and $t$-quark
are not accounted for. It is convenient to introduce in (\ref{E.4}) explicit
expression for $\delta\alpha_W +\delta\alpha_t$:
\begin{equation}
\bar{\alpha}=\alpha_0[1-\Pi_{\gamma}(m_Z^2)-2\frac{s}{c}\Pi_{\gamma Z}(0)
-\delta\alpha_W -\delta\alpha_t] \;\; ,
\label{E.44}
\end{equation}
where in accordance with eq.(\ref{B.3})
\begin{equation}
\delta\alpha_W +\delta\alpha_t
=\tilde{\Pi}_{\gamma}(m_Z^2)-\Pi_{\gamma}(m_Z^2) +\Sigma'_{\gamma}(0)
-\tilde{\Sigma}'_{\gamma}(0) \;\; .
\label{E.45}
\end{equation}

In the case of ``New Physics" one should add to eq.(\ref{E.44}) the term
$\delta\alpha_{NP}$. Our first basic equation is eq.(\ref{E.44}).

The  second basic equation is:
\begin{equation}
m_Z^2 =m_{Z0}^2 [1-\Pi_Z(m_Z^2)] =m_{W0}^2 /c_0^2
[1-\Pi_Z(m_Z^2)]\;\;.
\label{E.5}
\end{equation}
A similar equation holds for $m_W^2$:
\begin{equation}
m_W^2 =m_{W0}^2 [1-\Pi_W(m_W^2)] \;\;,
\label{E.6}
\end{equation}
where $\Pi_i(q^2) = \Sigma_i(q^2)/m_i^2$, $i=W, Z$.

Finally, the third basic equation is
\begin{equation}
G_{\mu} =\frac{g_0^2}{4\sqrt{2}m_{W0}^2} [1+\Pi_W(0)+D] \;\;,
\label{E.7}
\end{equation}
where $\Pi_W(0) = \Sigma_W(0)/m_W^2$ comes from the propagator of
$W$,
while $D$ is the contribution of the box and the vertex diagrams
(minus the electromagnetic corrections to the four-fermion
interaction) to the muon decay amplitude. According to Sirlin
\cite{62},
\begin{equation}
D=\frac{\bar{\alpha}}{4\pi
s^2}(6+\frac{7-4s^2}{2s^2} \log c^2 +4\Delta_W) \;\;, \label{E.8}
\end{equation}
where
\begin{equation}
\Delta_W \equiv \Delta(m_W) =\frac{2}{4-D} +\log 4\pi -\gamma
-\log(m_W^2 /\mu^2) \;\; .
\label{E.9}
\end{equation}

Now we are able to express $f_0$ and $g_0$ in terms of $\bar{\alpha}$,
$G_{\mu}$, $m_Z$ and the loop corrections. From (\ref{E.2}), (\ref{E.5}) and
(\ref{E.7}) we obtain:
\begin{equation}
f_0^2 =4\sqrt{2} G_{\mu} m_Z^2[1-\Pi_W(0) +\Pi_Z(m_Z^2)-D] \;\; .
\label{E.10}
\end{equation}

From (\ref{E.1}), (\ref{E.44}) and (\ref{E.10}) we get:
\begin{equation}
c_0 \equiv \frac{g_0}{f_0} =c\left[1+\frac{s^2}{2(c^2 -s^2)}
(-2\frac{s}{c}\Pi_{\gamma Z}(0) -\Pi_{\gamma}(m_Z^2)-\delta\alpha_W
-\delta\alpha_t +\Pi_Z(m_Z^2) -\Pi_W(0)-D)\right] \;\; .
\label{E.11}
\end{equation}

The next step is to express $m_W/m_Z$, $g_A$ and $g_V$ through $c$, $s$ and
loop corrections. Let us start with $m_W/m_Z$. From (\ref{E.6}) and
(\ref{E.5}) we get:
\begin{equation} m_W/m_Z =c_0 [1-\frac{1}{2}\Pi_W(m_W^2)
+\frac{1}{2}\Pi_Z(m_Z^2)] \;\; .
\label{E.12}
\end{equation} Substituting
$c_0$ given by (\ref{E.11}) we obtain:  \begin{eqnarray} \frac{m_W}{m_Z} & =
& c+\frac{cs^2}{2(c^2 -s^2)} \left(
\frac{c^2}{s^2}[\Pi_Z(m_Z^2)-\Pi_W(m_W^2)]+\Pi_W(m_W^2)-\Pi_W(0) - \right.
\nonumber \\
& - & \left. \Pi_{\gamma}(m_Z^2) -2\frac{s}{c}\Pi_{\gamma Z}(0)
-D-\delta\alpha_W -\delta\alpha_t \right) \;\; .
\label{E.13}
\end{eqnarray}

In order to obtain expression for $g_A$ we should recall that it is
proportional to $f_0$ and take into account the $Z$ boson wave function
renormalization and $Z\bar{l}l$ vertex loop correction:
\begin{equation}
g_A =-\frac{1}{2}-\frac{1}{4}[\Pi_Z(m_Z^2) -\Pi_W(0)-D-\Sigma'_Z(m_Z^2) -8 cs
F_A] \;\; ,
\label{E.14}
\end{equation}
where $F_A$ originates from the vertex correction.

The last quantity is the ratio $g_V/g_A$. One-loop corrections come from
$s_0^2 \equiv 1-c_0^2$ (eq.(\ref{E.11})), from vector and axial $Z\bar{l}l$
vertices and from $Z\to\gamma$ transition which contributes to $g_V$ only:
\begin{eqnarray}
\frac{g_V}{g_A} &=& 1-4s^2 -\frac{4c^2 s^2}{c^2 -s^2} [2\frac{s}{c}
\Pi_{\gamma Z}(0) +\Pi_{\gamma}(m_Z^2) +\delta\alpha_W +\delta\alpha_t -
\nonumber \\
&-& \Pi_Z(m_Z^2) +\Pi_W(0)+D] -4csF_V +4csF_A(1-4s^2)- \nonumber \\
&-& 4cs \Pi_{\gamma Z}(m_Z^2) \;\; .
\label{E.15}
\end{eqnarray}

Formulas (\ref{E.13}), (\ref{E.14}) and (\ref{E.15}) are derived in this
 Appendix according to the
``Five Steps" procedure described in Section 4.6.
They describe finite one-loop corrections to hadronless observables.

It is easy to evaluate the contribution of $t$-quark to physical observables
in the approximation $\sim\alpha_W m_t^2$. In this approximation
$\Pi_W(m_W^2) = \Pi_W(0)$, $\Pi_Z(m_Z^2) =\Pi_Z(0)$, $\Pi_Z(0) -\Pi_W(0)
=3\bar{\alpha}/16\pi s^2 c^2 \; t$ and from eq.(\ref{E.13})-(\ref{E.15})
we get:
\begin{equation}
\frac{m_W}{m_Z} \approx c+\frac{3\bar{\alpha}c}{32\pi(c^2 -s^2)s^2}t \;\; ,
\label{E.18}
\end{equation}
\begin{equation}
g_A \approx -\frac{1}{2}(1+\frac{3\bar{\alpha}}{32\pi s^2 c^2}t) \;\; ,
\label{E.19}
\end{equation}
\begin{equation}
\frac{g_V}{g_A} \approx 1-4s^2+\frac{3\bar{\alpha}}{4\pi(c^2 -s^2)}t
\;\; .
\label{E.20}
\end{equation}

The corrections proportional to $m_t^2$ were first pointed out by Veltman
\cite{19a}, who emphasized the appearance of such corrections for a large
difference $m_t^2 -m_b^2$ which violates the isotopic symmetry. In this
review the coefficients in front of the factors $t$ in equations
(\ref{E.18})-(\ref{E.20}) are used as coefficients for normalized radiative
corrections $V_i$ (see Sections 4.2 and 4.3).

\newpage

\setcounter{equation}{0}
\renewcommand{\theequation}{E.\arabic{equation}}
\begin{center}
{\large\bf Appendix E.}\\
\vspace{3mm}
{\Large\bf Radiators ${\bf R_{Aq}}$ and
${\bf R_{Vq}}$.}
\end{center}
\vspace{2mm}

For decays to light quarks $q = u,d,s$, we neglect the
quark masses and take into account the gluon exchanges in the final
state up to terms $\sim\alpha^3_s$ \cite{65a} - \cite{65d}, and also
one-photon exchange in the final state and the interference of
the photon and the gluon exchanges \cite{65e}. These corrections
are slightly different for the vector and the axial channels.

For decays to quarks we have
\begin{equation}
\Gamma_q = \Gamma(Z \to q\bar{q}) = 12[g^2_{Aq} R_{Aq} + g^2_{Vq}
R_{Vq}]\Gamma_0
\label{F.1}
\end{equation}
where the factors $R_{A,V}$ are responsible for the interaction in
the final state. According to \cite{65a} - \cite{65d}:
\begin{eqnarray}
R_{Vq} &=& 1 +
\frac{\hat{\alpha}_s}{\pi} + \frac{3}{4} Q^2_q
\frac{\bar{\alpha}}{\pi} - \frac{1}{4} Q^2_q \frac{\bar{\alpha}}{\pi}
\frac{\hat{\alpha}_s}{\pi} + \nonumber \\
&+& [1.409 + (0.065 + 0.015 \log t)\frac{1}{t}](\frac{\hat{\alpha}_s}
{\pi})^2 - \nonumber \\
&-& 12.77(\frac{\hat{\alpha}_s}{\pi})^3
+12 \frac{\hat{m}^2_q}{m^2_Z} \frac{\hat{\alpha}_s}{\pi} \delta_{vm}
\label{F.2}
\end{eqnarray}
$$
R_{Aq} = R_{Vq} - (2T_{3q})[I_2(t)(\frac{\hat{\alpha}_s}{\pi})^2 +
I_3(t)(\frac{\hat{\alpha}_s}{\pi})^3]
-
$$
\begin{eqnarray}
-12 \frac{\hat{m}^{2}_{q}}{m^{2}_{Z}} \frac{\hat{\alpha}_s}{\pi} \delta_{vm} -
6\frac{\hat{m}^2_q}{m^2_Z} \delta^1_{am}
- 10 \frac{\hat{m}^2_q}{m^2_t}(\frac{\hat{\alpha}_s}{\pi})^2 \delta^2_{am}
\;\; ,
\label{F.3}
\end{eqnarray}
where $\hat{m}_q$ is the running quark mass (see below),
\begin{eqnarray}
\delta_{vm} = 1+8.7
(\frac{\hat{\alpha}_s}{\pi})+45.15(\frac{\hat{\alpha}_s}{\pi})^2,
\label{F.4}
\end{eqnarray}
\begin{eqnarray}
\delta^1_{am} = 1 + 3.67(\frac{\hat{\alpha}_s}{\pi}) +(11.29 - \log t)
(\frac{\hat{\alpha}_s}{\pi})^2,
\label{F.5}
\end{eqnarray}
\begin{eqnarray}
\delta^2_{am} = \frac{8}{81}+\frac{\log t}{54},
\label{F.6}
\end{eqnarray}
\begin{equation}
I_2(t) = -3.083 - \log t + \frac{0.086}{t} + \frac{0.013}{t^2}\;\;,
\label{F.7}
\end{equation}
\begin{eqnarray}
I_3(t) &=& -15.988 - 3.722 \log t + 1.917 \log^2 t\;\;, \\ \nonumber
t &=& m^2_t/m^2_Z\;\;.
\label{F.8}
\end{eqnarray}
Terms of the order of $(\hat{\alpha}_s/\pi)^3$ caused
by the diagrams with three gluons in intermediate state were
calculated in \cite{65d}. For $R_{Vq}$ they are numerically very
small, $\sim 10^{-5}$; for this reason, we dropped them from
formula (\ref{F.2}).

For the $Z \to b\bar{b}$ decay, the $b$-quark mass is not
negligible; it reduces $\Gamma_b$ by about 1 MeV ($\sim 0.5\%$).
The gluon corrections result in a replacement of the pole mass
$m_b \simeq 4.7$ GeV by the running mass at
$q^2 =m_Z^2\;:\; m_b \to \hat{m}_b(m_Z)$. We express $\hat{m}_b(m_Z)$
in terms of $m_b$, $\hat{\alpha}_s(m_Z)$ and $\hat{\alpha}_s(m_b)$
using standard two-loop equations in the $\overline{MS}$ scheme (see
\cite{18}).

For the $Z \to c\bar{c}$ decay, the running mass $\hat{m}_c(m_Z)$
is of the order of
 $0.8$ GeV
and the corresponding contribution
to $\Gamma_c$ is of the order of $0.05$ MeV. We have included
this tiny term in the LEPTOP code, since it is taken
into account in other codes (see, for example, \cite{118}).

We need to remark in connection with $\Gamma_c$ that the term
$I_2(t)$, given by equation (\ref{F.7}), contains
interference terms $\sim(\hat{\alpha}_s/\pi)^2$. These terms
 are related to three types of final states:
 one quark pair, a quark pair and
a gluon, two quark pairs. This last contribution comes to about 5\%
of $I_2$ and is below the currently
achievable experimental accuracy. Nevertheless, in principle
these terms require special consideration, especially if these
quark pairs are of different flavors, for example, $b\bar{b}c\bar{c}$.
Such mixed quark pairs must be discussed separately.

Note that $\hat{\alpha}_s$ stands for the strong interaction constant
in the $\overline{MS}$ subtraction scheme, with $\mu^2 =m^2_Z$.

\newpage

\setcounter{equation}{0}
\renewcommand{\theequation}{F.\arabic{equation}}
\begin{center}
{\large \bf Appendix F.}\\

\vspace{3mm}

{\Large\bf ${\bf \alpha_W^2 t^2}$ corrections from reducible diagrams.}

\end{center}

\vspace{3mm}

In \cite{16} when deriving equations for physical observables
we systematically took into account corrections which contained first power
of polarization operators and neglected terms $\sim(\Pi_{W,Z})^2$. This
procedure was correct at one loop, but since $\Pi_{W,Z}$ contain terms of the
order of $\alpha_W t$ we evidently lost $\alpha_W^2 t^2$ terms.
To restore them let us repeat the procedure implemented in \cite{16}
this time taking squares of $\Pi_W$ and $\Pi_Z$ (reducible two-loop diagrams)
into account.

Our starting point are three basic equations for quantities $m_Z$, $G_{\mu}$
and $\bar{\alpha} =\alpha(m_Z^2)$.  Since terms $\sim\alpha_W t$ are absent
in $\Pi_{\gamma}$, $\Pi_{\gamma Z}$ and $D$ functions, we will not
consider these functions in this Section. Equation for $m_Z$ is the same as in
Appendix D, eqs.(\ref{E.5}), (\ref{E.2}):
\begin{equation}
m_Z^2 = \frac{1}{4}f_0^2\eta^2
[1-\Pi_Z(m_Z^2)] \;\; ,
\label{4.1}
\end{equation}
while for $G_{\mu}$ we have:
\begin{equation}
G_{\mu} =
\frac{g_0^2}{\sqrt{2}g_0^2 \eta^2 [1-\Pi_W(0)]} =
\frac{1}{\sqrt{2}\eta^2(1-\Pi_W(0)]} \;\; ,
\label{4.2}
\end{equation}
and we keep $\Pi_W(0)$ in the denominator to avoid loosing of the
$\Pi_W^2(0)$ term
(compare with eq.(\ref{E.7})). From these two equations we get:
\begin{equation}
f_0^2 = 4\sqrt{2}G_{\mu} m_Z^2 \frac{1-\Pi_W(0)}{1-\Pi_Z(m_Z^2)} =
4\sqrt{2}G_{\mu}m_Z^2 \frac{1-\Pi_W(0)}{1-\Pi_Z(0)} \;\; ,
\label{4.3}
\end{equation}
where we use equality $\Pi_Z(m_Z^2)=\Pi_Z(0)$ which is valid for
the leading term $\sim\alpha_W t$.

Finally dividing the equation for
the running electromagnetic coupling constant,
which in our approximation is simply:
\begin{equation}
e^2(m_Z^2)=4\pi\bar{\alpha} =g_0^2(1-\frac{g_0^2}{f_0^2}) \;\; ,
\label{4.4}
\end{equation}
by (\ref{4.3}), we obtain:
\begin{equation}
\frac{g_0^2}{f_0^2}(1-\frac{g_0^2}{f_0^2})
=\frac{\pi\bar{\alpha}}{\sqrt{2}G_{\mu}m_Z^2}[1-\delta] \;\; ,
\label{4.5}
\end{equation}

\begin{equation}
\delta \equiv 1-\frac{1-\Pi_Z(0)}{1-\Pi_W(0)} = \frac{\Pi_Z(0) -\Pi_W(0)}
{1-\Pi_W(0)} \;\; .
\label{4.6}
\end{equation}

Considering $\delta$ as a small parameter and solving equation (\ref{4.5})
perturbatively, we get:
\begin{equation}
\frac{g_0^2}{f_0^2} =c^2 [1+\frac{s^2}{c^2 -s^2}\delta -\frac{c^2
s^4}{(c^2 -s^2)^3}\delta^2] \;\; ,
\label{4.7}
\end{equation}
where we keep terms linear and quadratic in $\delta$.
For definitions of $c$ and $s$ see eq.(\ref{23}).

Our next step should be the calculation of the $\delta^2$ corrections to the
functions $V_i$. But first let us discuss the expression for
$\delta$ as given by eq.(\ref{4.6})
which contains factor $1-\Pi_W(0)$ in denominator. At one loop,
corrections proportional to $\delta$ appear in physical observables. They
should be carefully calculated in order not to induce extra $\alpha_W^2 t^2$
terms. Fortunately this can be done straightforwardly, using the following
chains of equalities:
\begin{equation}
\Pi_Z(0) \equiv \frac{4}{f_0^2 \eta^2}f_0^2
[\Sigma_Z(0)/f_0^2] = 4\sqrt{2}G_{\mu}(1-\Pi_W(0))
[\Sigma_Z(0)/f_0^2] \;\; ,
\label{4.9}
\end{equation}
\begin{equation}
\Pi_W(0) \equiv \frac{4}{g_0^2 \eta^2}g_0^2
[\Sigma_W(0)/g_0^2] =
4\sqrt{2}G_{\mu}(1-\Pi_W(0))
[\Sigma_W(0)/g_0^2] \;\; ,
\label{4.10}
\end{equation}
where expressions in square brackets contain self-energies
\underline{without} coupling constants ($\Sigma_Z/f_0^2$ and
$\Sigma_W/g_0^2$ respectively) and
eq.(\ref{4.2}) is used to express $\eta$
through $G_{\mu}$. Substituting
eq.(\ref{4.9}) and eq.(\ref{4.10}) into eq.(\ref{4.6})
we get:
\begin{eqnarray}
\delta &=& 4\sqrt{2}G_{\mu}[\Sigma_Z(0)/f_0^2 - \Sigma_W(0)/g_0^2] =
4\sqrt{2}G_{\mu}m_Z^2 \frac{[\Sigma_Z(0)/f_0^2 - \Sigma_W(0)/g_0^2]}
{m_Z^2} = \nonumber \\
&=& \frac{3\bar{\alpha}}{16\pi s^2 c^2}(\frac{m_t}{m_Z})^2 \;\; .
\label{4.11}
\end{eqnarray}

Now everything is prepared for the calculation of $\delta^2$ corrections to
physical observables. Let us start from the $W$-boson mass. For the ratio of
the squares of vector boson masses we have:
\begin{equation}
\frac{m_W^2}{m_Z^2} = \frac{g_0^2}{f_0^2} \frac{1-\Pi_W(m_W^2)}
{1-\Pi_Z(m_Z^2)} \;\; .
\label{4.12}
\end{equation}

Taking the ratio of bare coupling constants from equation (\ref{4.7}) we get:
\begin{equation}
\frac{m_W}{m_Z} = c\sqrt{\frac{1-\Pi_W(0)}{1-\Pi_Z(0)}}[1+\frac{s^2}{2(c^2
-s^2)}\delta +\frac{s^6 -5s^4 c^2}{(c^2 -s^2)^3}\frac{\delta^2}{8}] \;\; .
\label{4.13}
\end{equation}

It is easy to see that:
\begin{equation}
\sqrt{\frac{1-\Pi_W(0)}{1-\Pi_Z(0)}} =
\frac{1}{\sqrt{\frac{1-\Pi_Z(0)}{1-\Pi_W(0)}}} = \frac{1}{\sqrt{1-\delta}}
\;\; .
\label{4.14}
\end{equation}

Resulting formula for the correction to the ratio $m_W/m_Z$ is presented in
Section 7.1.

The next step is the correction to axial coupling of $Z$ boson to charged
leptons. Axial coupling is proportional to $f_0$, and from eq.(\ref{4.3})
and eq.(\ref{4.14}) we immediately obtain:
\begin{equation}
f_0 \sim \frac{1}{\sqrt{1-\delta}} = 1+\frac{1}{2}\delta
+\frac{3}{8}\delta^2 \;\; .
\label{4.18}
\end{equation}
The final formula for the correction to $g_{Al}$ is presented in Section 7.1.

For the ratio of vector to axial constants in our approximation we
have:
\begin{equation}
g_{Vl}/g_{Al} = 1-4s_0^2 = 1-4(1-\frac{g_0^2}{f_0^2})
\label{4.22}
\end{equation}

The expression for the correction to $g_{Vl}/g_{Al}$ through physical
parameters is presented in Section 7.1 as well.
\newpage

\setcounter{equation}{0}
\renewcommand{\theequation}{G.\arabic{equation}}
\begin{center}
{\large \bf Appendix G.}\\

\vspace{3mm}

{\Large\bf Oblique corrections from new generations and SUSY.}

\end{center}

  In this Appendix we collect analytical formulas for different oblique corrections.

  For the degenerate case the contribution of additional quark and lepton  to $\delta^4 V_i$
  are given by
 (  \cite{bb}):
\begin{eqnarray}
\delta^4 V_m &=&\frac{4}{9}N_c\left\{[(1-l)F(l)-(1-l/c^2)F(l/c^2)]
+2 s^2
[(1-l/c^2) F(l/c^2)- \right. \nonumber \\
&-&\left.(1+2l)F(l)]+ 4s^4 (Q^2_U +Q^2_D)[(1+2l)F(l)
-\frac{1}{3}]\right\}
\label{1.3}
\end{eqnarray}
\begin{eqnarray}
\delta^4 V_A &=&\frac{4}{9}N_c\left\{[1-l+(6l^2
-3l)F(l)] +
[4 s^4(Q_U^2 +Q_D^2) - \right.\nonumber \\
&-&\left.2 s^2][2l+1 - 12l^2 F(l)]\right\}/(1-4l)
 \;\; ,
\label{1.4}
\end{eqnarray}
\begin{equation}
\delta^4 V_R = -\frac{4}{9}N_c\left\{3lF(l)-4s^2 c^2(Q^2_U
+Q^2_D)[(1+2l)F(l)-\frac{1}{3}]\right\} \;\; ,
\label{1.5}
\end{equation}
where  $N_c =3$, $Q_U =2/3$, $Q_D =-1/3$ for quark doublet;
$N_c =1$, $Q_U =0$, $Q_D =-1$ for lepton
doublet;
$$
\ell = m^2_Q/m^2_Z \quad{\rm for~quarks}~,
\quad\quad \ell = m^2_L/m^2_Z
\quad{\rm for~leptons}~,
$$
and the function $F(l)$ is defined in Appendix B, eqs.(B8), (B.10).

 For different up- and down- quark (and lepton)
masses analytical expressions for
$\delta^4 V_i$ are given in by
\begin{eqnarray}
\frac{1}{n}\delta^4 V_m &=&
(\frac{64}{27}s^4 -\frac{16}{9}s^2)[(1+2u)F(u)+(1+2d)F(d)-\frac{2}{3}] +
\nonumber \\
&+& \frac{8}{9}[(1-u)F(u)+(1-d)F(d)-\frac{2}{3}]+\frac{4}{3}\frac{s^2}{c^2}
[u+d-\frac{2ud}{u-d}\log\frac{u}{d}] +
\nonumber \\
&+& \frac{8}{9}(1-\frac{s^2}{c^2})[\frac{u-d}{2}\log\frac{u}{d}+(u+d)+(c^2
-\frac{u+d}{2})\frac{u+d}{u-d}\log\frac{u}{d}-\frac{4}{3}c^2 -
\nonumber \\
&-&(2c^2 -u-d-\frac{(u-d)^2}{c^2})F(m_W^2, m_U^2, m_D^2)] \;\; ;
\label{1.9}
\end{eqnarray}
\begin{eqnarray}
\frac{1}{n}\delta^4 V_A&=&
\frac{4}{9}\left\{(\frac{16}{3}s^4 -4s^2 -1)[2u F(u)-(1+2u)F'(u)+2dF(d)
-(1+2d)F'(d)] + \right.
\nonumber \\
&+& \left. 3[u+d -\frac{2ud}{u-d}\log\frac{u}{d} -F'(u) -F'(d)]\right\} \;\; ;
\label{1.10}
\end{eqnarray}
\begin{equation}
\frac{1}{n}\delta^4 V_R =
-\frac{8}{3}[u F(u)+dF(d)+\frac{ud}{u-d}\log\frac{u}{d} -\frac{u+d}{2}]
 +\frac{64}{27}s^2 c^2 [(1+2u)F(u)+(1+2d)F(d)
-\frac{2}{3}] \;\; ,
\label{1.11}
\end{equation}
where $n$ is the number of generations and
$m_N=m_U$, $m_E=m_D$, $u= m_U^2/m_Z^2$, $d=m_D^2/m_Z^2$;
$F'(u) = -u(d/du)F(u)$ and
$F(s,m_1^2, m_2^2)$ is defined in Appendix B, eq.(B.6).

The formulae that describe the enhanced SUSY
corrections to the functions $V_i$ have the following form \cite{88}:

\begin{equation}
\delta_{SUSY}^{LR} V_A = \frac{1}{m_Z^2} [c_u^2 g(m_1, m_{\tilde{b}})
+ s_u^2 g(m_2, m_{\tilde{b}}) - c_u^2 s_u^2 g(m_1, m_2)] \;\; ,
\label{8.12}
\end{equation}

\begin{equation}
\delta_{SUSY}^{LR} V_R = \delta_{SUSY}^{LR}V_A + \frac{1}{3} Y_L
[c_u^2 \log(\frac{m_1^2}{m^2_{\tilde{b}}})
+ s_u^2 \log(\frac{m_2^2}{m^2_{\tilde{b}}})] - \frac{1}{3} c_u^2 s_u^2
h(m_1, m_2) \;\; ,
\label{8.13}
\end{equation}

\begin{eqnarray}
\delta_{SUSY}^{LR} V_m &=& \delta_{SUSY}^{LR}V_A + \frac{2}{3} Y_L
s^2[c_u^2 \log(\frac{m_1^2}{m^2_{\tilde{b}}})
+ s_u^2 \log(\frac{m_2^2}{m^2_{\tilde{b}}})] + \nonumber \\
&+& \frac{c^2 -s^2}{3}
[c_u^2 h(m_1, m_{\tilde{b}}) +
 s_u^2 h(m_2, m_{\tilde{b}})] - \frac{c_u^2 s_u^2}{3} h(m_1, m_2)
\;\; ,
\label{8.14}
\end{eqnarray}
where
\begin{equation}
g(m_1, m_2) = m_1^2 +m_2^2 -2\frac{m_1^2 m_2^2}{m_1^2 - m_2^2} \log
(\frac{m_1^2}{m_2^2}) \;\; ,
\label{8.15}
\end{equation}

\begin{equation}
h(m_1, m_2) = -\frac{5}{3} +
\frac{4m_1^2 m_2^2}{(m_1^2 -m_2^2)^2} + \nonumber \\
 \frac{(m_1^2 +m_2^2)(m_1^4 -4m_1^2 m_2^2 +m_2^4)}{(m_1^2 -m_2^2)^3}
\log (\frac{m_1^2}{m_2^2}) \;\; ,
\label{8.16}
\end{equation}
and $Y_L = Q_t +Q_b =1/3$ is the hypercharge of the left doublet.

\newpage

\setcounter{equation}{0}
\renewcommand{\theequation}{H.\arabic{equation}}
\begin{center}
{\large \bf Appendix H.}\\

\vspace{3mm}

{\Large\bf Other parametrizations of radiative corrections.}

\end{center}

\vspace{3mm}

Here we will present formulas which connect our functions $V_i$ with two
other sets of parameters widely used in the literature to describe
electroweak radiative corrections. All formulas of this Appendix are valid at
one electroweak loop approximation.

A set of three parameters $\varepsilon_1$, $\varepsilon_2$, $\varepsilon_3$
has been suggested by Altarelli, Barbieri and Jadach \cite{63} for
phenomenological analysis of New Physics:
\begin{equation}
\varepsilon_1 =\Delta\rho \;\; ,
\label{T.1}
\end{equation}
\begin{equation}
\varepsilon_2 = c^2\Delta\rho +\frac{s^2}{c^2 -s^2}\Delta r_W -
2s^2 \Delta k' \;\; ,
\label{T.2}
\end{equation}
\begin{equation}
\varepsilon_3 = c^2\Delta\rho +(c^2 -s^2)\Delta k' \;\; ,
\label{T.3}
\end{equation}
where $\Delta\rho$ describes the correction to $g_A$, $\Delta k'$ to $g_V$
and $\Delta r_W$ to $m_W/m_Z$:
\begin{equation}
g_A =-\frac{1}{2}(1+\frac{1}{2}\Delta\rho) \;\; ,
\label{T.4}
\end{equation}
\begin{equation}
g_V /g_A =1-4s^2(1+\Delta k') \;\; ,
\label{T.5}
\end{equation}
\begin{equation}
m_W /m_Z =c[1-s^2\Delta r_W/2(c^2 -s^2)] \;\; .
\label{T.6}
\end{equation}

By comparing these definitions with the definitions of $V_A$, $V_R$ and $V_m$
we obtain:
\begin{equation}
\Delta\rho =\frac{3\bar{\alpha}}{16\pi} \frac{V_A}{s^2 c^2} \;\; ,
\label{T.7}
\end{equation}
\begin{equation}
\Delta k' =-\frac{3\bar{\alpha}}{16\pi} \frac{V_R}{(c^2 -s^2)s^2} \;\; ,
\label{T.8}
\end{equation}
\begin{equation}
\Delta r_W =-\frac{3\bar{\alpha}}{16\pi} \frac{V_m}{s^4} \;\; .
\label{T.9}
\end{equation}

Hence:
\begin{equation}
\varepsilon_1 =\frac{3\bar{\alpha}}{16\pi s^2 c^2} V_A \;\; ,
\label{T.10}
\end{equation}
\begin{equation}
\varepsilon_2 =\frac{3\bar{\alpha}}{16\pi(c^2 - s^2)s^2}
[(V_A -V_m)-2s^2(V_A -V_R)] \;\; ,
\label{T.11}
\end{equation}
\begin{equation}
\varepsilon_3 =\frac{3\bar{\alpha}}{16\pi s^2}(V_A -V_R) \;\; .
\label{T.12}
\end{equation}

As is evident from the last two formulas, the virtue of $\varepsilon_2$ and
$\varepsilon_3$ is that they do not contain the term $t$. So, at the time
when top-quark mass was not measured at Tevatron the corresponding
uncertainties in $\varepsilon_2$ and $\varepsilon_3$ were diminished.

Another set of parameters, $S$, $T$, $U$ was introduced
 a few years earlier by Peskin and Takeuchi \cite{64}.
These parameters were proposed to describe only the so-called
oblique corrections due to the physics beyond the Standard Model.
Using
the definitions of $S$, $T$, $U$ from \cite{64} and designating
New Physics contributions to $\varepsilon_i$ as $\delta\varepsilon_i$
we obtain:
\begin{equation}
\delta\varepsilon_1 =\bar{\alpha}T \;\; ,
\label{T.13}
\end{equation}
\begin{equation}
\delta\varepsilon_2 =-\bar{\alpha}U/4s^2 \;\; ,
\label{T.14}
\end{equation}
\begin{equation}
\delta\varepsilon_3 =\bar{\alpha}S/4s^2 \;\; .
\label{T.15}
\end{equation}

From (\ref{T.10}) - (\ref{T.12}) we get:
\begin{equation}
T=\frac{3}{16\pi s^2 c^2}\delta V_A \;\; ,
\label{T.16}
\end{equation}
\begin{equation}
U=-\frac{3}{4\pi(c^2 -s^2)}[(\delta V_A -\delta V_m)-2s^2(\delta V_A -\delta
V_R)] \;\; ,
\label{T.17}
\end{equation}
\begin{equation}
S=\frac{3}{4\pi}(\delta V_A -\delta V_R) \;\; ,
\label{T.18}
\end{equation}
where $\delta V_i$ are New Physics contributions to $V_i$.

According to \cite{64}
\begin{equation}
S= 16\pi [\Sigma'_{33}(0)-\Sigma'_{3Q}(0)]=16\pi [\Sigma'_A(0)-\Sigma'_V(0)]
\;\; ,
\label{T.19}
\end{equation}
\begin{equation}
T= \frac{4\pi}{s^2m^2_W} [\Sigma_{11}(0)-\Sigma_{33}(0)] \;\; ,
\label{T.20}
\end{equation}
\begin{equation}
U= 16\pi [\Sigma'_{11}(0)-\Sigma'_3(0)] \;\; ,
\label{T.21}
\end{equation}
where $\Sigma'(0)=d\Sigma(q^2)/dq^2|_{q^2 =0}$ and $\Sigma$'s are defined by
the corresponding currents (isotopic, 1 and 3, and electromagnetic, $Q$,
vector, $V$, and axial, $A$) with coupling constants being extracted. Thus
$S$ characterizes the degree of chiral symmetry breaking, while both $T$ and
$U$ that of isotopic symmetry. Note that in eqs.(\ref{T.19}) - (\ref{T.21})
only the contribution of New Physics should be considered.
Since new particles should be heavy it is reasonable to take into account
only values of self-energies at $q^2 =0$ and their first derivatives (higher
derivatives are power suppressed) \cite{64}. Altogether we have eight
parameters ($\Sigma_{WW}(0)$, $\Sigma_{ZZ}(0)$, $\Sigma_{\gamma Z}(0)$,
$\Sigma_{\gamma\gamma}(0)$, $\Sigma'_{WW}(0)$, $\Sigma'_{ZZ}(0)$,
$\Sigma'_{Z\gamma}(0)$, $\Sigma'_{\gamma\gamma}(0)$), two of which are equal
zero ($\Sigma_{\gamma\gamma}(0)$ and $\Sigma_{\gamma Z}(0)$), while three
combinations can be absorbed in the definition of $\alpha$, $G_{\mu}$ and
$m_Z$. The remaining three combinations enter $S$, $T$ and $U$ (or
$\delta\varepsilon_i$, $i=1,2,3$).

\newpage


\noindent {\bf Figure Captions}

Figure 1: Muon decay in the tree approximation.

Figure 2: The $Z$ boson as a resonance in $e^+ e^-$ annihilations.

Figure 3: Photon polarization of the vacuum, resulting in the
          logarithmic running of the electromagnetic charge 
          $e$ and the `fine structure constant' 
          $\alpha \equiv \frac{e^2}{4\pi}$,
          as a function of $q^2$, where $q$ is the 4-momentum 
          of the photon (a). 
          Some of the diagrams that contribute to the 
          self-energy of the $W$-boson (b)-(g).
          Some of the diagrams that contribute to the self-energy of the
          $Z$-boson (h)-(n). 
          Some of the diagrams that contribute to the
          $Z \leftrightarrow \gamma$ transition (o)-(r).

Figure 4: Vertex triangular diagrams in the
          $Z \to l\bar{l}$ decay (a), (b), (c). 
          Loops that renormalize the lepton wavefunctions in the
          $Z \to l\bar{l}$ decay. 
          (Of course, antilepton have similar loops.)  (d), (e). 
          Types of diagrams that renormalize the $Z$ boson wavefunction 
          in the $Z \to l\bar{l}$ decay  (f), (g). 
          Virtual particles in the loops are those that we were discussing
          above. 

Figure 5: Virtual $t$-quarks (a) and $W$ bosons (b), (c) in
          the photon polarization of the vacuum.

Figure 6: Gluon corrections to the electroweak quark loop of the Z-boson 
          self-energy.

Figure 7. $V_m$ as a function of $m_t$ for three values of $m_H$.
The dotted parabola corresponds to Veltman approximation: $V_m = t$.
Solid horizontal line traces the experimental value of  $V_m$
while the dashed horizontal lines give its upper and lower limits
at the $1\sigma$ level.

Figure 8. $V_A$ as a function of $m_t$.
The dotted parabola corresponds to Veltman approximation: $V_A = t$.
Solid horizontal line traces the experimental value of  $V_A$
while the dashed horizontal lines give its upper and lower limits
at the $1\sigma$ level.

Figure 9: $V_R$ as a function of $m_t$.
The dotted parabola corresponds to Veltman approximation: $V_R = t$.
Solid horizontal line traces the experimental value of  $V_R$
while the dashed horizontal lines give its upper and lower limits
at the $1\sigma$ level.

Figure 10: $m_t - m_H$ exclusion plots with assumptions of:
a) $m_t = 150(5)$ GeV;
b) $m_t=175(5)$ GeV; c) $m_t = 200\pm 5$ GeV.

Figure 11: The vertex electroweak diagrams involving t-quark and
 contributing to
the $Z \to b \bar{b}$ decay. Diagram (b) represents gluon corrections
to  diagram (a).

Figure 12: Some of Feynman diagrams that give $\alpha_W^2 t^2$ corrections.

Figure 13: $\chi^2$ {\it vs.} $m_H$ curves.

Figure 14: The 2-dimensional exclusion plot for the case of $N$
 extra generations and for the choice  $m_D=130$ GeV -- the lowest allowed
value for new quark mass from Tevatron search \cite{13},
using $m_H > 90$ GeV at 95 \% C.L. from LEP-2 \cite{lep2}.
Little cross corresponds to $\chi^2$ minimum; lines show one sigma,
two sigma, etc allowed domains.

Figure 15: Contribution of $\tilde{t}$ and $\tilde{b}$ squarks into
  W- and Z bosons self-energy.

Figure 16: Values of $\delta V_A$, $\delta V_R$ and $\delta V_m$ at
$m_{\tilde{b}} = 200$ GeV.


\newpage
\begin{figure}
\begin{center}
\epsfysize= 100pt
\epsfbox{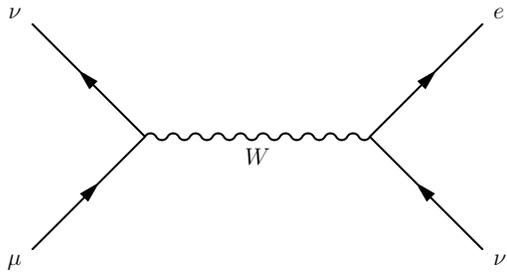}
\end{center}
\caption{Muon decay in the tree approximation}
\end{figure}

\begin{figure}
\begin{center}
\epsfysize= 100pt
\epsfbox{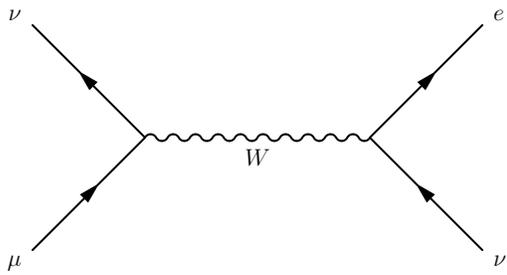}
\end{center}
\caption{The Z-boson as a resonance in $e^+e^-$ annihilations.}
\end{figure}


\newpage
\begin{figure}
\begin{center}
\epsfysize= 460pt
\epsfbox{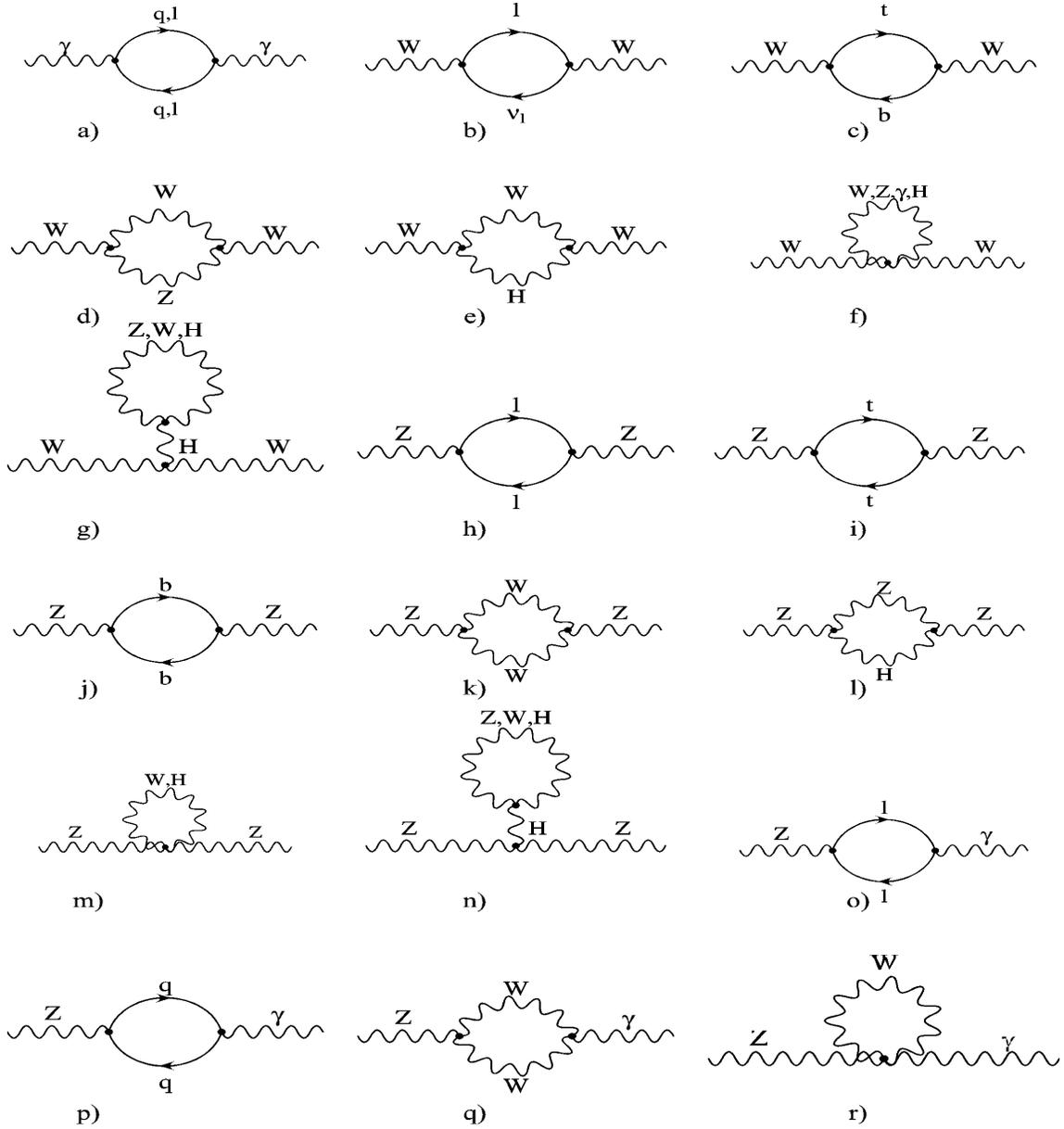}
\end{center}
\caption{ Photon polarization of the vacuum, resulting in the
logarithmic running of the electromagnetic charge 
$e$ and the `fine structure constant' $\alpha \equiv
\frac{e^2}{4\pi}$, as a function of $q^2$, where $q$ is the 4-momentum 
of the photon (a). 
Some of the diagrams that contribute to the 
self-energy of the $W$-boson (b)-(g).
Some of the diagrams that contribute to the self-energy of the
$Z$-boson (h)-(n). 
Some of the diagrams that contribute to the $Z \leftrightarrow
\gamma$ transition (o)-(r).}
\end{figure}

\newpage
\begin{figure}
\begin{center}
\epsfysize= 480pt
\epsfbox{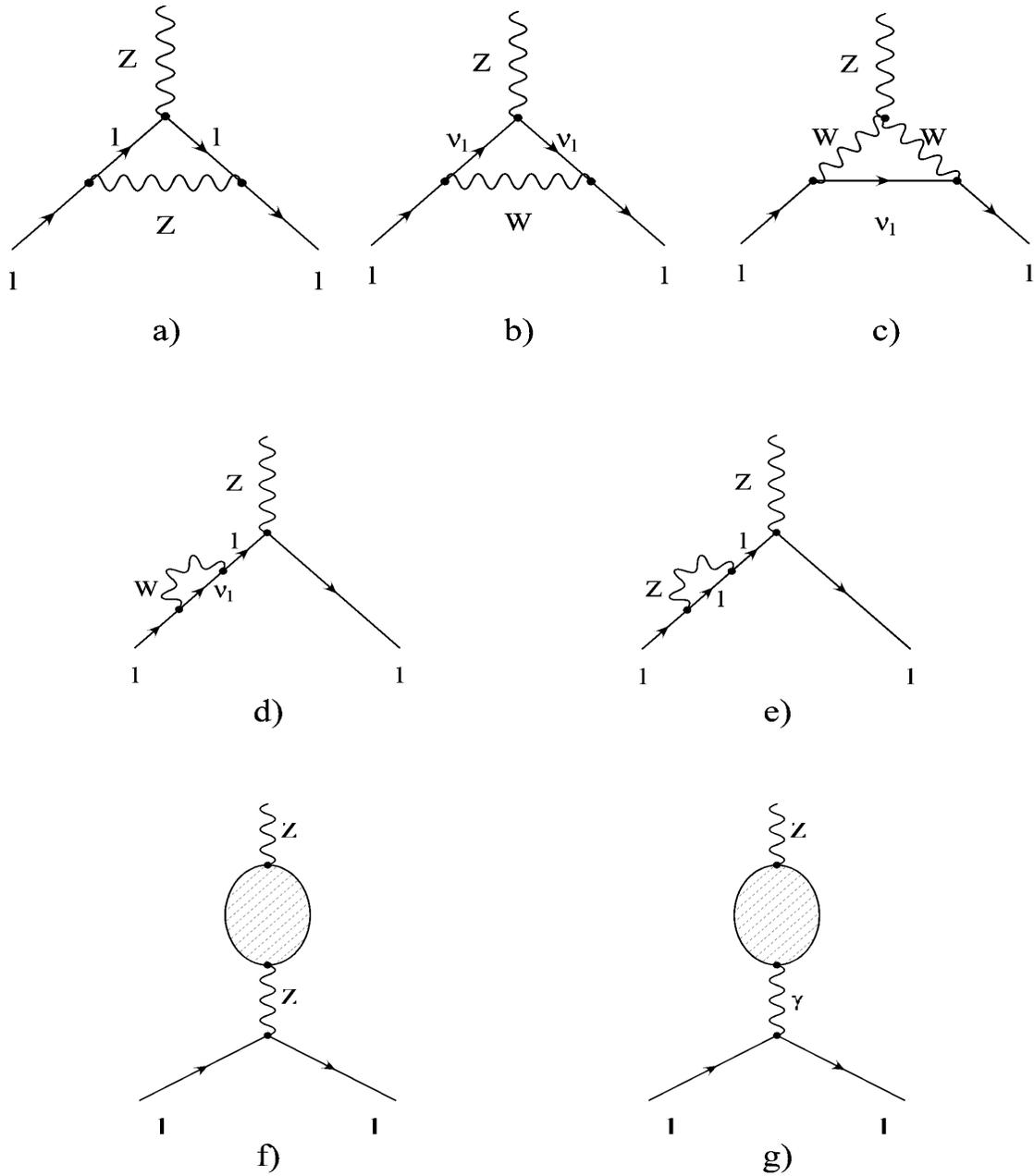}
\caption{Vertex triangular diagrams in the $Z \to
l\bar{l}$ decay (a), (b), (c). 
Loops that renormalize the lepton wavefunctions in the $Z \to l\bar{l}$
decay. 
(Of course, antilepton have similar loops.)  (d), (e). 
Types of diagrams that renormalize the $Z$ boson wavefunction 
in the $Z \to l\bar{l}$ decay  (f), (g). 
Virtual particles in the loops are those that we were discussing above.}
\end{center}
\end{figure}

\newpage
\begin{figure}
\begin{center}
\epsfysize= 240pt
\epsfbox{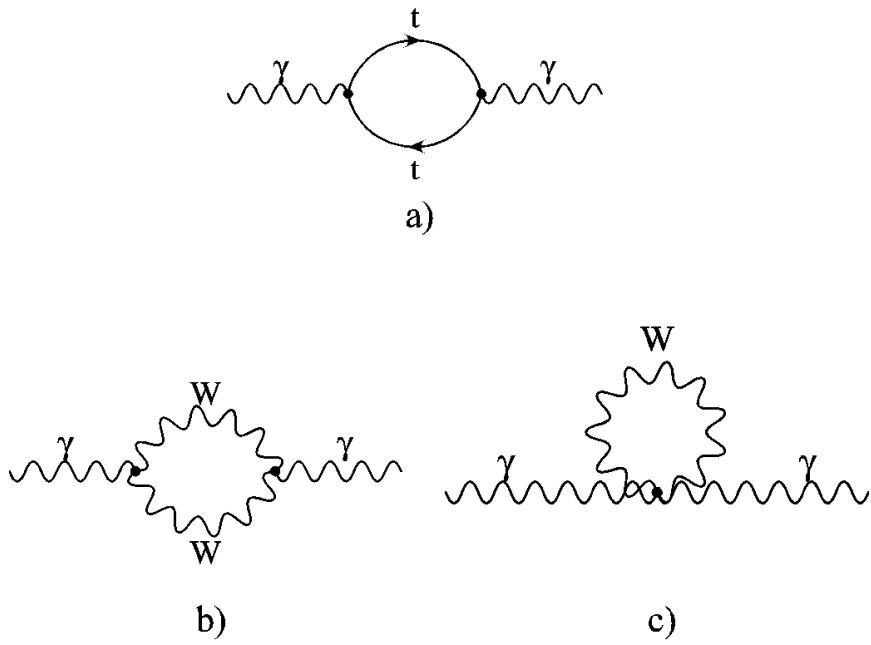}
\caption{ Virtual $t$-quarks (a) and $W$ bosons (b), (c) in
the photon polarization of the vacuum.}
\end{center}
\end{figure}

\begin{figure}
\begin{center}
\epsfysize= 600pt
\epsfbox{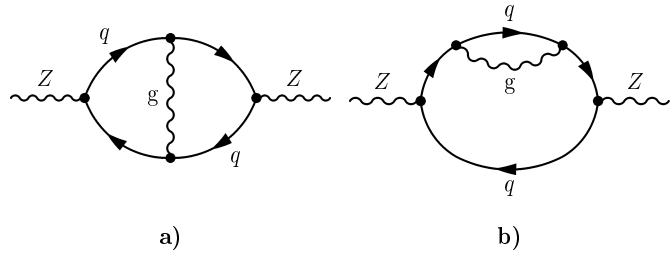}
\caption{Gluon corrections to the electroweak quark loop of the Z-boson 
self-energy.}
\end{center}
\end{figure}

\begin{figure}
\begin{center}
\epsfysize= 480pt
\epsfbox{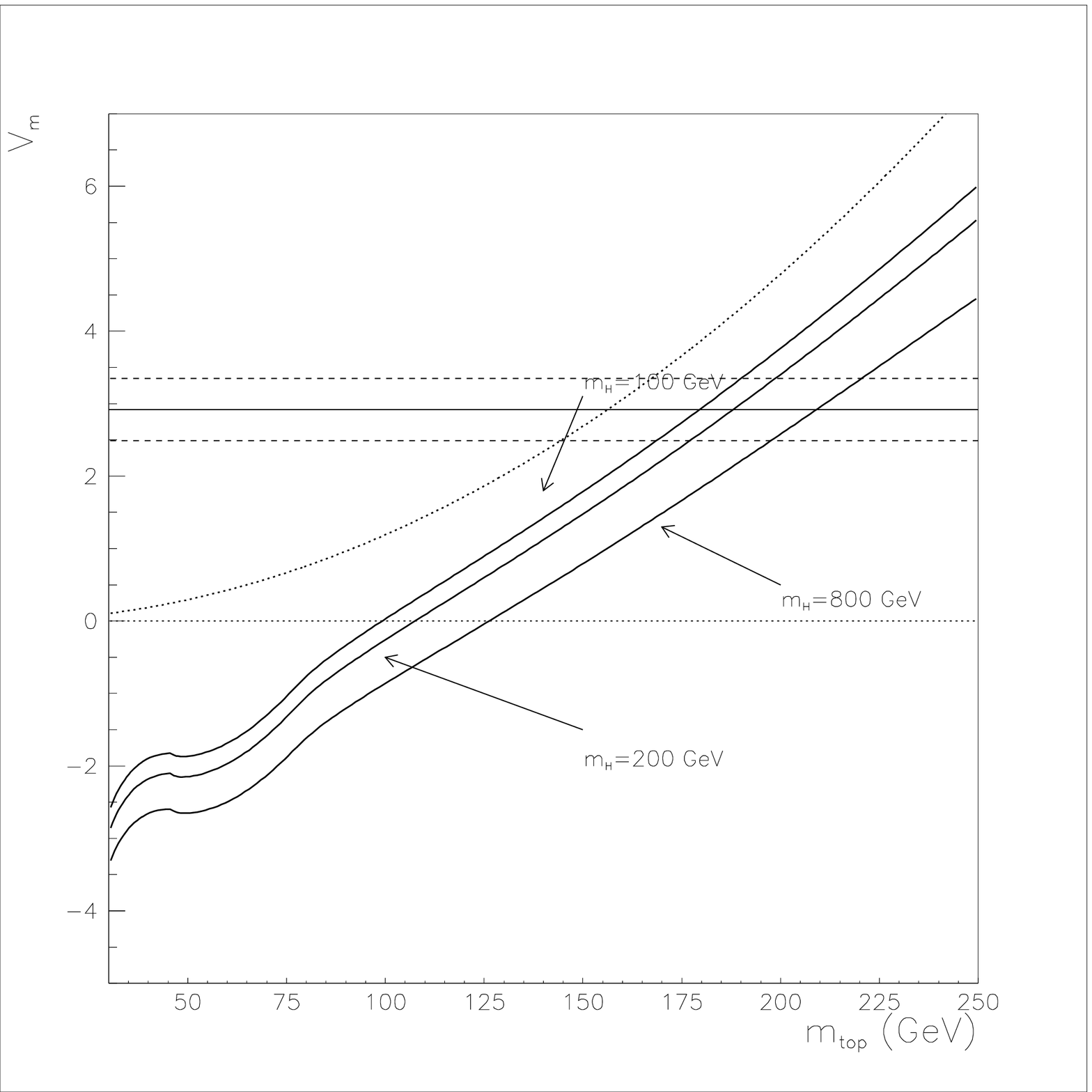}
\caption{$V_m$ as a function of $m_t$ for three values of $m_H$.
The dotted parabola corresponds to Veltman approximation: $V_m = t$.
Solid horizontal line traces the experimental value of  $V_m$
while the dashed horizontal lines give its upper and lower limits
at the $1\sigma$ level.
}
\end{center}
\end{figure}

\begin{figure}
\begin{center}
\epsfysize=480pt
\epsfbox{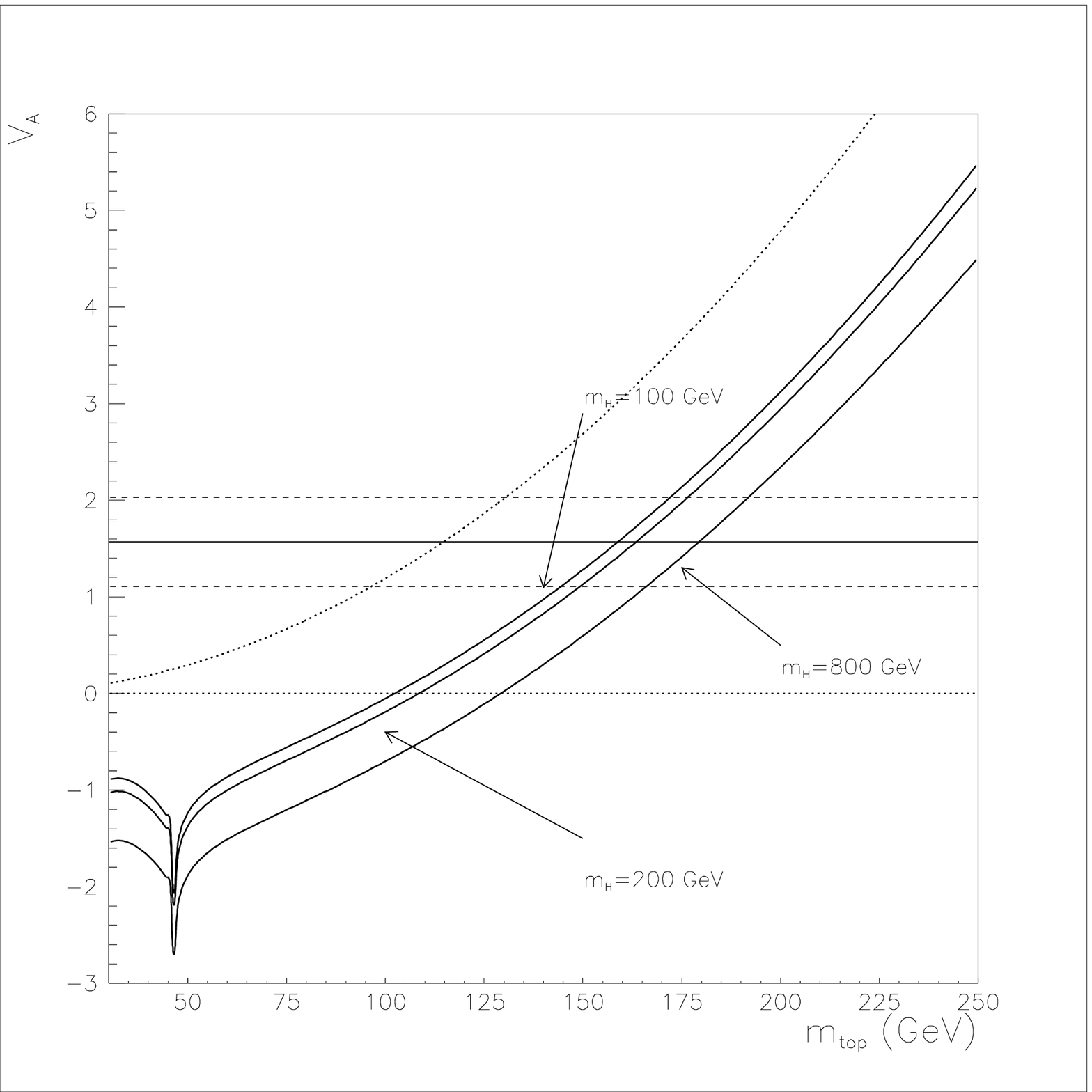}
\caption{$V_A$ as a function of $m_t$.
The dotted parabola corresponds to Veltman approximation: $V_A = t$.
Solid horizontal line traces the experimental value of  $V_A$
while the dashed horizontal lines give its upper and lower limits
at the $1\sigma$ level.
}
\end{center}
\end{figure}

\begin{figure}
\begin{center}
\epsfysize=480pt
\epsfbox{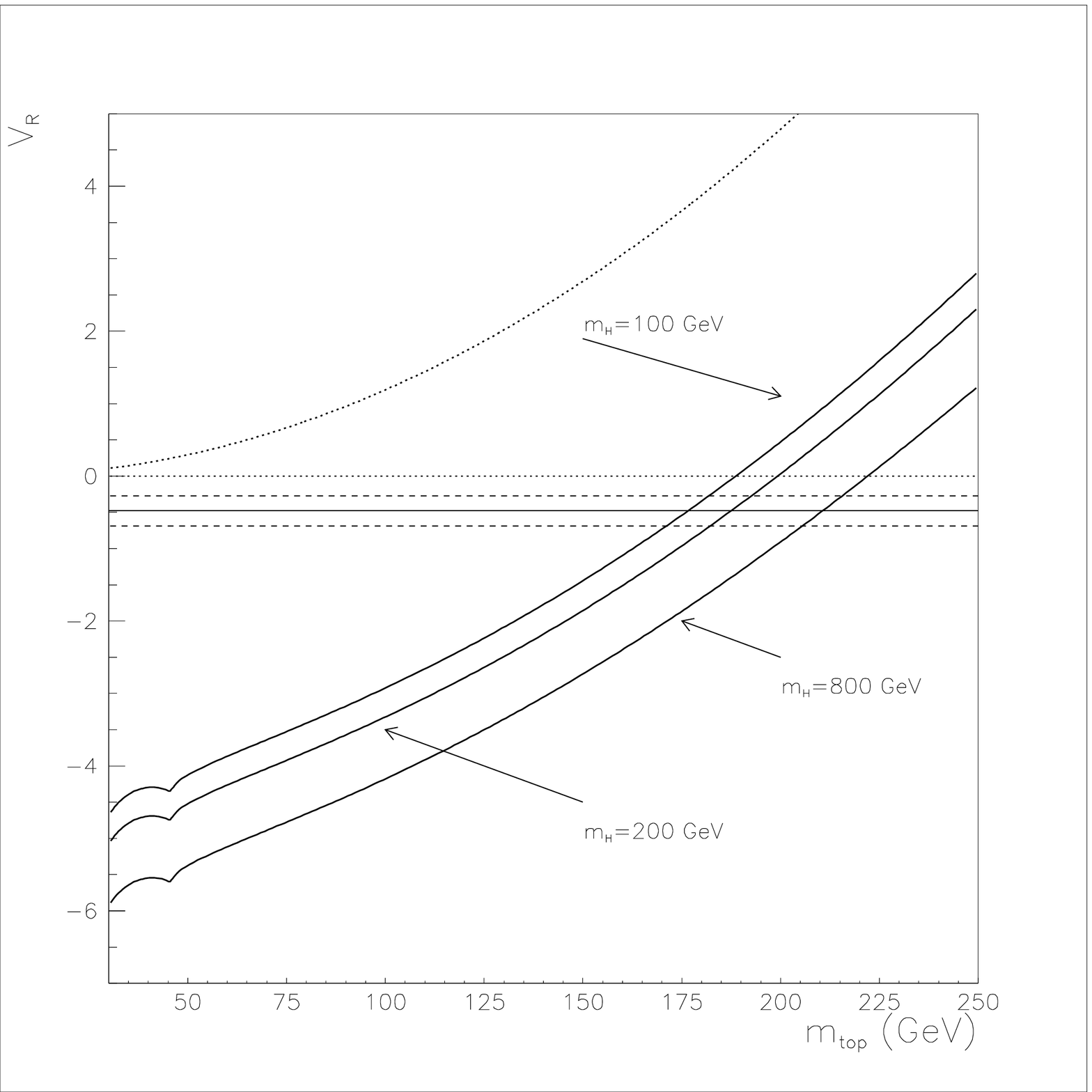}
\caption{$V_R$ as a function of $m_t$.
The dotted parabola corresponds to Veltman approximation: $V_R = t$.
Solid horizontal line traces the experimental value of  $V_R$
while the dashed horizontal lines give its upper and lower limits
at the $1\sigma$ level.
}
\end{center}
\end{figure}

\begin{figure}
\begin{center}
\subfigure[]{
\epsfysize=230pt
\epsfbox{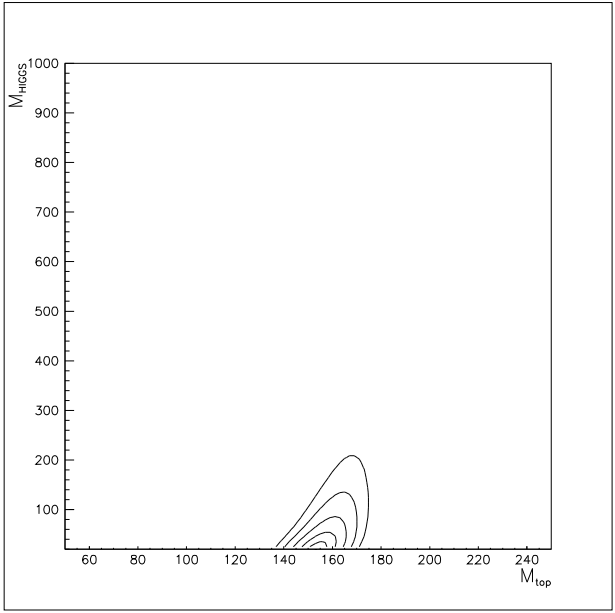}
}
\subfigure[]{
\epsfysize=230pt
\epsfbox{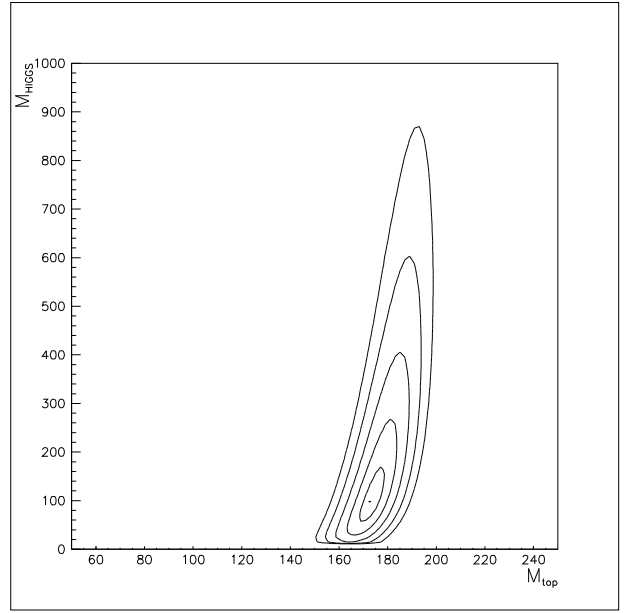}
}
\subfigure[]{
\epsfysize=230pt
\epsfbox{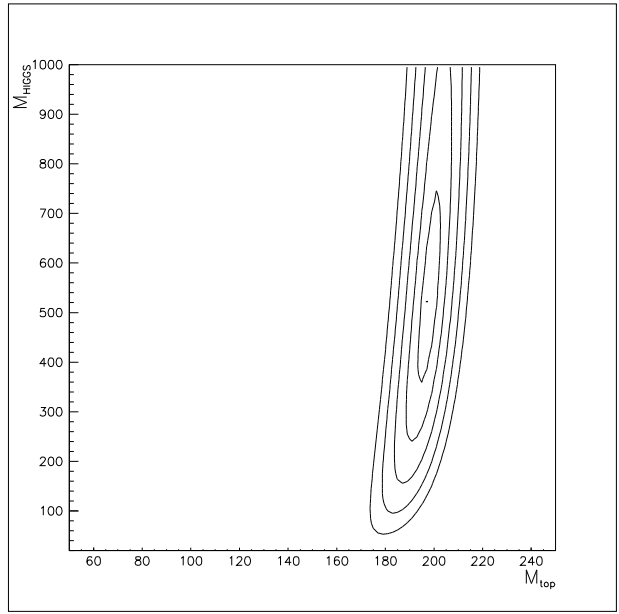}
}
\caption{ $m_t-m_H$ exclusion plots with assumptions of:
a)$m_t=150(5)$ GeV; b) $m_t=175(5)$ GeV; c)$m_t=200(5)$ GeV.}
\end{center}
 \end{figure}

\begin{figure}
\begin{center}
\epsfysize=600pt
\epsfbox{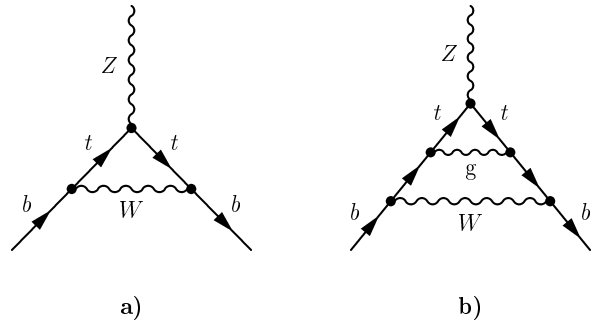}
\caption{The vertex electroweak diagrams involving t-quark and contributing to
the $Z \to b\bar{b}$ decay. Diagram (b) represents gluon corrections
to  diagram (a).}
\end{center}
\end{figure}

\begin{figure}
\begin{center}
\epsfysize=600pt
\epsfbox{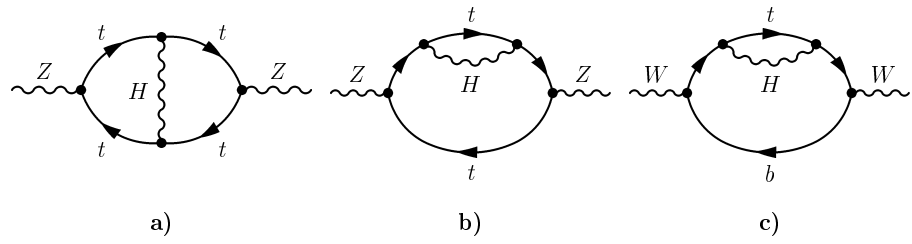}
\caption{Some of Feynman diagrams that give $\alpha_W^2 t^2$ corrections.
}
\end{center}
\end{figure}

\begin{figure}
\begin{center}
\epsfysize=480pt
\epsfbox{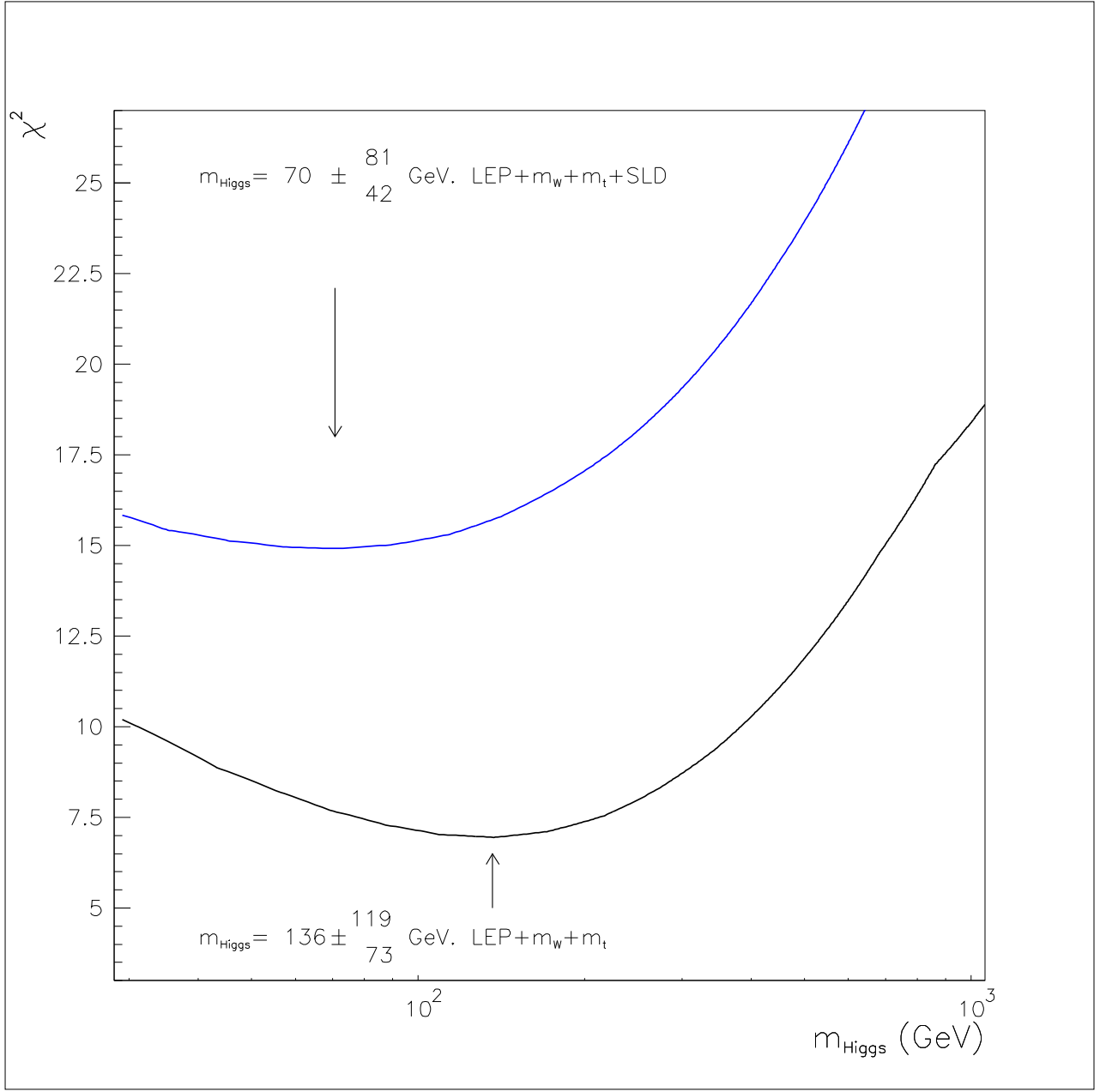}
\caption{$\chi^2$ {\it vs.} $m_H$ curves.}
\end{center}
\end{figure}

\begin{figure}
\begin{center}
\epsfysize=480pt
\epsfbox{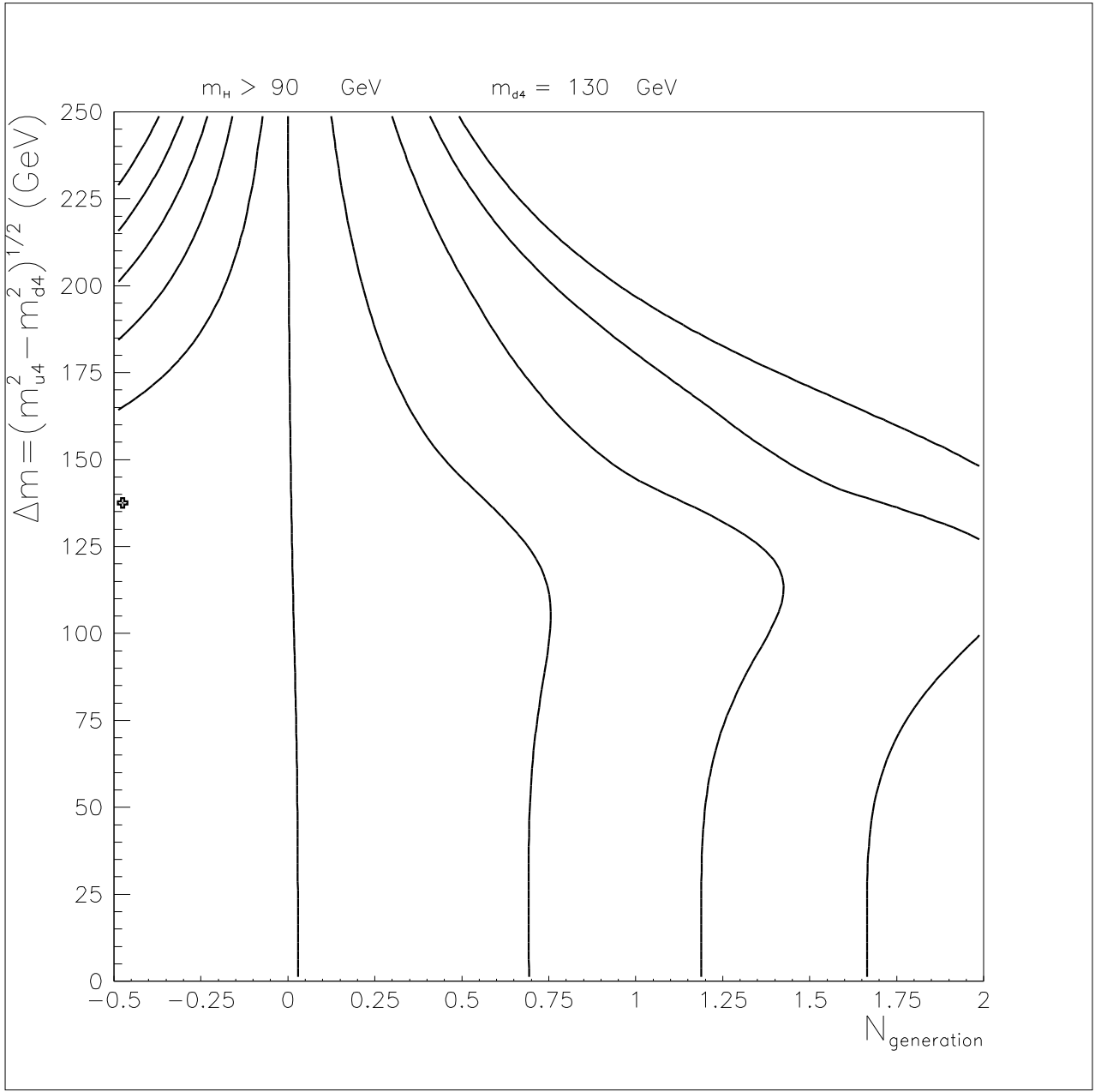}
\caption{ The 2-dimensional exclusion plot for the case of $N$
 extra generations and for the choice  $m_D=130$ GeV -- the lowest allowed
value for new quark mass from Tevatron search(\cite{13}),
using $m_H > 90$ GeV at 95 \% C.L. from LEP-2 \cite{lep2}.
Little cross corresponds to $\chi^2$ minimum; lines show one sigma,
two sigma, etc allowed domains.
}
\end{center}
\end{figure}

\begin{figure}
\begin{center}
\epsfysize=600pt
\epsfbox{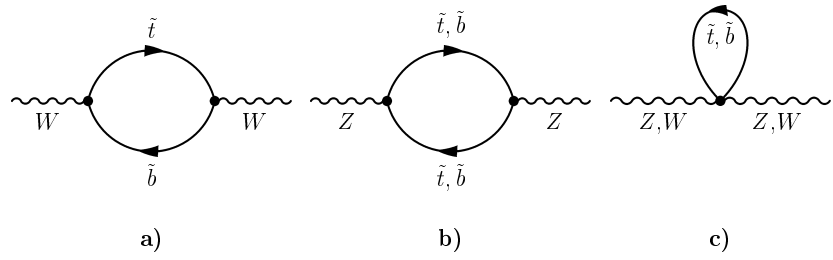}
\caption{Contribution of $\tilde{t}$ and $\tilde{b}$ squarks into
  W- and Z bosons self-energy.
}
\end{center}
\end{figure}

\begin{figure}
\epsfysize=190pt
\epsfbox{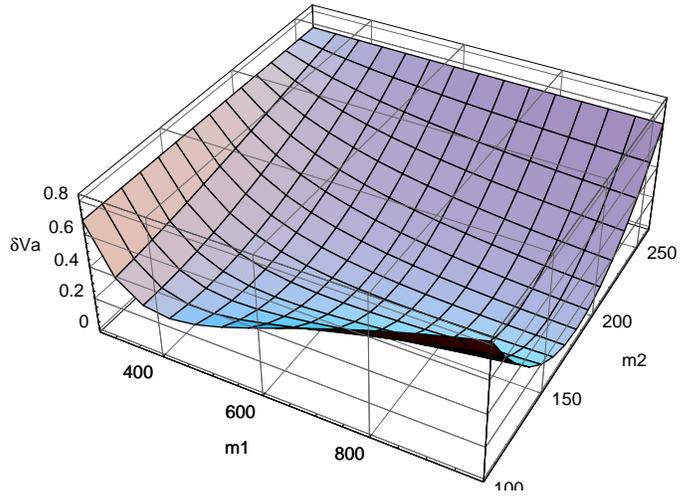}
\epsfysize=190pt
\epsfbox{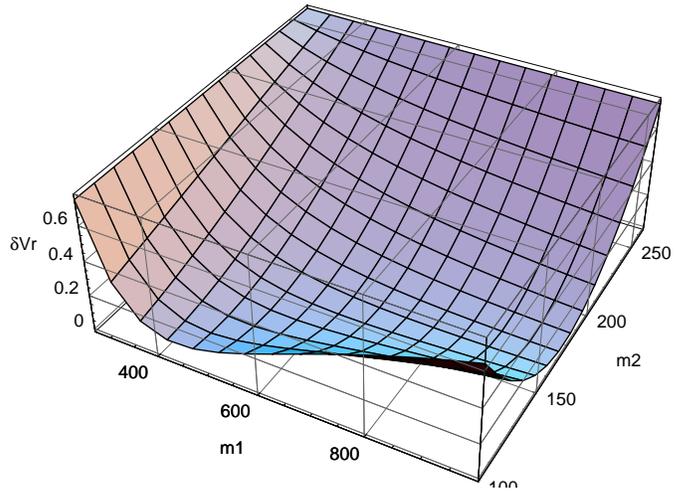}
\epsfysize=190pt
\epsfbox{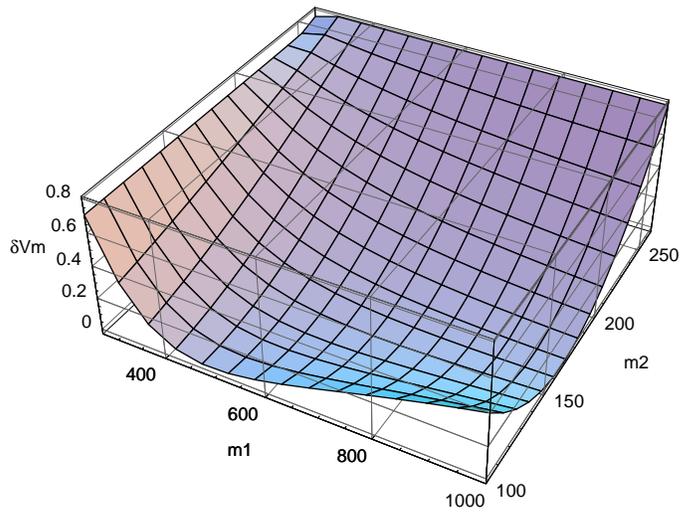}
\caption{Values of the
$\delta V_A$, $\delta V_R$ and $\delta V_m$ at $m_{\tilde{b}} = 200$ GeV.}
\end{figure}

\end{document}